\documentclass[11pt,oneside,english]{book}
\usepackage[T1]{fontenc}
\usepackage[latin1]{inputenc}
\usepackage{geometry}
\usepackage{subfigure}
\usepackage{amsmath}
\usepackage{graphicx}
\usepackage{setspace}
\usepackage{amssymb}

\makeatletter

\providecommand{\LyX}{L\kern-.1667em\lower.25em\hbox{Y}\kern-.125emX\@}
\let\SF@@footnote\footnote
\def\footnote{\ifx\protect\@typeset@protect
    \expandafter\SF@@footnote
  \else
    \expandafter\SF@gobble@opt
  \fi
}
\expandafter\def\csname SF@gobble@opt \endcsname{\@ifnextchar[
  \SF@gobble@twobracket
  \@gobble
}
\edef\SF@gobble@opt{\noexpand\protect
  \expandafter\noexpand\csname SF@gobble@opt \endcsname}
\def\SF@gobble@twobracket[#1]#2{}

\usepackage{babel}
\makeatother
\begin{document}

\author{By Ross O'Connell, Advisor Professor William Loinaz}

\title{A Foray into Quantum Dynamics%
\footnote{Submitted to the Department of Physics of Amherst College in partial
fulfillment of the requirements for the degree of Bachelor of Arts
with Departmental Honors.%
}}

\maketitle

\chapter*{Acknowledgments}

First, I thank my parents for being endless sources of love and support.
I hope that they appreciate all of the shades of {}``I wouldn't be
here if it weren't for you.''

After family, friends. You know who you are. You've all lent color
and depth to my time here, especially this year.

I must, of course, thank my advisor, Will, for helping me see this
problem from all sides, for optimism and enthusiasm. You've helped
keep this problem from getting me down.

And, finally, Sarah. Because as much as I like my thesis, I go to
sleep and wake up thinking of you. As much as I enjoyed working on
it, you make me happy.

\chapter*{Abstract}

\section*{An Explanation of my Thesis in Words of One Syllable}

\emph{With apologies to Paul Boolos.}\\
\\
My goal for the year was to learn what small things do. If you have
a small thing it's not a point, it's a blob -- we knew that from the
start. But does it stay a blob? No, of course not. It breaks up, but
at some time it comes back. At times less than that time, we find
small blobs, too. When it's not made of small blobs, the small thing
is most odd. It forms lots of hills and troughs. I have ways in here
to find out when the blob comes back, where and when the small blobs
are, how long it takes for the blobs to break up, where the hills
and troughs are, and how many of them there ought to be.

\section*{The Same Explanation with Longer Words}

The dynamics of a quantum mechanical particle in a time-independent
potential are found to contain many interesting phenomena. These are
direct consequences of the (typical) existence of more than one time-scale
governing the problem. This gives rise to full revivals of initial
wavepackets, fractional revivals (multiple wavepackets appearing at
fractions of the revival time), and the striking quantum carpets.
A variety of analytic techniques are used to consider the interference
that gives rise to these phenomena while skirting calculations involving
cross-terms. Novel results include a new theorem on the weighting
coefficients $a_{m}$ that govern fractional revivals, a demonstration
that $\Psi _{cl}$, the function that governs the distribution and
features of these fractional revivals, really does behave classically,
a treatment of wavepacket dephasing in the infinite square well by
means of the Poisson summation formula, and a correct analysis of
the spatial distribution of intermode traces. Also, this work presents
a coherent treatment of these phenomena, which before now did not
exist.

\tableofcontents{}

\chapter{Introduction}

\newcommand{\Erf}[1]{\textrm{Erf}\left(#1\right)}

\newcommand{\re}[1]{\textrm{Re}\left\{ #1\right\} }

\newcommand{\im}[1]{\textrm{Im}\left\{ #1\right\} }

\newcommand{\row}[2]{\begin{array}[t]{c}
 \begin{array}[t]{r}
 \left(#1\right.\\
 \left.#2\right).\end{array}
\end{array}
}

\newcommand{\rowc}[2]{\begin{array}[t]{r}
 \left(#1\right.\\
 \left.#2\right),\end{array}
}

\newcommand{\Z}{\in \mathbb{Z}}

\newcommand{\p}[1]{\left(#1\right)}

\section{A Hitchhiker's Guide to the Infinite Square Well}

\begin{figure}[hbtp]
\begin{center}\includegraphics[  height=0.45\textheight]{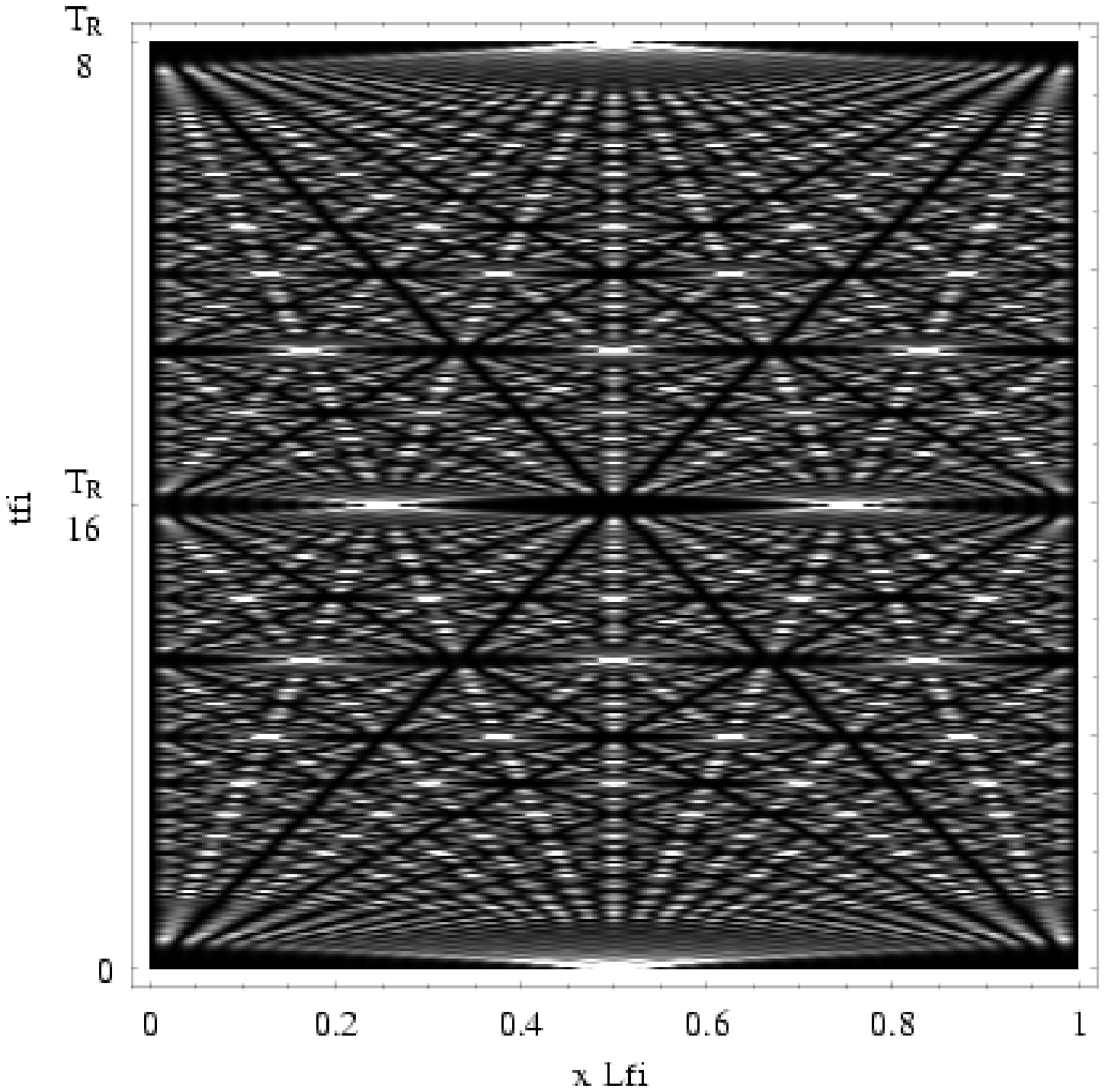}\end{center}

\caption{\label{fig:first}A particle in the infinite square well. Lighter
areas indicate greater probability density. Our initial wavefunction
is a Gaussian packet, centered on $x=L/2$, with $\sigma _{x}=0.003L$.
For the definition of $T_{R}$, see Chapter \ref{ch: rev}.}
\end{figure}
The infinite square well is treated in almost every introductory quantum
mechanics course. It is the essence of simplicity, a potential bereft
of features, its eigenfunctions are simple sine waves. Exact solutions
abound. We build an intuition for how quantum mechanical objects should
behave, and we move on -- after all, the infinite square well is just
too simple.

Figure \ref{fig:first} was enough to convince me that I didn't really
understand the square well. I had no idea that its probability density
was so structured -- it hadn't even occurred to me that it was periodic.
So I began the year of study that culminated in this thesis, following
much the same pattern as before: look at the square well, find an
interesting phenomenon, study it, then see if it occurs in other types
of systems%
\footnote{I focused on time-independent potential wells, in one dimension.%
}. While there are applications for most of these phenomena, I won't
pretend that my goal was to find more of them -- my goal was to gain
a deeper understanding of a fundamental part of quantum mechanics.

\subsection{Classical Oscillation and Dispersion}

\vspace{0.3cm}
\begin{center}\includegraphics[  width=0.80\textwidth]{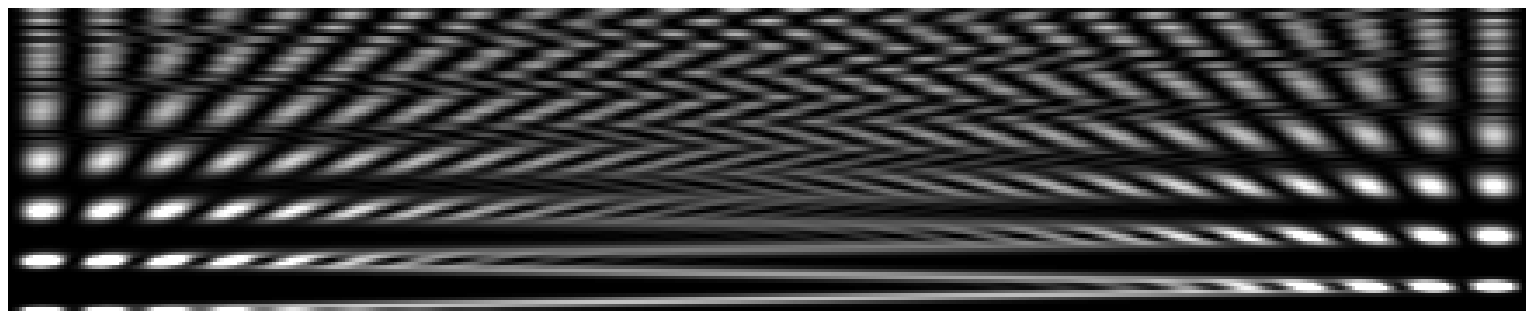}\end{center}
\vspace{0.3cm}

The first interesting phenomenon is that of the classical oscillation
of wavepackets. Given a well-localized wavepacket, or even a wavefunction
with several distinct peaks, those peaks will undergo something like
classical oscillation on a time scale that will be studied in Chapter
\ref{ch: rev}. This is not particularly surprising if we consider
Ehrenfest's Theorem%
\footnote{See \cite[17]{misc:Griffiths1}.%
},\begin{equation}
\frac{d\left\langle p\right\rangle }{dt}=\left\langle -\frac{\partial V}{\partial x}\right\rangle .\end{equation}
So long as the expectation value of $p$ coincides with the value
of $p$ at the center of the wavepacket, the packet's oscillation
will be classical. Of course, this is rarely exactly true and virtually
never stays true -- when the expectation value of $p$ and the value
of $p$ at the center of the wavepacket do not coincide, the wavepacket
will begin to disperse, and more characteristically quantum mechanical
phenomena will begin to appear. An obvious question to ask is, {}``can
we quantify when this dispersion occurs?'' A related question is
{}``does this dispersion occur simultaneously for the whole wavefunction,
or does it depend on which point in the wavefunction we are considering?''
This question will be considered for a zero-dimensional {}``quantum
beats'' system in Chapter \ref{ch: beats}, and in more dimensions
in Chapter \ref{ch: rev}.

\subsection{Revivals}

\vspace{0.3cm}
\begin{center}\includegraphics[  width=0.80\textwidth]{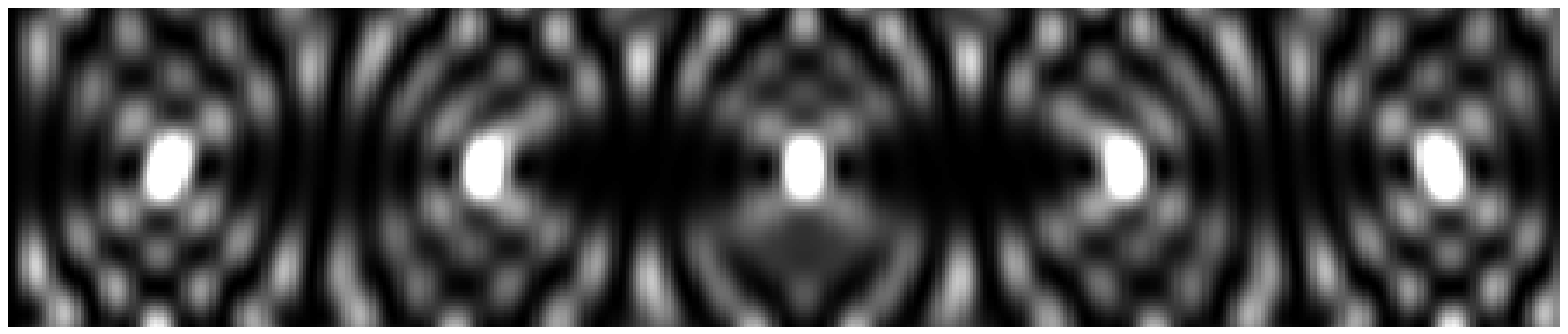}\end{center}
\vspace{0.3cm}

The next phenomenon that we will consider is closely related to the
previous one. It turns out that just as there is a time scale that
governs classical oscillation, there is another that governs a distinctly
quantum mechanical feature. At this time scale, the original configuration
of the wavefunction reappears, in some cases approximately, in others
exactly. That is, if we start with a localized wavepacket it will
disperse, interfere, and eventually reconstruct itself -- what I will
call a revival. As part of this process, it will also undergo the
same sort of initial classical oscillation that it undergoes immediately
after its formation. Stranger still are the existence of fractional
revivals, the appearance at times less than the revival time of multiple
wavepackets, each related to the original wavepacket. Curiously, both
full and fractional revivals became topics of study only recently,
even though their existence should be apparent from the $\exp \p{-iEt/\hbar }$
form of the time evolution terms of the wavefunction. This subject
is the focus of Chapter \ref{ch: rev}, and plays a part in Chapters
\ref{ch: beats} and \ref{ch: pois} as well.

\subsection{Quantum Carpets}

\begin{center}\includegraphics[  width=0.80\textwidth]{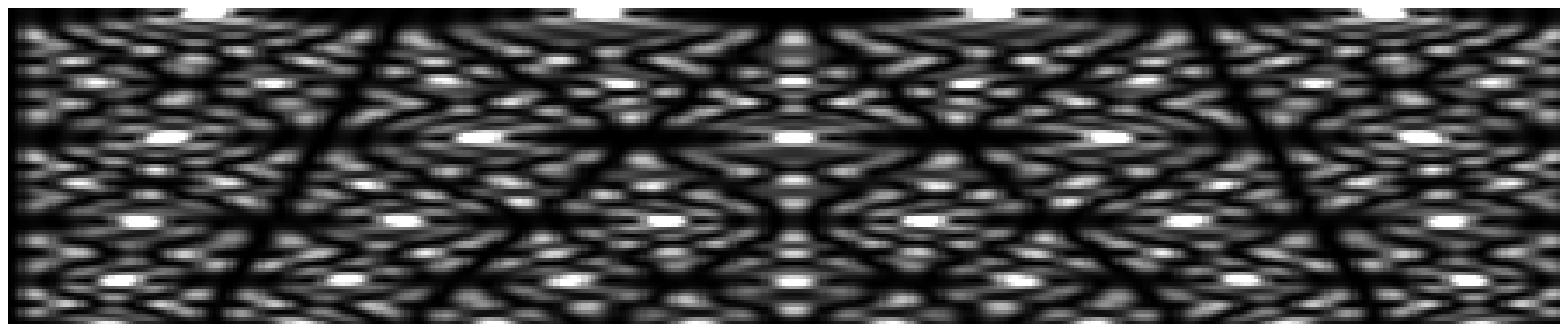}\end{center}

Perhaps the most striking phenomenon is the existence of easily-discerned
canals and ridges in the probability density, collections of which
have come to be called quantum carpets. While these follow straight
lines in the square well, they various curved lines in other potentials.
These lines look much like, but are distinct from, classical trajectories.
In Chapter \ref{ch: traces} we will see how the two are related.
Carpets happen to be, I think, the most difficult phenomena to study.
While the revivals problem, and with it the problem of classical oscillation
and dispersion, can be reduced to a clever shuffling of phases, it
is rather harder to avoid dealing with the entire parameter space
of spectrum, eigenfunctions, and particular weighting coefficients.
Considering this, it is remarkable that we are able to make interesting
statements at all.

\section{An Analytic Theme: Interference Without Cross-Terms}

Last Fall I asked a friend of mine, a math major, if he knew anything
about simplifying finite sums. {}``No,'' he replied, {}``I think
you just have to add those up.'' The finite sums that I was asking
about don't differ so much from the infinite sums that also appear
in this family of problems, and they're ugly. I was a bit intimidated.
One interesting aspect of these problems is that they all pertain
to features that we look for \emph{after} {}``solving'' the problem.
That is, we start with a set of eigenfunctions, a spectrum, and weighting
coefficients, then try to rewrite them to make the features we are
interested in manifest. Fortunately, we do have some analytic tools
to do this.

All of the phenomena that we are interested in are essentially interference
phenomena, produced by the multiplication of two sums. The critical
step in every technique presented here is the packaging of the interference
information in a form that doesn't involve cross-terms. These techniques
include the grouping of certain parts of the sum by phase (Chapter
\ref{ch: rev} and \ref{ch: traces}) and the application of a bit
of analysis, the Poisson summation formula (Chapters \ref{ch: beats}
and \ref{ch: pois}). Both of these techniques seem initially to make
things worse by introducing new sums or integrals, but those new sums
and integrals are exactly what give us manageable forms for enough
of the interference to make sense of things.

\section{Semiclassical Physics, Distinctively Quantum Behavior}

What exactly constitutes a {}``semiclassical limit'' is not entirely
clear. Most of the time, we will take it to mean that the distribution
of weighting coefficients $n$ is centered around some central value,
$\bar{n}$, with some characteristic width $\Delta n$, and that the
hierarchy $1\ll \Delta n\ll \bar{n}$ holds -- i.e. that the packet
is localized in physical space. In some cases, we will relax this
to the WKB assumption that the potential changes much more slowly
than the relevant eigenfunctions oscillate.

It appears that all of the phenomena that I have studied appear most
naturally in the study of semiclassical wavepackets. Although the
revivals results%
\footnote{In Chapter \ref{ch: beats} we will consider the case of quantum beats,
a zero-dimensional problem that is interesting even when a small number
of states interfere with each other.%
}are quite general, they are most striking when we begin with a well-localized
initial packet. Otherwise, we just get a mess that occasionally looks
like the mess that we started with. Though we are able to solve the
infinite square well exactly with each of these techniques (and there
$\Delta n\approx \bar{n}\gg 0$ is enough to produce a carpet), our
generalizations mostly rely on the WKB approximation. Of course, since
we are studying interference phenomena it is not surprising that our
most interesting results appear when we have a large number of states
that can interfere with each other. Nothing that we have done rules
out the appearance of similar phenomena in superpositions of low-energy
states, but we have not found any.

What is remarkable is that we find so many distinctly quantum mechanical
phenomena in systems that are \emph{supposed} to be becoming more
like classical ones. Although we start with wavepackets that are more
localized than would be a superposition of, say, the bottom three
states in a well, and seem more particle-like, they exhibit bizarre
behavior. What is remarkable is that, in the case of carpets, we can
use a semiclassical tool to study this.

\section{The Role of Pictures}

I was able to talk to a lot of people, before my thesis kept me really
busy. One evening I was talking with a friend-of-a-friend who happened
to be a Chemistry major. She had been through Quantum Chemistry, but
was still curious about what things like the uncertainty principle
really \emph{meant}. I spent a few minutes waving my hands, then went
back to my room and picked up one of my plots of the square well.
It proved to be a much more effective way of demonstrating how quantum
mechanics can be deterministic, yet still produce so much uncertainty
-- this pattern \emph{is} the particle in the box. It's obvious that
it doesn't really have a velocity because it doesn't even have a trajectory.
I think that just as quantum mechanics seems haphazard and riddled
with uncertainty until one realizes that the wavefunction is the fundamental
object in the theory, pictures seem ill-advised only until one realizes
which objects to plot.

I was also attracted to this subject because of a talk that included
pictures. There were very few pictures in my otherwise superb quantum
mechanics textbook, and as a consequence there were things that I
didn't understand. It didn't bother me because I didn't know that
I didn't understand them. The reason that this field developed so
recently, and not before, is that these phenomena are immediately
apparent in pictures and hopelessly hidden in eigenfunction expansions
-- and, of course, we couldn't generate those pictures without computers.
I've included a lot of pictures in this thesis because I've found
that while they're not worth a thousand equations, they're a fine
substitute for a paragraph of my prose.

\section{A Bit of Background}

Much of this thesis is based on three fine papers: the original, authoritative
article on fractional revivals, \cite{rev:Averbukh1}, a strong treatment
of quantum beats, \cite{beats:Leichtle1,beats:Leichtle2}, and a well-conceived
(if roughly executed) paper on intermode traces, \cite{trace:Kaplan2,trace:Kaplan1}.

There are a variety of other papers available. Most studied are revival
phenomena, which have been treated in a variety of ways in a variety
of systems. Though the paper mentioned above is remarkable for its
clarity and generality, other interesting works include \cite{rev:Aronstein1,rev:Aronstein2,rev:Bluhm1,rev:Chen1,rev:Jie1,rev:Knopse1,rev:Loinaz1,rev:Naqvi1,rev:Rozmej1}.
A variety of approaches to quantum carpets have been proposed, though
most are restricted to the infinite square well. Some of the more
interesting of these involve an analysis in terms of Wigner functions,
which I have not pursued. In any case, interesting carpet papers include
\cite{car:Friesch1,car:Grossman1,car:Hall1,car:Marzoli1,car:Marzoli2}.
For a somewhat dated discussion of some experimental aspects of these
phenomena, see \cite{misc:Averbukh1}.

One of the more intriguing applications of quantum carpet techniques
is in the field of Bose-Einstein Condensation. Although I had little
time to devote to them, the growing literature suggests that this
is a subject of current interest: \cite{bec:Choi1,bec:Schleich1,bec:Wright1}.

Papers have been written which study the fractal geometry of various
problems, including \cite{frac:Berry1,frac:Wojcik1,frac:Wojcik2}.
The latter two focus on preparing wavepackets that are variations
on the Weierstrass function (which everywhere continuous, nowhere
differentiable), while the former focuses on the fractal dimension
of a particle with a smooth spatial distribution in an n-dimensional
box. The most useful insight to come from any of this, I think, is
that fractals proper only emerge from initial states that are not
proper solutions to the Schrödinger equation. I still suspect that
there may be some treatment of carpets that exploits their obvious
self-similarity, but the difficulty of fractal proofs combined with
the difficulty of working with almost-fractals turned me away from
this approach.

There have been attempts to connect the carpet problem to quantum
chaos, \cite{chaos:Provost,chaos:Saif1}. There is a sort of similarity
between the phenomena of quantum carpets and chaotic scars (representation
of particular classical orbits in the probability densities of a chaotic
potential), though little has yet come of this. There is also one
paper purporting to connect all of this to quantum computing, \cite{comp:Harter1},
though I must confess that I was unable to follow it.

The final connection, which I would have liked to spend more time
exploring, is between the probability density of a quantum mechanical
particle in one spatial dimension and and electromagnetic wave propagating
in two spatial dimensions, in the paraxial approximation. In that
field, revivals are instances of the Talbot effect, and fractional
revivals instances of the fractional Talbot effect. Apparently, the
square well problem, a toy problem in quantum mechanics, translates
into a wave guide problem of greater practical importance. For the
interested reader, the primary papers on this subject seem to be \cite{opt:Berry1,opt:Berry2,opt:Dubra1,opt:Lock1}.

\chapter{\label{ch: rev}Quantum Revivals, Full and Fractional}

\section{In the Beginning}

Our quantum mechanics course began with the Schrödinger equation,
\begin{equation}
i\hbar \frac{\partial \Psi }{\partial t}=-\frac{\hbar ^{2}}{2m}\frac{\partial ^{2}\Psi }{\partial x^{2}}+V\Psi ,\end{equation}
and promptly separated it, assuming that the potential $V$ depended
only on $x$:\begin{eqnarray}
\Psi (x,t) & = & \psi (x)f(t),\\
i\hbar \frac{1}{f}\frac{df}{dt} & = & E,\label{eq:time-dep}\\
-\frac{\hbar ^{2}}{2m}\frac{d^{2}\psi }{dx^{2}}+V\psi  & = & E\psi .
\end{eqnarray}
Since we could solve Equation \ref{eq:time-dep}, we did, \begin{equation}
f(t)=e^{-i\frac{E}{\hbar }t},\end{equation}
 and turned our attention to the Time-Independent Schrödinger Equation.
We quickly found that for bound particles the possible energies $E$
were quantized, making the general solution to the Schrödinger Equation
a weighted sum of the solutions we had found: \begin{equation}
\Psi (x,t)=\sum _{n}c_{n}\psi _{n}(x)e^{-i\frac{E_{n}}{\hbar }t}.\end{equation}

This eigenfunction expansion is useful for algebraic purposes, and
admittedly we often want to do things with the wavefunction once we've
found it. Unfortunately, the eigenfunction expansion conceals the
various interesting things that happen during the time-evolution of
a wavefunction. We'll start our analysis of these phenomena with the
simplest, quantum revivals.

\section{Fun Revival Facts}

Given a well-localized distribution of weighting coefficients $c_{n}$
centered around some mean value $\bar{n}$, we can perform a Taylor
expansion%
\footnote{From here onward, I will assume that $\hbar ^{2}=2m=1$.%
} of $E_{n}$ around $\bar{n}$:\begin{equation}
E_{n}\simeq E_{\bar{n}}+E_{\bar{n}}^{\prime }(n-\bar{n})+\frac{1}{2!}E_{\bar{n}}^{\prime \prime }(n-\bar{n})^{2}+\frac{1}{3!}E_{\bar{n}}^{\prime \prime \prime }(n-\bar{n})^{3}+....\end{equation}
We now define \begin{equation}
\frac{2\pi }{T_{j}}=\frac{E_{\bar{n}}^{\p{j}}}{j!},\label{eq:perrev}\end{equation}
a set of time scales. From these we pick out two important time scales,
a classical period $T_{cl}=T_{1}$ and the revival time $T_{R}=T_{2}$.
It is the existence of more than one time scale that makes time-evolution
in time-independent potentials interesting.

We ignore the $j\geq 3$ terms for two reasons -- they make our calculations
harder%
\footnote{If we really want to, we could consider the so-called {}``superrevivals.''
These aren't so absurd as to be unobservable, but the big qualitative
change in behavior comes from introducing the second time scale, $T_{R}$.%
}, and in the semiclassical case, which we are most interested in,
they really are ignorable. There are two ways to see this. First,
we assume that our $E_{n}$ can be written as a finite sum of weighted
powers of $n$, \begin{equation}
E_{n}=\sum _{m=N}^{M}d_{m}n^{m}.\end{equation}
This won't always be the case, but I think it's fair to consider transcendental
spectra to be unusual. From looking at $E_{n}$, we see that we can
write the $j$th derivative as \begin{equation}
E_{n}^{\p{j}}=\sum _{k=N}^{M}c_{k}n^{k}\times \p{\frac{1}{n^{j}}\prod _{m=k-j+1}^{k}m}.\end{equation}
We know that the $j$th characteristic time scale ($T_{cl}$ is the
first, $T_{R}$ is the second) will be proportional to $1/\left|E_{\bar{n}}^{\p{j}}\right|$.
We also know that in the semiclassical case, $\bar{n}\gg 1$. If $\bar{n}$
is larger than $k$, it will force the $E_{\bar{n}}^{\p{j}}$ terms
toward zero, which will in turn force the higher time scales toward
infinity. Another way to see this is to realize that the limiting
spectrum of any deep potential well is at most $n^{2}$. You can think
of this as a consequence of the fact that no well can have walls harder
than the infinite square well, but for a more detailed explanation
you should consult \cite{misc:Nieto1}. Finally, we observe that in
the semiclassical limit we are typically concerned with a small region
of $n$-space, within which the first two terms of the Taylor expansion
should be an adequate approximation.

To the extent that $T_{cl}\ll T_{R}$ (which is true in most cases),
we can identify two distinct behaviors in the time-evolution of a
wavepacket. Writing out our approximate wavefunction%
\footnote{Note that we are writing our wavefunction as a function of $\vec{x}$
rather than as a function of $x$. This is because when considering
revival problems, we aren't particularly concerned about the dimensionality
of the wavefunction. We \emph{will} be concerned about this in subsequent
chapters, and when we are the arrow will come off of the $x$.%
}, \begin{equation}
\Psi (x,t)=\exp \p{-2\pi iE_{\bar{n}}t}\sum _{n}c_{n}\psi _{n}(x)\exp \left[-2\pi i\left(\frac{(n-\bar{n})t}{T_{cl}}+\frac{(n-\bar{n})^{2}t}{T_{R}}\right)\right],\end{equation}
we can see that when $t\approx T_{cl}$, the $t/T_{R}$ term will
make no significant contribution to the time-evolution, and the wavefunction
will oscillate classically with period $T_{cl}$. With time the $t/T_{R}$
term causes the wavepacket to disperse, and interference between the
various states produces a quantum carpet. However, as $t$ approaches
within a few $T_{cl}$ of $T_{R}$, the $t/T_{R}$ term contribution
again becomes small and the wavepacket returns to classical oscillation.
When $t=T_{R}$, a quantum revival occurs. If $T_{R}$ is an integer
multiple of $T_{cl}$, the wavefunction returns to its original configuration,
and if not then it returns to something very close to its original
configuration.

\begin{figure}[hbtp]
\begin{center}\includegraphics[  width=3in]{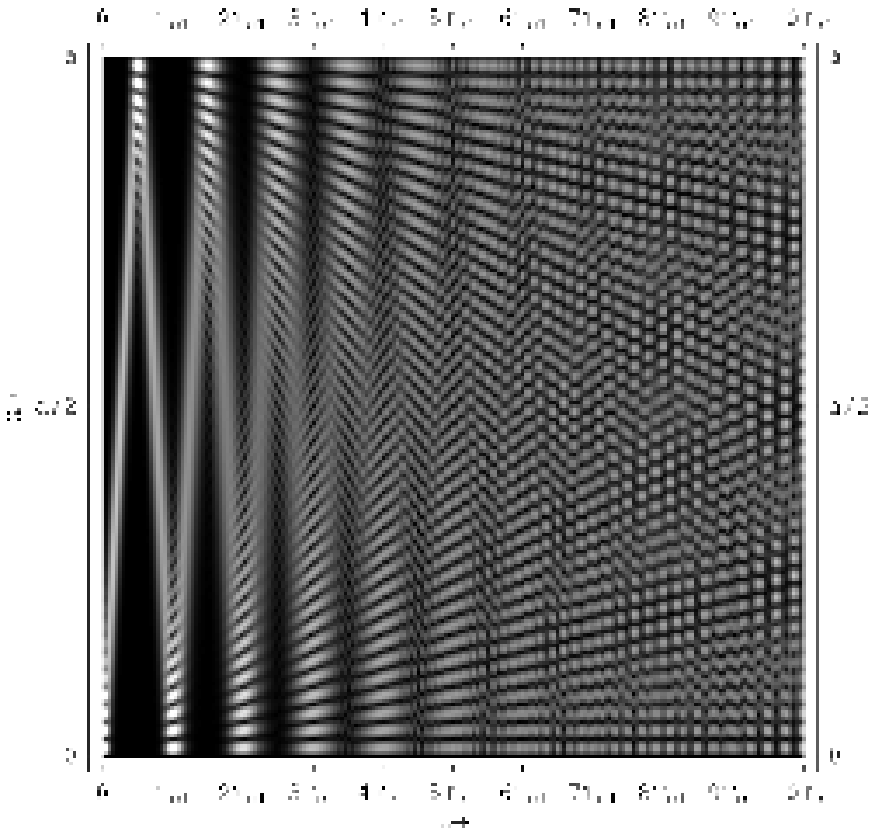}\hfill{}\includegraphics[  width=3in]{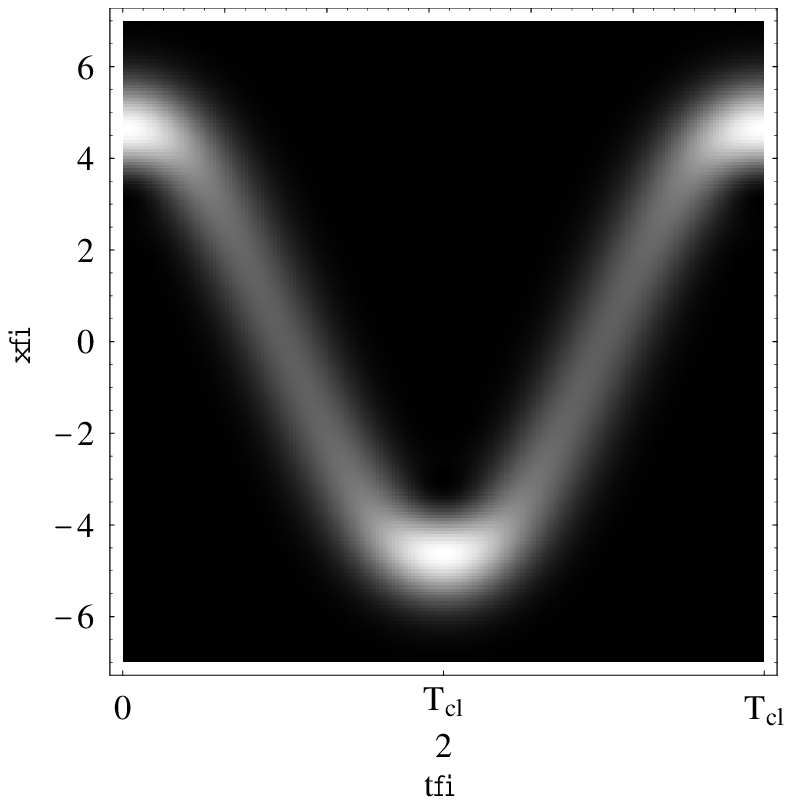}\end{center}

\caption{\label{fig:rev} $\left\Vert \Psi \right\Vert ^{2}$ for a particle
in an infinite square well, such that $T_{R}=80T_{cl}$. The distribution
of coefficients is Gaussian, with $\bar{n}=40$ and $\sigma _{n}=2$.
For the form of the square well that I am using, see Section \ref{pot: isw}.}

\caption{\label{fig:sho-pd} $\left\Vert \Psi \right\Vert ^{2}$ for a simple
harmonic oscillator. The distribution of coefficients is Gaussian,
with $\bar{n}=6$ and $\sigma _{n}=2$. For the version of the simple
harmonic oscillator that I am using, see Section \ref{pot: sho}.}
\end{figure}

This is illustrated in Figure \ref{fig:rev}, where we see the initial
evolution of a particle in an infinite square well of width $a$.
We use a Gaussian distribution of weighting coefficients with $\bar{n}=40$
and $\sigma _{n}=2$,\begin{equation}
c_{\bar{n}+k}=\frac{1}{\sigma _{n}\sqrt{2\pi }}e^{-k^{2}/2\sigma _{n}^{2}}.\end{equation}
 Referring to Equation \ref{eq:perrev} and Section \ref{pot: isw},
we can calculate the time scales for the square well,\begin{eqnarray}
T_{cl} & = & \frac{L^{2}}{\bar{n}\pi },\\
T_{R} & = & \frac{2L^{2}}{\pi },
\end{eqnarray}
and find that $T_{R}=80T_{cl}$. We can see clearly the initial classical
oscillation of the particle (the oscillations are straight lines,
as we would expect in the square well) and its subsequent dispersion
and interference. Note that the periods of the oscillations, so long
as they are well-defined, are $T_{cl}$. 

Finally, we may note that in cases where the spectrum is strictly
linear in $n$, such as the simple harmonic oscillator, we would (from
this analysis) expect no dispersion, little to no interference, no
quantum carpet -- just classical oscillation. Though this is not entirely
true, it is roughly true in virtually all cases%
\footnote{In Chapter \ref{ch: traces} we will learn the black art of {}``quadratizing
spectra,'' which will let us make wavefunctions with carpets in problems
with linear spectra.%
}. An example of classical oscillation is shown in Figure \ref{fig:sho-pd}.
There, $\bar{n}=5$ and $\sigma _{n}=2$ .%

\begin{figure}[hbtp]
\begin{center}\includegraphics[  width=7in]{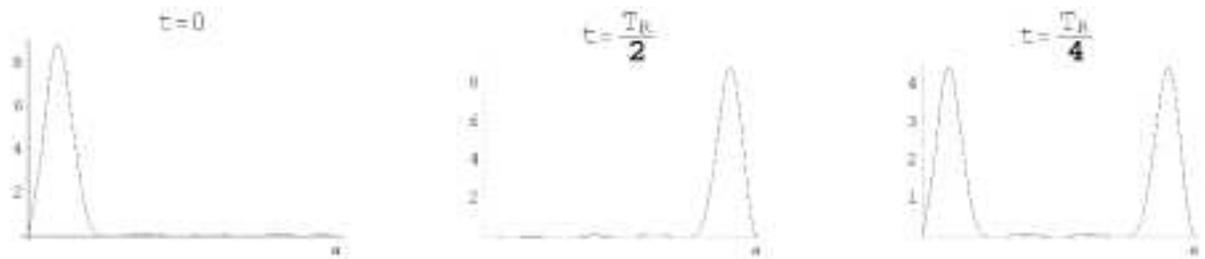}\end{center}

\caption{\label{fig:ISW-Cuts} Plots of $\left\Vert \Psi \right\Vert ^{2}$
for a particle in an infinite square well at $t=0$, $t=T_{R}/2$,
and $t=T_{R}/4$. Note that the scale is different for the rightmost
plot -- though conservation of probability shrinks our peaks, they
have the same shape as the peaks in the other two plots.}
\end{figure}

We also find revival-esque phenomena at $t<T_{R}$. If we examine
$\left\Vert \Psi \right\Vert ^{2}$ at a few particular times in the
infinite square well, as in Figure (insert figure ISW-Cuts)\ref{fig:ISW-Cuts},%
 we notice something interesting. At $t=T_{R}/2$, $\left\Vert \Psi (x,0)\right\Vert ^{2}$
has been reflected about the center of the well, and at $t=T_{R}/4$
there are two copies of the original probability distribution, one
reflected about the center of the well. Although this particular result
does not hold generally%
\footnote{For this result to hold we require eigenstates of definite parity
and a purely quadratic spectrum. For a discussion of these exact fractional
revivals, see \cite{rev:Loinaz1}.%
}, it may suggest to us that interesting things happen at rational
fractions of the revival time.

\section{\label{sec:FracRev}Fractional Revivals}

\begin{figure}[hbtp]
\begin{center}\includegraphics[  width=0.90\textwidth]{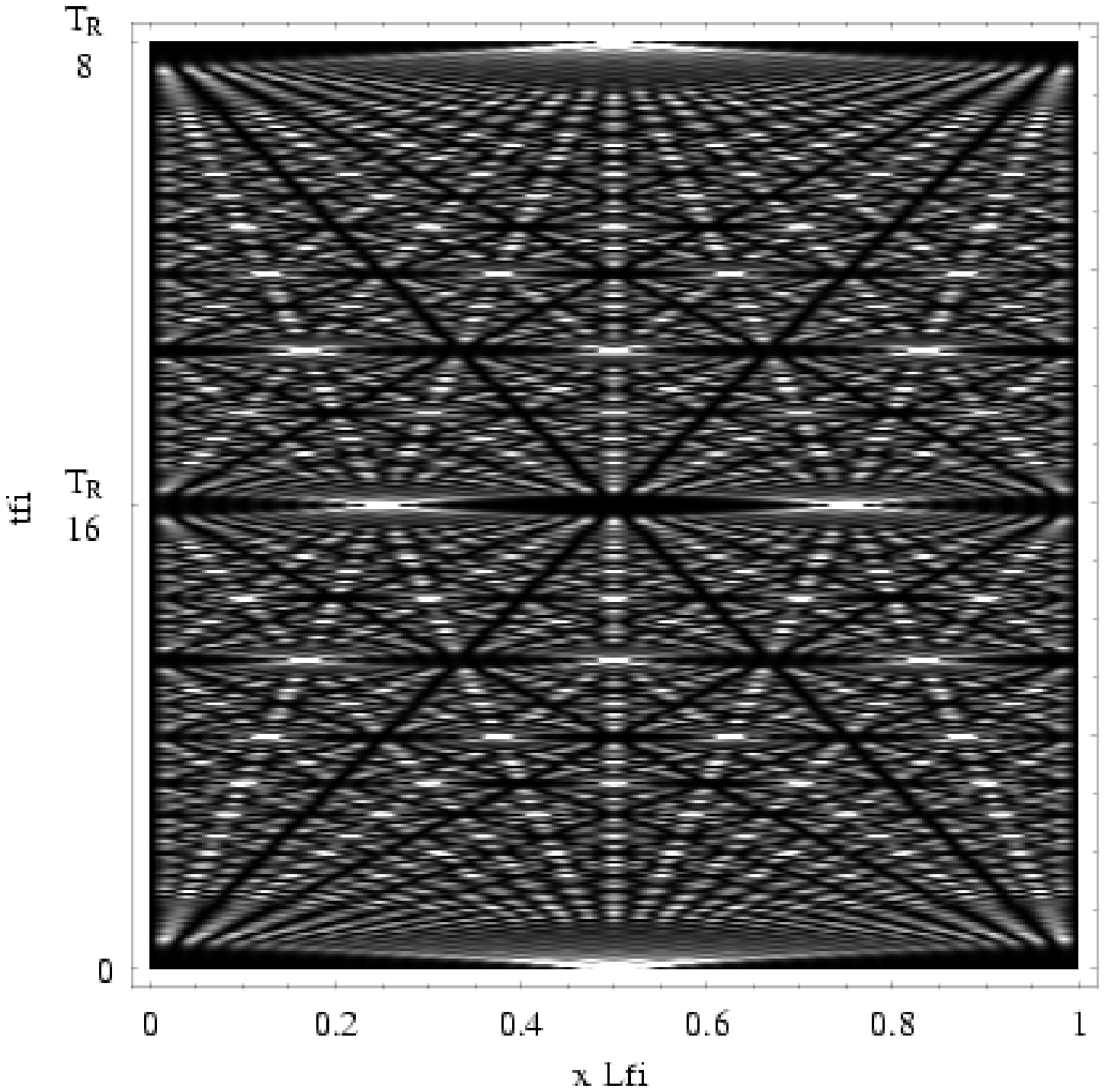}\end{center}

\caption{\label{fig: isw-frac}Fractional revivals in the infinite square
well. Our initial wavefunction is a Gaussian packet, centered on $x=L/2$,
with $\sigma _{x}=0.003L$. The symmetry of our wavepacket allows
it to have a full revival (modulo an overall phase factor) at $t=T_{R}/8$.
At times below that, though, we should be able to count the fractional
revivals.}
\end{figure}

We have seen the existence of revivals -- times when the original
wavefunction perfectly or near-perfectly reassembles itself. We have
also seen that in at least one case we can find revival-like phenomena
at $t\ll T_{R}$. Indeed, if we look at Figure \ref{fig: isw-frac},
we might suspect that something interesting happens at most rational
fractions of $T_{R}$. %
It turns out that while perfect revivals%
\footnote{See \cite{rev:Loinaz1} for a discussion of such revivals.%
} at times other than the revival time are rather unusual, an initial
wavepacket will give rise to arrays of other packets (which are not,
in general, identical to the original packet) at rational fractions
of $T_{R}$. The motivation for this is more algebraic than physical
-- it has to do with our ability at rational fractions of $T_{R}$
to group together states with the same phase. This phenomenon was
treated quite definitively in \cite{rev:Averbukh1}, and the next
section will closely follow their derivation.

\subsection{The General Case}

In this section, we will restrict ourselves to wavefunctions that
can be well-approximated by the following%
\footnote{We apologize in advance for the lower limit on the sum, which should
only go to $-\bar{n}$. The idea is that while all the $k<-\bar{n}$
terms should have weightings of \emph{exactly} zero, we can make the
approximation that their weighting coefficients are very small. For
somee sums and integrals these limits will greatly simplify our calculation.
We've also dropped the leading $\exp \p{-2\pi iE_{\bar{n}}t}$ term,
as it has no impact on the relative phase of the various eigenfunctions
and thus no physical significance.%
}:\begin{equation}
\Psi \p{\vec{x},t}=\sum _{k=-\infty }^{\infty }c_{\p{k+\bar{n}}}\psi _{\p{k+\bar{n}}}\p{\vec{x}}\exp -2\pi i\p{k\frac{t}{T_{cl}}+k^{2}\frac{t}{T_{R}}},\, k=n-\bar{n}\label{eq:basic-psi}\end{equation}
This is a fairly large class of problems - virtually any problem with
a well-localized distribution of $c_{n}$ around some mean value $\bar{n}$,
and any problem in a potential with a quadratic spectrum. This derivation
makes no assumptions about the dimensionality of the problem, but
does assume that the spectrum is a function of only one quantum number.
Though we don't believe the generalization to more variables to be
particularly problematic, we have not considered it here. 

We will look at the wavefunction at rational fractions of the revival
time, $t=\frac{p}{q}T_{R}$, where $p,\, q\Z $ and $p$ and $q$
are relatively prime%
\footnote{This relatively prime business is going to force the set of revivals
that we can resolve to fit a Farey sequence. For more on this, see
Appendix \ref{sec:Farey} or \cite{comp:Harter1} or \cite{misc:Farey}.%
}. At such times we might expect that the $t/T_{R}$ terms in the time
evolution will reduce to some set of rational numbers, and that many
values of $k$ will have the same phase contributions. We will begin
by looking only at exact times, but will find that our result is also
a good approximation at $t=\frac{p}{q}T_{R}+\Delta t$. 

If we now define \begin{equation}
\phi _{k}=\frac{p}{q}k^{2}\textrm{ mod }1,\label{eq:phi-k}\end{equation}
which characterizes the deviation in phase of each eigenfunction from
classical oscillation, we can write\begin{equation}
\Psi (\vec{x},t=\frac{p}{q}T_{R})=\sum _{k=-\infty }^{\infty }c_{k}\psi _{x}(\vec{x})\exp \left(-2\pi i\left(k\frac{p}{q}\frac{T_{R}}{T_{cl}}+\phi _{k}\right)\right).\label{eq:psi-phi}\end{equation}
Now, consider the conditions for some of these phases to match. If
we can collect the terms of the sum into a handful of equivalence
classes (defining two terms as equivalent when they have identical
$\phi _{k}$), we can turn the infinite sum over $k$ into a finite
sum over another variable%
\footnote{Well, that finite sum is itself over another infinite sum, but since
when did you get something for nothing?%
}. This is a question about the periodicity of $\phi _{k}$ which can
be easily answered. We are looking for the minimal $l$ such that
$p,q\in \mathbb{Z}$ and for all $k$,\begin{eqnarray}
\phi _{k} & = & \phi _{k+l}\nonumber \\
\frac{p}{q}k^{2}\textrm{mod }1 & = & \frac{p}{q}(k+l)^{2}\textrm{ mod }1.
\end{eqnarray}
 This leads us to two conditions on $l$,\begin{eqnarray}
\frac{2pl}{q} & \in  & \mathbb{Z},\label{eq:zcond-1}\\
\frac{pl^{2}}{q} & \in  & \mathbb{Z}.\label{eq:zcond-2}
\end{eqnarray}
There are two important solutions to Equation \ref{eq:zcond-1},
$l=q$ and $l=q/2$. Our task is to find out when the $l=q/2$ is
consistent with \ref{eq:zcond-2}, and when the minimal value of
$l$ is just $q$. There are three cases here for us to consider,

\subsubsection{Case 1: $q$ is odd}

For odd $q$, $l=q$. While $l=q/2$ would satisfy Equation \ref{eq:zcond-1},
$q/2$ is not an integer, as we require.

\subsubsection{Case 2: $q$ contains more than one power of 2}

When $q$ contains more than one power of 2, $l=q/2$. This obviously
satisfies Equation \ref{eq:zcond-1}, and Equation \ref{eq:zcond-2}
reduces to $pq/4\Z $, which is also true.

\subsubsection{Case 3: $q$ contains only one power of 2}

If $q$ contains only one power of 2, $l=q$. While $l=q/2$ satisfies
Equation \ref{eq:zcond-1}, it does not satisfy Equation \ref{eq:zcond-2}.
That would require that $pq/4\Z $, but since $p$ and $q$ are relatively
prime, $p$ must be odd, and by assumption $q/4$ is half-integer.\\

We have established that $\phi _{k}$ is periodic, and we will be
able to turn our sum over $k$ into a sum with $l$ terms. Note that
if we were to consider times that were not rational fractions of $T_{R}$,
we would be able to group the phase contributions from the $k^{2}$
term. With this relationship in hand, we may make a guess about how
we may rewrite the wavefunction. Let us first define\begin{equation}
\Psi _{cl}(\vec{x},t)=\sum _{k=-\infty }^{\infty }c_{k}\psi _{k}(\vec{x})\exp \left(-2\pi ik\frac{t}{T_{cl}}\right),\label{eq:psi-cl}\end{equation}
a version of the wavefunction with the {}``dispersion'' terms removed.
We postulate that we can then write the wavefunction as a weighted
sum of time-slices%
\footnote{Recall that the rational numbers form a dense subset of the real numbers.
This means that we can approximate the wavefunction arbitrarily well
at any value of $t$ with slices of $\Psi _{cl},$ which in turn implies
that the set of time slices of $\Psi _{cl}$ taken at rational times
form a basis for the Hilbert space of a particular problem. This was
first observed in \cite{rev:Aronstein1}.%
} of $\Psi _{cl}$,\begin{equation}
\Psi \p{\vec{x},t=\frac{p}{q}T_{R}}=\sum _{s=0}^{l-1}a_{s}\Psi _{cl}\left(\vec{x},\frac{p}{q}T_{R}+\frac{s}{l}T_{cl}\right).\label{eq:psi-cl_step1}\end{equation}
We will now demonstrate that Equation \ref{eq:psi-cl_step1} is valid
by explicit construction of the coefficients $a_{s}$. First, we plug
Equation \ref{eq:psi-cl} into Equation \ref{eq:psi-cl_step1}, \begin{eqnarray}
\Psi (x,t=\frac{p}{q}T_{R}) & = & \sum _{s=0}^{l-1}a_{s}\Psi _{cl}\left(\vec{x},t+\frac{s}{l}T_{cl}\right)\nonumber \\
 & = & \sum _{s=0}^{l-1}a_{s}\sum _{k=-\infty }^{\infty }c_{k}\psi _{k}(\vec{x})\exp \left(-2\pi ik\frac{t+(s/l)T_{cl}}{T_{cl}}\right)\nonumber \\
 & = & \sum _{k=-\infty }^{\infty }c_{k}\psi _{k}(\vec{x})\exp \left(-2\pi ik\frac{t}{T_{cl}}\right)\sum _{s=0}^{l-1}a_{s}\exp \left(-2\pi i\frac{s}{l}k\right).\label{eq:psi-cl_step2}
\end{eqnarray}
We now compare the result with Equation \ref{eq:psi-phi} and arrive
at the condition

\begin{equation}
\exp \left(-2\pi i\phi _{k}\right)=\sum _{s=0}^{l-1}a_{s}\exp \left(-2\pi i\frac{s}{l}k\right),\label{eq:phiperiod}\end{equation}
for $k=0,1,...,l-1$.

We now need to find an explicit expression for the constants $a_{s}$.
If we multiply Equation \ref{eq:phiperiod} by $\exp \left(2\pi i\frac{m}{l}k\right)$,
where $m$ is an integer, and sum over $k$, we can use the completeness
relation of the Fourier sum to extract the coefficient:

\begin{eqnarray}
\sum _{k=0}^{l-1}e^{-2\pi i\left(\phi _{k}-\frac{m}{l}k\right)} & = & \sum _{k=0}^{l-1}\sum _{s=0}^{l-1}a_{s}e^{-2\pi i\frac{s-m}{l}k}\nonumber \\
 & = & l\sum _{s=0}^{l-1}a_{s}\delta _{s,m},\\
a_{m} & = & \frac{1}{l}\sum _{k=0}^{l-1}e^{-2\pi i\left(\phi _{k}-\frac{m}{l}k\right)}.\label{eq:coeffsum}
\end{eqnarray}
We have constructed the $a_{s},$ but we can rewrite equation (\ref{eq:coeffsum})
in a form more conducive to computation. First, we define how we will
step through the possible values of $m$:\begin{equation}
m^{\prime }=(m+2\frac{pl}{q})\textrm{ mod }l.\label{eq:m-recurr}\end{equation}
With this in place, we are prepared to derive the following:

\begin{eqnarray}
a_{m^{\prime }}\exp \left(-2\pi i\left(\frac{m}{l}+\frac{p}{q}\right)\right) & = & \frac{1}{l}\exp \left(-2\pi i\left(\frac{m}{l}+\frac{p}{q}\right)\right)\sum _{k=0}^{l-1}\exp \left(-2\pi i\left(\frac{p}{q}k^{2}-\frac{m}{l}k-\frac{p}{q}2k\right)\right)\nonumber \\
 & = & \frac{1}{l}\sum _{k=0}^{l-1}\exp \left(-2\pi i\left(\frac{p}{q}(k-1)^{2}-\frac{m}{l}(k-1)\right)\right),
\end{eqnarray}
from which, shifting the summation index by 1, exploiting the periodicity
of $\phi _{k}$, and rearranging the equation for convenience, we
conclude \begin{equation}
a_{m^{\prime }}=a_{m}e^{2\pi i\left(\frac{m}{l}+\frac{p}{q}\right)}.\label{eq:coeffrecur}\end{equation}

It is easily confirmed that when $q$ is odd or contains more than
one power of 2, the $a_{m}$'s with $m=l$ and $m=2\frac{pl}{q}$
will have opposite parity, and all of the $a_{m}$ will have the same
modulus. When $q$ contains only one power of two, both $a_{m}$ for
$m=l$ and $m=2\frac{pl}{q}$ will be even, and the coefficients with
even and odd indices will have different moduli. As it happens, the
coefficients with even indices are all zero.

Now, let us compare our result with the actual wavefunction at $t=\frac{p}{q}T_{R}+\Delta t$.
The actual wavefunction is \begin{equation}
\Psi \p{\vec{x},\frac{p}{q}T_{R}+\Delta t}=\sum _{k=-\infty }^{\infty }c_{k}\psi _{k}(\vec{x})\exp -2\pi i\left(\frac{p}{q}\frac{T_{R}}{T_{cl}}k+\frac{\Delta t}{T_{cl}}k+\frac{p}{q}k^{2}+\frac{\Delta t}{T_{R}}k^{2}\right).\end{equation}
Using Equations \ref{eq:psi-cl_step2} and \ref{eq:phiperiod}, we
find that \begin{eqnarray}
\Psi \left(\vec{x},t=\frac{p}{q}T_{R}+\Delta t\right) & = & \sum _{s=0}^{l-1}a_{s}\Psi _{cl}\left(\vec{x},\frac{p}{q}T_{R}+\Delta t+\frac{s}{l}T_{cl}\right)\\
 & = & \sum _{k=-\infty }^{\infty }\sum _{s=0}^{l-1}a_{s}\exp \left(-2\pi i\frac{s}{l}k\right)c_{k}\psi _{k}(\vec{x})\exp -2\pi i\left(\frac{p}{q}\frac{T_{R}}{T_{cl}}k+\frac{\Delta t}{T_{cl}}k\right)\\
 & = & \sum _{k=-\infty }^{\infty }c_{k}\psi _{k}(\vec{x})\exp -2\pi i\left(\frac{p}{q}\frac{T_{R}}{T_{cl}}k+\frac{\Delta t}{T_{cl}}k+\frac{p}{q}k^{2}\right),
\end{eqnarray}
which will be a good approximation so long as $k\Delta t/T_{R}\ll \Delta t/T_{cl}$,
where we must consider the largest {}``relevant'' $k$. This is,
perhaps, a closer correspondence than we should really expect -- remember
that at the start of this derivation we only required that our rewriting
of the wavefunction be equivalent to it \emph{at} $t=\p{p/q}T_{R}$.

We may ask what this means for $\left\Vert \Psi \right\Vert ^{2}$.
At a rational fraction of the revival time, we have\begin{equation}
\left\Vert \Psi \left(x,t=\frac{p}{q}T_{R}\right)\right\Vert ^{2}=\sum _{s,z=0}^{l-1}a_{s}a_{z}^{*}\Psi _{cl}\left(\vec{x},\frac{p}{q}T_{R}+\frac{s}{l}T_{cl}\right)\Psi _{cl}^{*}\left(\vec{x},\frac{p}{q}T_{R}+\frac{z}{l}T_{cl}\right),\end{equation}
a product of several time-slices of $\Psi _{cl}$. In cases where
we didn't have a well-defined wavepacket to begin with, or in the
case where $\Psi _{cl}$ doesn't preserve this wavepacket, we should
not expect discernible fractional revivals. However, to the extent
that each slice of $\Psi _{cl}$ contains a peak that has no significant
overlap with any of the other time-slices in question, the cross-terms
in the sum will make a negligible contribution and we will have several
well-defined wavepackets - a fractional revival. To see this, consult
Figure \ref{fig: isw-frac}, where you should be able to count the
fractional revivals.

\begin{figure}[hbtp]
\begin{center}\includegraphics[  width=0.40\textwidth]{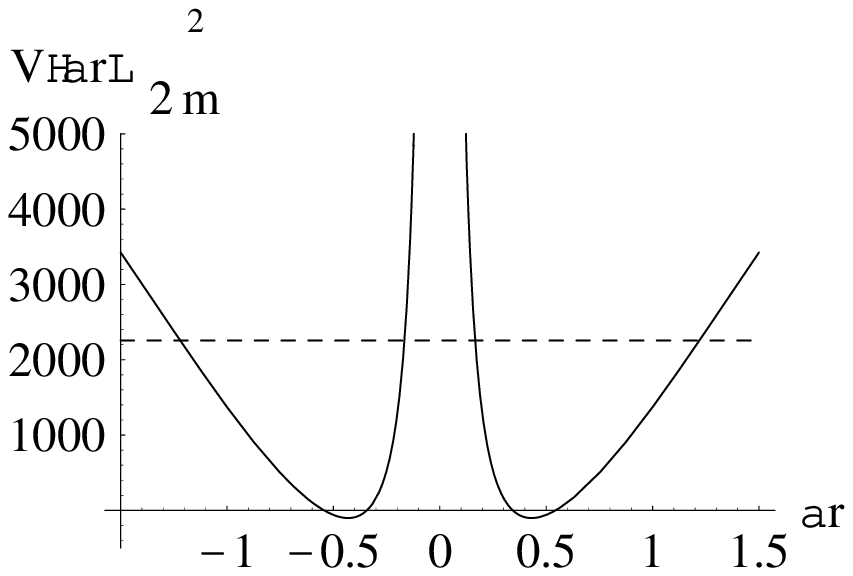}\hfill{}\includegraphics[  width=0.40\textwidth]{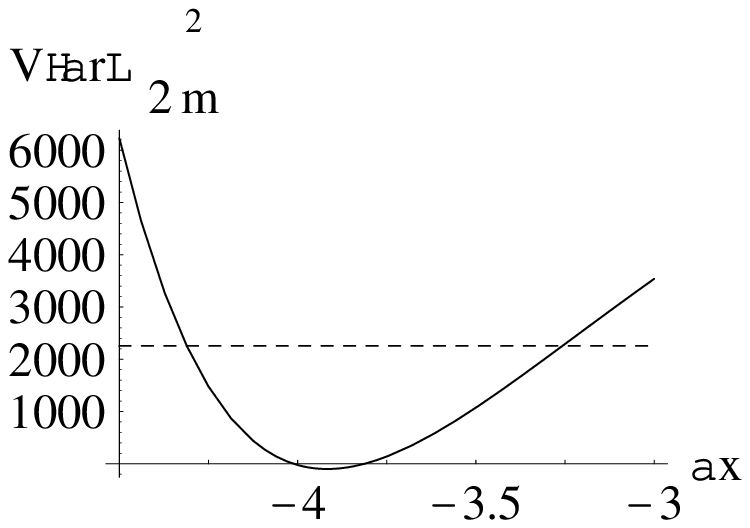}\\
\includegraphics[  width=0.40\textwidth]{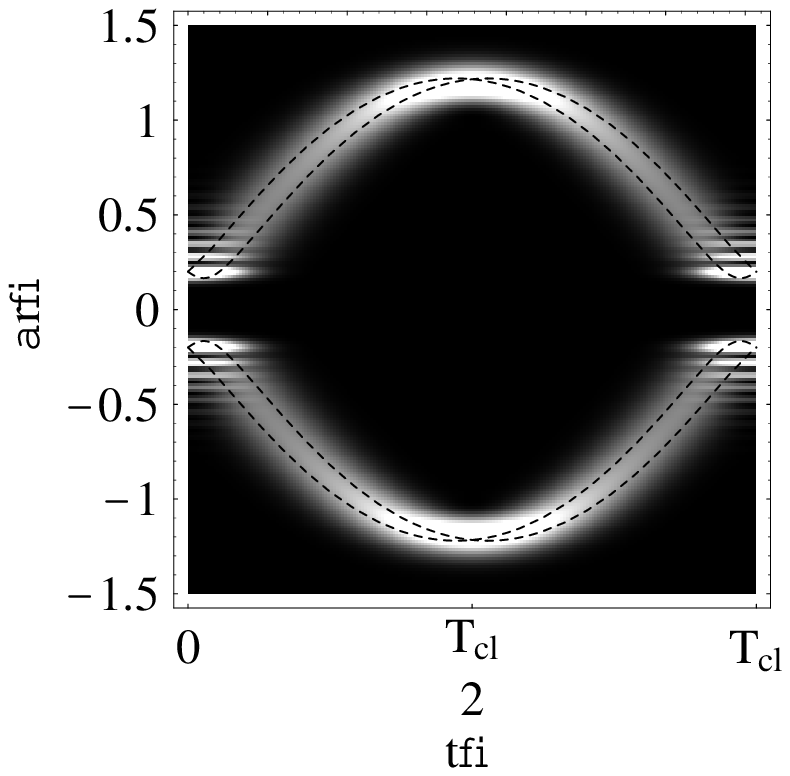}\hfill{}\includegraphics[  width=0.40\textwidth]{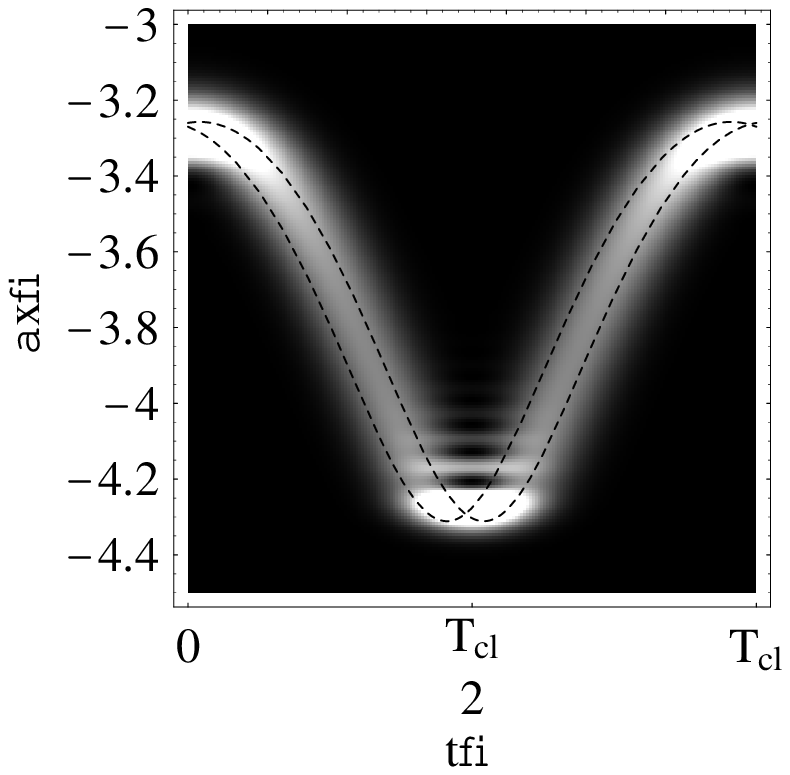}\end{center}

\caption{\label{fig: PT-psicl}$V(\alpha r)$ (above) and $\left\Vert \Psi _{cl}\right\Vert ^{2}$
(below) for a particle in the Pöshcl-Teller potential (see Appendix
\ref{Ap: Potentials}). The dotted line in the above plot indicates
$E_{\bar{n}}$, in the lower plot they indicate four different classical
paths of energy $E_{\bar{n}}$. The Pöschl-Teller potential has a
purely quadratic spectrum. For more details on it, see Section \ref{pot: p-t}.}

\caption{\label{fig: MR-psicl}$V(\alpha x)$ (above) and $\left\Vert \Psi _{cl}\right\Vert ^{2}$
(below) for a particle in the Morse potential (see Appendix \ref{Ap: Potentials}).
The dotted line in the above plot indicates $E_{\bar{n}}$, in the
lower plot they indicate four different classical paths of energy
$E_{\bar{n}}$. The Morse potential has a purely quadratic spectrum.
For more details on it, see Section \ref{pot: mrs}.}
\end{figure}

Of course, we have (as of yet) no guarantee that $\Psi _{cl}$ is
really free of dispersion, or that wavepackets really follow their
classical paths. Though we have attached the label of {}``classical''
to $\Psi _{cl}$ and $T_{cl}$, we have yet to prove anything about
them -- that will have to wait for Section \ref{sec:psi-cl}, when
we've developed a technique for analyzing interference in quantum
carpets. Before then, though, we can present some highly suggestive
pictures, Figures \ref{fig: PT-psicl} and \ref{fig: MR-psicl}.%

\subsection{A Closer Look at $\Psi _{cl}$ and $a_{m}$}

If you take a moment to write down the particular $a_{m}$ for some
$p/q$, you will find that all values in question appear twice%
\footnote{While the explicit examples that they calculated reflected the results
of this section, Averbukh and Perelman did not state or use them.%
}. For example, at $p/q=1/5$ we find \begin{eqnarray}
a_{0} & = & a_{0},\\
a_{2} & = & a_{0}\exp \p{-2\pi i\frac{1}{5}},\\
a_{4} & = & a_{0}\exp \p{-2\pi i\frac{4}{5}},\\
a_{1} & = & a_{0}\exp \p{-2\pi i\frac{4}{5}},\\
a_{3} & = & a_{0}\exp \p{-2\pi i\frac{1}{5}}.
\end{eqnarray}
Of course, we can find a reason for this. Let us suppose that from
Equation \ref{eq:m-recurr} we can construct a function $m(n)$ such
that $m(0)\equiv 0$ and $m(-n)\equiv m(l-n)$. If this is true, we
can further simplify our expression for fractional revivals. Obviously,
$m(n)$ would depend on whether $l$ was odd, a multiple of more than
one power of two, or an odd multiple of two, so we will have to perform
this analysis for three different cases. Fortunately, the proof that
these $a_{m}$'s come in pairs follows the same outline in each case.
First, we will confirm that $\left(m(n)+m(-n)\right)\mod l=0$, which
will tell us that $a_{m(n)}$ and $a_{m(-n)}$ connect $\Psi _{cl}\left(\vec{x},t+\Delta t\right)$
and $\Psi _{cl}\left(\vec{x},t-\Delta t\right)$ (see Equation \ref{eq:psi-cl_step1}).
Second, we will prove that $a_{m(n)}=a_{m(-n)}$, to see if we could
further simplify the sum in Equation \ref{eq:psi-cl_step1}. This
is to be done by induction - since $a_{m(0)}=a_{m(-0)}$, if $a_{m(n)}=a_{m(-n)}$
implies $a_{m(n+1)}=a_{m(-n-1)}$ then $a_{m(n)}=a_{m(-n)}$ for all
$n$. 

From Equation \ref{eq:coeffrecur}, we know that \begin{equation}
a_{m(n+1)}=a_{m(n)}\exp 2\pi i\left(\frac{m(n)}{l}+\frac{p}{q}\right),\end{equation}
and if we change the sign of $n$ and shift the indices by one we
can rearrange this to read\begin{equation}
a_{m(-n-1)}=a_{m(-n)}\exp -2\pi i\left(\frac{m(-n-1)}{l}+\frac{p}{q}\right).\end{equation}
Using our inductive hypothesis, $a_{m(n)}=a_{m(-n)}$, we need only
show that \begin{equation}
\frac{-m(-n-1)}{l}=\frac{m(n)}{l}+2\frac{p}{q}.\label{eq:m-induct}\end{equation}
If all of this holds, we can define \begin{equation}
h\equiv \left\{ \begin{array}{l}
 \frac{l-1}{2},\, l\textrm{ odd},\\
 \frac{l}{2},\, l\textrm{ even},\end{array}
\right.\end{equation}
and rewrite Equation \ref{eq:psi-cl_step1} as\begin{equation}
\Psi \left(\vec{x},t=\frac{p}{q}T_{R}\right)=\sum _{s=0}^{h}a_{s}\left(\Psi _{cl}\left(\vec{x},\frac{p}{q}T_{R}+\frac{s}{l}T_{cl}\right)+\Psi _{cl}\left(\vec{x},\frac{p}{q}T_{R}-\frac{s}{l}T_{cl}\right)\right).\label{eq:fracrev-h}\end{equation}

Let's examine the three cases:

\subsubsection{Case 1: $q$ is odd}

If $q$ is odd, then $l=q,$ and $m^{\prime }=(m+2p)\textrm{ mod }q$,
from which we can deduce \begin{eqnarray}
m(n) & = & 2pn\textrm{ mod }q,\\
a_{m(n)} & = & a_{0}\exp 2\pi i\left(n\frac{p}{q}+\sum _{N=0}^{n-1}\frac{2pn\textrm{ mod }q}{q}\right).
\end{eqnarray}
The definition $m(-n)\equiv -2pn\mod q$ is consistent with our condition
that $m(-n)=m(l-n)$. Moreover, under this definition $m(n)+m(-n)=0,$
satisfying our first condition. Turning our attention to Equation
\ref{eq:m-induct}, we see that \begin{eqnarray}
\frac{-m(-n-1)}{l} & = & \frac{-(-2pn-2p)\textrm{ mod }q}{q}\nonumber \\
 & = & \frac{2pn}{q}\textrm{ mod }q+\frac{2p}{q}\nonumber \\
 & = & \frac{m(n)}{l}+2\frac{p}{q},
\end{eqnarray}
and our theorem holds.

\subsubsection{Case 2: $q$ contains more than one power of 2}

If $q$ contains more than one power of 2, then $l=q/2$, and $m^{\prime }=(m+p)\textrm{ mod }q/2$.
We may then say that\begin{eqnarray}
m(n) & = & pn\textrm{ mod }q/2,\\
a_{m(n)} & = & a_{0}\exp 2\pi i\left(n\frac{p}{q}+\sum _{N=0}^{n-1}\frac{pn\textrm{ mod }q/2}{q/2}\right).
\end{eqnarray}
The definition $m(-n)\equiv -pn\mod q/2$ is consistent with our condition
that $m(-n)=m(l-n)$. Moreover, under this definition $m(n)+m(-n)=0,$
satisfying our first condition. Turning our attention to Equation
\ref{eq:m-induct}, we see that \begin{eqnarray}
\frac{-m(-n-1)}{l} & = & \frac{-(-pn-p)\textrm{ mod }q}{q/2}\nonumber \\
 & = & \frac{pn}{q/2}\textrm{ mod }q+\frac{2p}{q}\nonumber \\
 & = & \frac{m(n)}{l}+2\frac{p}{q},
\end{eqnarray}
and our theorem holds again.

\subsubsection{Case 3: $q$ contains only one power of 2 }

If $q$ contains only one power of two, then again $l=p$. The principle
difference from the $q$-odd case is that $a_{0}=0,$ while $a_{1}$
is non-zero. The proof in the $q$-odd case holds for the odd $m$,
while for even $m$, $a_{m}=0$, and we may easily define $a_{l-m}=0$.
Thus our theorem holds in this final case.

\subsection{The Infinite Square Well}

\begin{figure}[hbtp]
\begin{center}\includegraphics{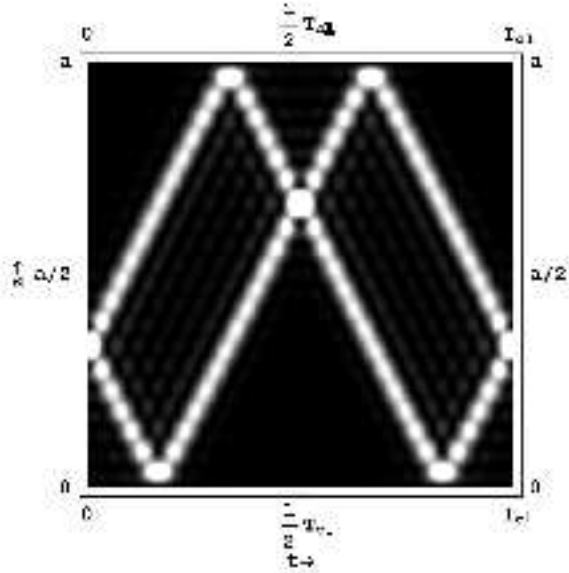}\end{center}

\caption{\label{fig:ISW-Psi-cl}$\Psi _{cl}$ for a roughly Gaussian spatial
distribution, centered at $x=a/3$.}
\end{figure}

Equation \ref{eq:fracrev-h} is particularly useful when we can find
some connection between the paired slices of $\Psi _{cl}$. The eigenfunctions
of the infinite square well are harmonic functions, particularly easy
to combine with the time evolution term in both $\Psi $ and $\Psi _{cl}$.
Perhaps the first thing to do is plot a particular example -- Figure
\ref{fig:ISW-Psi-cl} shows what looks like two wave packets bouncing
off the walls of the well.%

We start out from equation (\ref{eq:psi-cl}) and fill in the specifics
of the problem:\begin{equation}
\Psi _{cl}(x,t)=\sum _{k=-\infty }^{\infty }c_{\p{k+\bar{n}}}\sqrt{\frac{2}{L}}\sin \left(\frac{\p{k+\bar{n}}\pi }{L}x\right)\exp \left(-2\pi ik\frac{t}{T_{cl}}\right).\end{equation}
We then rewrite the sine terms as a sum of exponentials,\begin{eqnarray}
\Psi _{cl}(x,t) & = & \sum _{k=-\infty }^{\infty }c_{k}\sqrt{\frac{2}{L}}\frac{i}{2}\left(\exp \left(i\frac{k\pi }{L}x\right)-\exp \left(-i\frac{k\pi }{L}x\right)\right)\exp \left(-2\pi ik\frac{t}{T_{cl}}\right)\nonumber \\
 & = & \sum _{k=-\infty }^{\infty }c_{k}\sqrt{\frac{2}{L}}\frac{i}{2}\left(\exp i\frac{k\pi }{L}\left(x-\frac{2L}{T_{cl}}t\right)-\exp -i\frac{k\pi }{L}\left(x+\frac{2L}{T_{cl}}t\right)\right)\nonumber \\
 & = & \sum _{k=-\infty }^{\infty }c_{k}\sqrt{\frac{2}{L}}\frac{1}{2}\begin{array}[t]{l}
 \left(\sin \frac{k\pi }{L}\left(x-\frac{2L}{T_{cl}}t\right)+\sin \frac{k\pi }{L}\left(x+\frac{2L}{T_{cl}}t\right)\right.\\
 \left.+i\cos \frac{k\pi }{L}\left(x-\frac{2L}{T_{cl}}t\right)-i\cos \frac{k\pi }{L}\left(x+\frac{2L}{T_{cl}}t\right)\right).\end{array}
\label{eq:psi-cl-isw}
\end{eqnarray}
 Note that $\frac{2L}{T_{cl}}$ is the velocity associated with classical
motion in the square well. If we plug in $t=\frac{p}{q}T_{R}\pm \frac{s}{l}T_{cl}$,
we find that we can rewrite the arguments of the trigonometric functions
as $\left(x+\frac{2Lp}{q}\frac{T_{R}}{T_{cl}}\right)\pm 2L\frac{s}{l}$.
Now, inserting Equation \ref{eq:psi-cl-isw} into Equation \ref{eq:fracrev-h},
we find\begin{eqnarray}
\Psi (\vec{x},t=\frac{p}{q}T_{R}) & = & \sum _{s=0}^{h}a_{s}\sum _{k=-\infty }^{\infty }c_{k}\sqrt{\frac{2}{L}}\left(\sin \frac{k\pi }{L}\left(x-\frac{2L}{T_{cl}}\frac{s}{l}T_{cl}\right)+\sin \frac{k\pi }{L}\left(x+\frac{2L}{T_{cl}}\frac{s}{l}T_{cl}\right)\right)\nonumber \\
 & = & \sum _{s=0}^{h}a_{s}\left(\Psi (x-\Delta x_{s},0)+\Psi (x+\Delta x_{s},0)\right),
\end{eqnarray}
where we have defined $\Delta x_{s}=2L\frac{s}{l}$.

It is rather surprising that in the infinite square well, at any rational
fraction of the revival time, the wavefunction can be treated as a
weighted sum of translations of the initial wavefunction. Note that
$\Psi (-x,0)=\Psi (2L-x,0)=-\Psi (x,0)$, which ensures that each
pair of translations will meet the boundary conditions of the square
well - indeed, these translations can be thought of as disturbances
in a dispersionless string, with the leftward (rightward) translation
contributing the $\pi $-phase-shifted reflection of the rightward
(leftward) translation.

The result that all fractional revivals in the square well consist
of translated copies of the original wavefunction was stated but not
proven in \cite{rev:Aronstein1}. There, the analogy with the dispersionless
string is simply asserted, and the translations derived from that.

This impressively simple result is, of course, a direct consequence
of the overwhelming simplicity of the infinite square well. That does
not mean, however, that we cannot gain important physical insight
from this result. We can think of $\Psi _{cl}$ as not changing shape
during its time-evolution because there are no features on the bottom
of the well. We might then guess that the degree of dispersion, and
thus the degree to which the fractional revivals do not resemble the
original wavepacket, depends on the features of the potential.

\subsection{A Final Note on $\Psi _{cl}$}

Obviously, $\Psi _{cl}$ does not satisfy the Schrödinger equation
-- can we find a similar equation that it \textit{does} satisfy? We
can examine a few derivatives, and see if we can reconstruct something
analogous to the Schrödinger equation. First, referring to equation
\ref{eq:psi-cl}, defining $E\equiv \left|E_{\bar{n}}^{\prime }\right|$we
can calculate the following derivatives:\begin{eqnarray}
\frac{\partial \Psi _{cl}}{\partial t} & = & \sum _{k}c_{k}\psi _{k}\left(-iEk\right)e^{-iEkt}\\
\frac{\partial ^{2}\Psi _{cl}}{\partial x^{2}} & = & \sum _{k}c_{k}\frac{\partial ^{2}\psi _{k}}{\partial x^{2}}e^{-iEkt}.
\end{eqnarray}
Invoking the Time-Independent Schrödinger Equation, $E_{k}\psi _{k}=-\frac{\partial ^{2}\psi _{k}}{\partial t^{2}}+V\psi _{k}$,
we can find\begin{eqnarray}
\frac{\partial ^{2}\Psi _{cl}}{\partial x^{2}} & = & \sum _{k}c_{k}\left(V-E_{k}\right)\psi _{k}e^{-iEkt}\nonumber \\
 & = & V\Psi _{cl}-\sum _{k}c_{k}E_{k}\psi _{k}e^{-iEkt}\nonumber \\
 & = & V\Psi _{cl}-i\frac{\partial \Psi _{cl}}{\partial t}-\sum _{k}c_{k}\left(E_{k}-Ek\right)\psi _{k}e^{-iEkt},
\end{eqnarray}
which we can rearrange to the following Schrödinger-like equation:\begin{equation}
i\frac{\partial \Psi _{cl}}{\partial t}=-\frac{\partial ^{2}\Psi _{cl}}{\partial x^{2}}+V\Psi _{cl}-\sum _{k}c_{k}\left(E_{k}-Ek\right)\psi _{k}e^{-iEkt}.\label{sch-like}\end{equation}
Brief examination of Equation \ref{sch-like} should reveal why this
did not prove to be a successful vein of analysis. We have lost linearity,
one of the most attractive features of the Schrödinger equation.

\section{Summary}

One of the principal differences between classical and quantum mechanics
is that quantum systems have more than one time scale. Because of
this, we can find many interesting phenomena in their time-evolution.
In this chapter, we saw that the phenomenon of full and fractional
revivals are quite general consequences of weighted sums of complex
exponentials. We have also seen that we can express fractional revivals
of a wavefunction as a weighted sum of a {}``classicized'' wavefunction,
$\Psi _{cl}$ -- regardless of the original spectrum, we give $\Psi _{cl}$
a linear spectrum, greatly simplifying \emph{its} evolution. A new
relationship between the weighting coefficients in that sum, $a_{m}$
and $a_{-m}$, was proved, leading to a proof that fractional revivals
in the infinite square well are simply translations and reflections
of the original wavefunction.

\chapter{\label{ch: beats}Quantum Beats}

The object of consideration in this chapter%
\footnote{Our treatment of quantum beats is based largely on that in \cite{beats:Leichtle1,beats:Leichtle2}.%
} is a weighted sum of exponentials, \begin{equation}
f(t)=\sum _{n}P_{n}e^{iE(n)t}.\end{equation}
As one might expect, these sorts of sums exhibit revival phenomena
just as in the previous chapter, when the $P_{n}$ were functions.
In the literature, this sort of sum is said to produce {}``quantum
beats,'' and they are important because many laboratory measurements
are of signals that can be written in this form -- for example, the
fluorescence of an ensemble of atoms, each excited to one of a few
almost identitical energy levels. This is a useful enough phenomenon
to have given rise to techniques such as quantum beat spectroscopy%
\footnote{For more on the experimental aspects of quantum beats, see \cite{beats:Silverman1}.%
}. Because this is effectively a zero dimensional problem, there are
fewer dynamical features for us to consider. Because of the weighted
exponential form, we will see revivals and fractional revivals. What
is interesting about this case is that we will not only note that
they exist, but find a way to characterize how long the smooth revivals
live before dephasing.

\begin{figure}[hbtp]
\begin{center}\includegraphics[  width=0.45\textwidth]{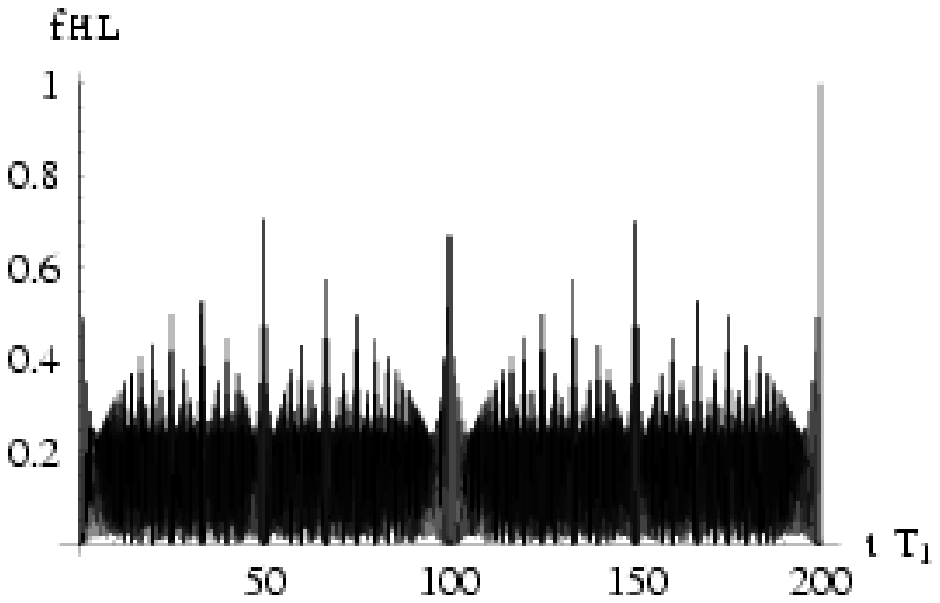}\hfill{}\includegraphics[  width=0.45\textwidth]{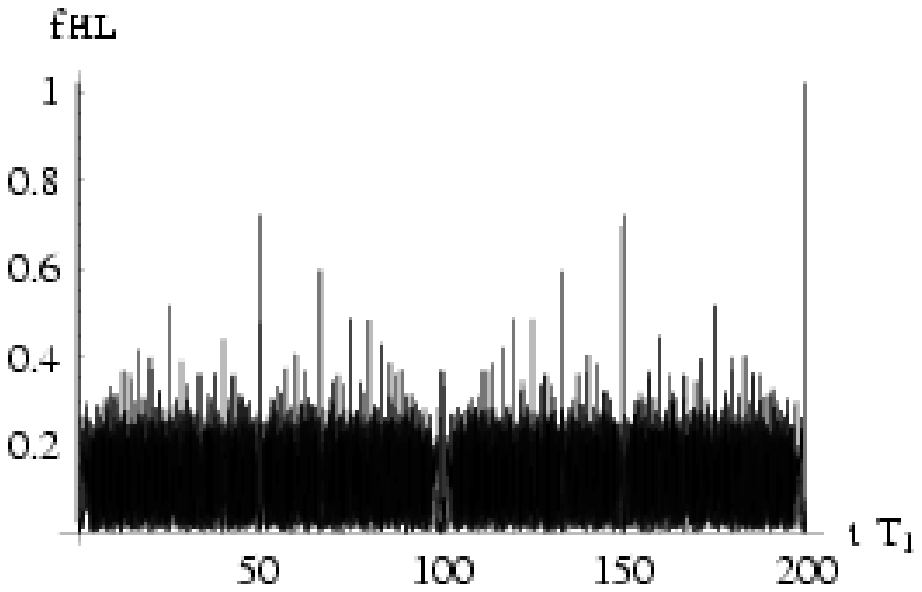}\end{center}

\includegraphics[  width=0.45\textwidth]{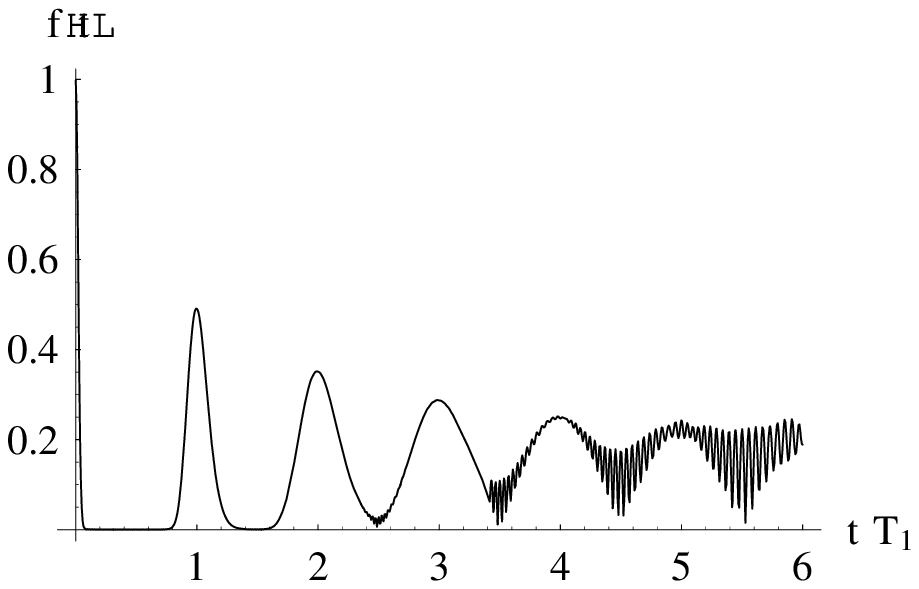}\hfill{}\includegraphics[  width=0.45\textwidth]{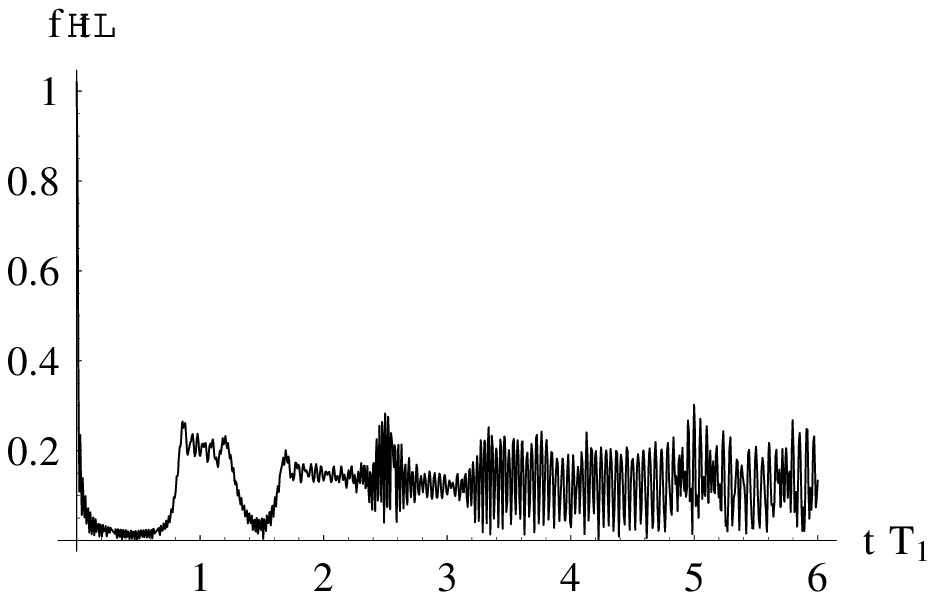}

\begin{center}\includegraphics[  width=0.45\textwidth]{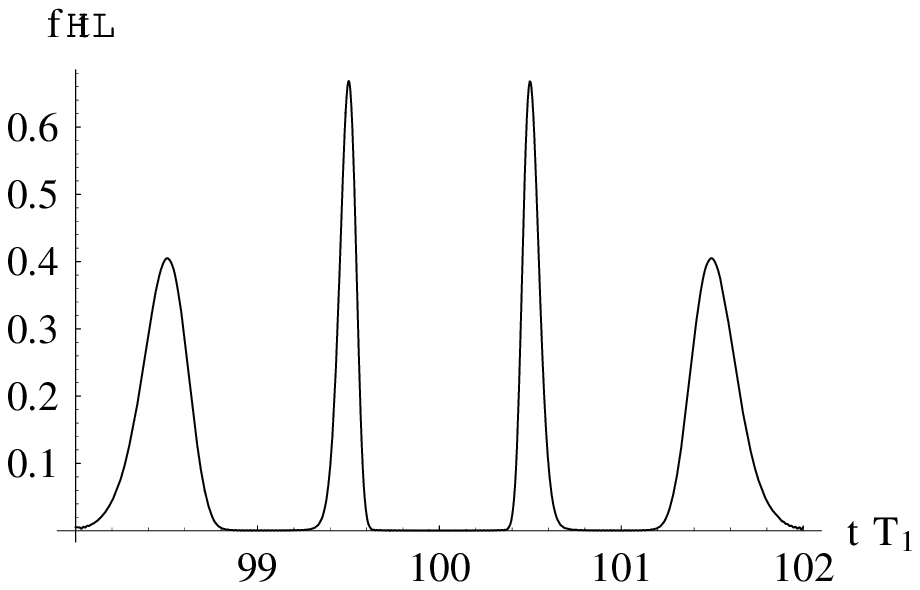}\hfill{}\includegraphics[  width=0.45\textwidth]{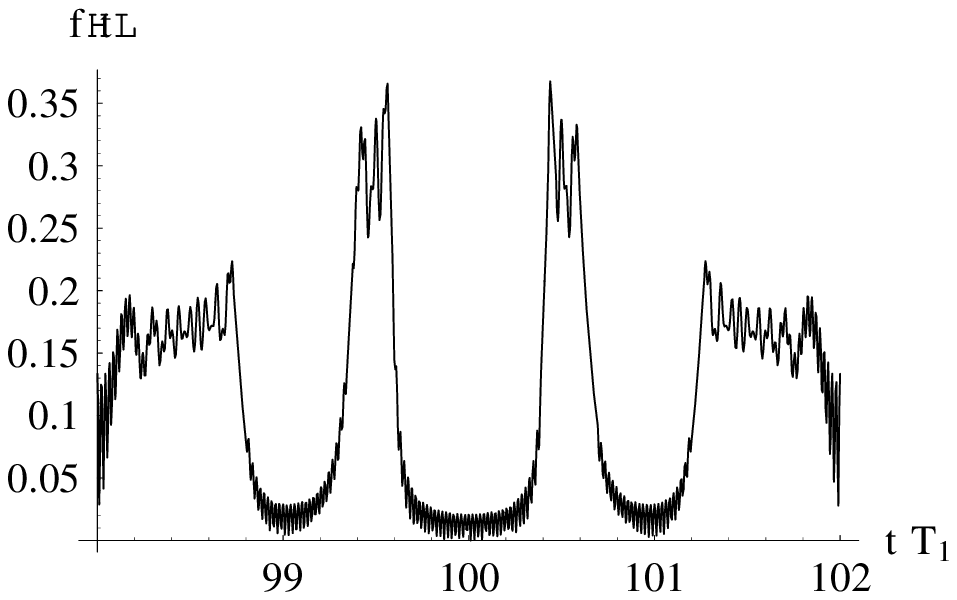}\end{center}

\includegraphics[  width=0.45\textwidth]{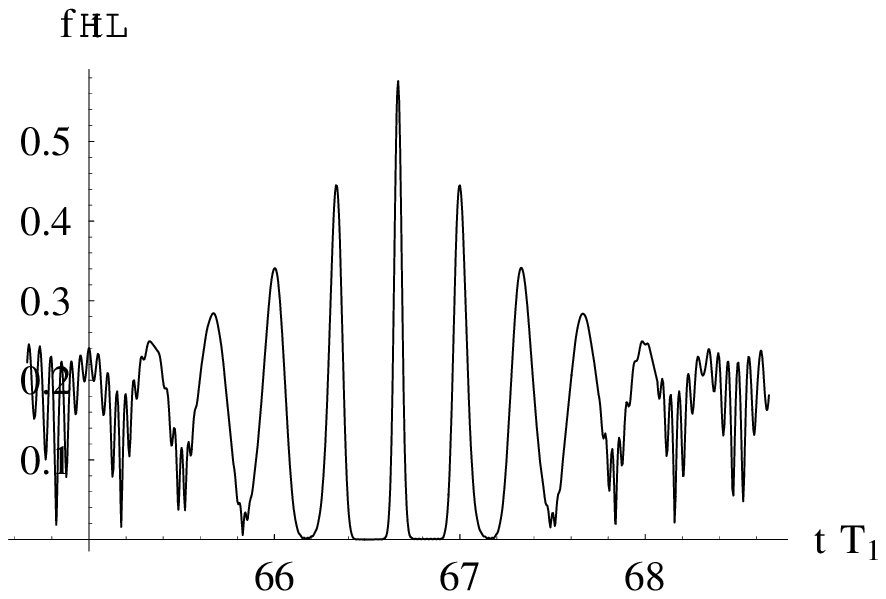}\hfill{}\includegraphics[  width=0.45\textwidth]{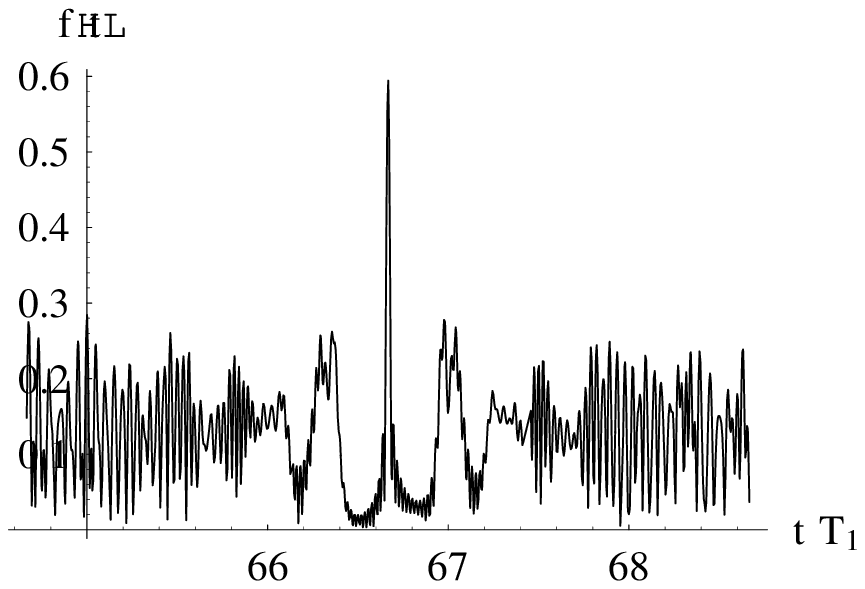}

\caption{\label{fig: qbeats-g} A Gaussian distribution of the $P_{k}$, with
$T_{2}/T_{1}=200$ and $\sigma _{n}=8$. In the plot of the early
evolution, by Equation \ref{eq:gauss-dephase} dephasing should occur
at $t\approx 3.125T_{1}$.}

\caption{\label{fig: qbeats-t} A Gaussian distribution of the $P_{k}$, with
$T_{2}/T_{1}=200$ and $N=8$. In the plot of the early evolution,
by Equation \ref{eq:qbeats-tophat-dephase} dephasing should occur
at $t\approx 3.125T_{1}$.}
\end{figure}

As in the previous chapter, we are primarily interested in distributions
of the weighting constants $P_{n}$ centered around some mean value,
$\bar{n}$, with spread $\Delta n$, such that $1\ll \Delta n\ll \bar{n}$.
In that case we can define $k=n-\bar{n}$, perform a Taylor expansion
of $E(n)$ around $\bar{n}$, and define \begin{equation}
\frac{2\pi }{T_{j}}=\frac{E^{(j)}(\bar{n})}{j!},\end{equation}
we can write $f(t)$ as follows:\begin{equation}
f(t)=\exp \left(iE_{\bar{n}}t\right)\sum _{k}P_{\bar{n}+k}\exp 2\pi i\left(\frac{t}{T_{1}}k+\frac{t}{T_{2}}k^{2}+\frac{t}{T_{3}}k^{3}+...\right).\label{eq:qbeats-basic}\end{equation}
Henceforth, we will ignore the leading phase term, as it has no impact
on $\left\Vert f(t)\right\Vert ^{2}$. Note that the $T_{j}$ are
not necessarily positive. When we interpret them as characteristic
time scales, we will consider their absolute values, but allowing
them to be negative simplifies our formalism. We will use this form
of $f(t)$ for the remainder of this chapter. For examples of two
different sets of $P_{n}$, one a Gaussian distribution and the other
a {}``top hat'' distribution, both with $T_{2}/T_{1}=200$, see
Figures \ref{fig: qbeats-g} and \ref{fig: qbeats-t} . These specific
examples will be treated in detail later in this chapter.%

\section{The Early Phase of the Evolution}

The convenience introduced by considering $P_{n}$ instead of $\psi _{n}(x)$
is that we can use the Poisson summation formula to recast the sum
in Equation \ref{eq:qbeats-basic} as a sum of time-separated signals.
The Poisson summation formula%
\footnote{For a derivation, see \cite{beats:Hilbert1}.%
} is \begin{equation}
\sum _{l=-\infty }^{\infty }\int _{-\infty }^{\infty }g(k)\exp (-2\pi ilk)dk=\sum _{k=-\infty }^{\infty }g_{k},\end{equation}
\label{eq:poisson}where $g(k)$ is an arbitrary continuous function
such that $g(k)=g_{k}$. Using this formula, we can rewrite Equation
\ref{eq:qbeats-basic} as\begin{equation}
f(t)=\sum _{l=-\infty }^{\infty }\int _{-\infty }^{\infty }dkP(k)\exp 2\pi i\left(\left(\frac{t}{T_{1}}-l\right)k+\frac{t}{T_{2}}k^{2}+\frac{t}{T_{3}}k^{3}+...\right),\label{eq:qbeats-early}\end{equation}
where we interpret each integral term as a signal in time. Of course,
this only constitutes a useful simplification when each integral term
has a width in time that is less than the distance from its neighboring
signals. Also note that our choice of $P(k)$ has a significant impact
on the tractability of the integral. Fortunately, the only restriction
on $P(k)$ imposed by the Poisson summation formula is that it be
continuous and that $P(k)=P_{k}$.

\section{\label{sec: beat-fracs}Fractional Revivals}

It would be surprising if we were not able to combine our use of the
Poisson summation formula in the previous section with the fractional
revival technique of the previous chapter. As there, we might begin
by assuming that it is the influence of $T_{2}$ that will dominate
most of the behavior beyond the scale of $T_{1}$. Our definition
of a fractional revival must change a bit -- here we mean the appearance
of smooth oscillation at times less than $T_{2}$. To that end, we
will start near a time that is an integer multiple of $T_{1}$ and
also close to a rational fraction of $T_{2}$, \begin{equation}
t_{p/q}\equiv lT_{1}+\Delta t=\frac{p}{q}T_{2}+\epsilon _{p/q}T_{1}+\Delta t.\end{equation}
Note that $\left|\epsilon _{p/q}\right|\leq 1/2$. At $t=t_{p/q}$
we can rewrite Equation \ref{eq:qbeats-basic} as follows:\begin{equation}
f\left(t=t_{p/q}\right)=\sum _{k=-\infty }^{\infty }P_{\bar{n}+k}\exp 2\pi i\left(\frac{p}{q}k^{2}\right)\exp 2\pi i\left(\frac{\Delta t}{T_{1}}k+\left(\epsilon _{p/q}+\frac{\Delta t}{T_{1}}\right)\frac{T_{1}}{T_{2}}k^{2}+\left(l+\frac{\Delta t}{T_{1}}\right)\frac{T_{1}}{T_{3}}k^{3}+...\right).\label{eq:qbeats-frac1}\end{equation}
If we define \begin{eqnarray}
w_{k} & = & \exp 2\pi i\left(\frac{p}{q}k^{2}\right),\\
s_{k}\left(\Delta t\right) & = & P_{\bar{n}+k}\exp 2\pi i\left(\frac{\Delta t}{T_{1}}k+\left(\epsilon _{p/q}+\frac{\Delta t}{T_{1}}\right)\frac{T_{1}}{T_{2}}k^{2}+\left(l+\frac{\Delta t}{T_{1}}\right)\frac{T_{1}}{T_{3}}k^{3}+...\right),
\end{eqnarray}
we can rewrite the sum as \begin{equation}
f\left(t=t_{p/q}\right)=\sum _{k=-\infty }^{\infty }w_{k}s_{k}\left(\Delta t\right).\label{eq:qbeats-ws}\end{equation}
Note that $w_{k}$, the weighting factor, is related to $\phi _{k}$
(Equation \ref{eq:phi-k}) from the previous chapter -- $w_{k}=\exp \p{-2\pi i}\phi _{k}$.
In analogy with that, we define\[
j=\left\{ \begin{array}{l}
 q,\, q\textrm{ contains one or zero powers of }2,\\
 \frac{q}{2},\, q\textrm{ contains more than one power of }2,\end{array}
\right.\]
 the minimum period for $w_{k}$, just as we defined $l$, the minimum
period of $\phi _{k}$ in Section \ref{sec:FracRev}. Since we can
write $w_{k}=w_{k+mj}$, where $m$ is an integer, we can use the
formula \begin{equation}
\sum _{k=-\infty }^{\infty }a_{k}=\sum _{r=0}^{j-1}\sum _{m=-\infty }^{\infty }a_{r+mj}\end{equation}
to simplify the sum in Equation \ref{eq:qbeats-ws}. The periodicity
of $w_{k}$ lets us extract the weighting factor, and rewrite Equation
\ref{eq:qbeats-ws} as\begin{equation}
f\left(t=t_{p/q}\right)=\sum _{r=0}^{j-1}w_{r}\sum _{m=-\infty }^{\infty }s_{r+mj}\left(\Delta t\right).\end{equation}
This is a form to which we can apply the Poisson summation formula,
\begin{equation}
f\left(t=t_{p/q}\right)=\sum _{r=0}^{j-1}w_{r}\sum _{k=-\infty }^{\infty }\int _{-\infty }^{\infty }s\left(r+mj,\Delta t\right)\exp \left(-2\pi imk\right)dm,\end{equation}
where $s\left(r+mj,\Delta t\right)$ is, again, a continuous extension
of $s_{r+mj}\left(\Delta t\right)$%
\footnote{We just applied the Poisson summation formula to a sum of functions
instead of a simple sum. Fortunately, we can think of each $\Delta t$
as indexing a separate sum, and thus what we have done is a shorthand
for applying the Poisson summation formula to a large (uncountably
infinite) number of different sums.%
}. This is still not so simple a form as we might like, so we introduce
a change of variables, $x=r+mj$, and finally complete our serious
manipulations, \begin{equation}
f\left(t=t_{p/q}\right)=\frac{1}{j}\sum _{r=0}^{j-1}w_{r}\sum _{k=-\infty }^{\infty }\exp \left(2\pi i\frac{r}{j}k\right)\int _{-\infty }^{\infty }s\left(x,\Delta t\right)\exp \left(-2\pi i\frac{x}{j}k\right)dx.\end{equation}
We may, of course, rearrange some of these terms in order to make
this equation look like more of an improvement over Equation \ref{eq:qbeats-frac1}.
We may group all of the time-independent terms together into one new
weighting coefficient, \begin{equation}
W_{k}=\frac{1}{j}\sum _{r=0}^{j-1}\exp 2\pi i\left(r^{2}\frac{p}{q}+r\frac{k}{j}\right),\end{equation}
and a time-dependent term, \begin{equation}
S_{k}\left(\Delta t\right)=\int _{-\infty }^{\infty }dxP\left(\bar{n}+x\right)\exp 2\pi i\left(\left(\frac{\Delta t}{T_{1}}-\frac{k}{j}\right)x+\left(\epsilon _{p/q}+\frac{\Delta t}{T_{1}}\right)\frac{T_{1}}{T_{2}}x^{2}+\left(l+\frac{\Delta t}{T_{1}}\right)\frac{T_{1}}{T_{3}}x^{3}+...\right),\end{equation}
and arrive at our final form for this sum, \begin{equation}
f\left(t=t_{p/q}\right)=\sum _{k=-\infty }^{\infty }W_{k}S_{k}\left(\Delta t\right).\end{equation}

The form of $S_{k}\left(\Delta t\right)$ should evoke the integral
in Equation \ref{eq:qbeats-early}. In fact, it arises by making
the substitutions\begin{eqnarray}
\left(\frac{t}{T_{1}}-l\right) & \rightarrow  & \left(\frac{\Delta t}{T_{1}}-\frac{k}{q}\right),\label{eq:frac-conv1}\\
\frac{t}{T_{2}} & \rightarrow  & \left(\epsilon _{p/q}+\frac{\Delta t}{T_{1}}\right)\frac{T_{1}}{T_{2}},\label{eq:frac-conv2}\\
\frac{t}{T_{j}} & \rightarrow  & \left(l+\frac{\Delta t}{T_{1}}\right)\frac{T_{1}}{T_{j}},\, j\geq 3.
\end{eqnarray}
Where the original solutions are centered on $t/T_{1}=-l$, the fractional
revivals are centered on $\Delta t/T_{1}=k/q$, and the $T_{2}$ and
$T_{j}$ terms are shifted relative to the $T_{1}$ term, though,
due to the hierarchy $\left|T_{1}\right|\ll \left|T_{2}\right|\ll \left|T_{3}\right|\ll ...$,
the shifting should be small. Because of this, the fractional revivals
should look very much like the early evolution, further justifying
our calling them {}``Fractional revivals.'' That one is able to
arrive at the equations for the fractional revivals just by making
substitutions, without doing any new integrals, is quite a computational
nicety. Finally, note that though each $S_{k}$ constitutes a fractional
revival, as in the previous section these fractional revivals are
only distinguishable when they don't overlap significantly.

\section{A Special Case: The Gaussian Distribution}

The integral that will be used heavily in this subsection is \begin{equation}
\int _{-\infty }^{\infty }e^{-ax^{2}+bx}=\sqrt{\frac{\pi }{a}}e^{b^{2}/4a},\, \textrm{Re}\left\{ a\right\} >0.\end{equation}
We will define $P(x)$ to be \begin{equation}
P(x)=\frac{1}{\sqrt{2\pi \Delta n^{2}}}\exp \left[-\frac{\left(x-\bar{n}\right)^{2}}{2\Delta n^{2}}\right],\end{equation}
 a Gaussian distribution of $P_{n}$. In order to simplify the integrals,
we will assume%
\footnote{For a treatment of a Gaussian distribution of coefficients with non-infinite
time scales $T_{1},\, T_{2},\, \textrm{and }T_{3}$, see \cite{beats:Leichtle1}.
The solutions are in terms of Airy functions, and while the results
are quite interesting, plenty of interesting results come from an
case with just two time scales.%
} $T_{j}=\infty ,\, \textrm{for }j\geq 3$.

\subsection{Early Evolution}

We are able to evaluate the integral in Equation \ref{eq:qbeats-early},
since it is a product of two Gaussian functions:\begin{eqnarray}
 & = & \int _{-\infty }^{\infty }dk\frac{1}{\sqrt{2\pi \Delta n^{2}}}\exp \left(-\frac{k^{2}}{2\Delta n^{2}}\right)\exp 2\pi i\left(\left(\frac{t}{T_{1}}-l\right)k+\frac{t}{T_{2}}k^{2}\right)\nonumber \\
 & = & \int _{-\infty }^{\infty }dk\frac{1}{\sqrt{2\pi \Delta n^{2}}}\exp \left(2\pi i\left(\frac{t}{T_{1}}-l\right)k-\left(-2\pi i\frac{t}{T_{2}}+\frac{1}{2\Delta n^{2}}\right)k^{2}\right)\nonumber \\
 & = & \frac{1}{\sqrt{2\pi \Delta n^{2}}}\sqrt{\frac{\pi }{-2\pi i\frac{t}{T_{2}}+\frac{1}{2\Delta n^{2}}}}\exp \left(\frac{-4\pi ^{2}\left(\frac{t}{T_{1}}-l\right)^{2}}{4\left(-2\pi i\frac{t}{T_{2}}+\frac{1}{2\Delta n^{2}}\right)}\right)\nonumber \\
 & = & \frac{1}{\sqrt{1-4\pi i\Delta n^{2}t/T_{2}}}\exp \left(-\frac{2\pi ^{2}\Delta n^{2}}{1-4\pi i\Delta n^{2}t/T_{2}}\left(\frac{t}{T_{1}}-l\right)^{2}\right).\label{eq:qbeats-ugly}
\end{eqnarray}
We can write this as a product of two Gaussians, one of real and one
of imaginary argument, by rewriting the argument of the exponential,\begin{eqnarray}
\frac{a}{1-bi} & = & \frac{a}{1+b^{2}}+\frac{abi}{1+b^{2}},\\
\frac{2\pi ^{2}\Delta n^{2}}{1-4\pi i\Delta n^{2}t/T_{2}} & = & \frac{2\pi ^{2}\Delta n^{2}}{1+16\pi ^{2}\Delta n^{4}t^{2}/T_{2}^{2}}+i\frac{8\pi ^{3}\Delta n^{4}t/T_{2}}{1+16\pi ^{2}\Delta n^{4}t^{2}/T_{2}^{2}},
\end{eqnarray}
defining the (time-dependent) widths,\begin{eqnarray}
\sigma _{r}^{2}(t) & = & \frac{1}{2}\left[\frac{1}{2\pi ^{2}\Delta n^{2}}+8\Delta n^{2}\frac{t^{2}}{T_{2}^{2}}\right],\\
\sigma _{i}^{2}(t) & = & \frac{1}{2}\left[\frac{1}{8\pi ^{3}\Delta n^{4}t/T_{2}}+\frac{2}{\pi }\frac{t}{T_{2}}\right],
\end{eqnarray}
and writing $f(t)$ as \begin{equation}
f(t)=\sum _{l=-\infty }^{\infty }\frac{1}{\sqrt{1-4\pi i\Delta n^{2}t/T_{2}}}\exp \left(-\frac{\left(\frac{t}{T_{1}}-l\right)^{2}}{2\sigma _{r}^{2}(t)}\right)\exp \left(-i\frac{\left(\frac{t}{T_{1}}-l\right)^{2}}{2\sigma _{i}^{2}(t)}\right).\label{eq:qbeats-gauss}\end{equation}

Equation \ref{eq:qbeats-gauss} holds exactly so long as our assumptions
are true (note that a Gaussian distribution of coefficients is always
going to be truncated, so long as there is some minimum $n$, introducing
an error function). $f(t)$ consists of a sum of Gaussian packets
separated in time and expanding as time increases, multiplied by a
phase factor. These Gaussian packets begin to interfere significantly
when their widths becomes comparable to their separation, $2\times 2\sigma _{r}(t)=1$,\begin{eqnarray}
\frac{4}{\sqrt{2}}\sqrt{\frac{1}{2\pi ^{2}\Delta n^{2}}+8\Delta n^{2}\frac{t^{2}}{T_{2}^{2}}} & = & 1,\nonumber \\
16\Delta n^{2}\left(\frac{t}{T_{2}}\right)^{2} & = & \frac{1}{4}-\frac{1}{\pi ^{2}\Delta n^{2}},\nonumber \\
\frac{t}{T_{2}} & = & \frac{1}{8\Delta n}\sqrt{1-\frac{1}{\pi ^{2}\Delta n^{2}}},\nonumber \\
\frac{t}{T_{1}} & = & \frac{T_{2}}{T_{1}}\frac{1}{8\Delta n}\sqrt{1-\frac{1}{\pi ^{2}\Delta n^{2}}}\nonumber \\
 & \approx  & \frac{T_{2}}{T_{1}}\frac{1}{8\Delta n}\left(1-\frac{1}{2\pi ^{2}\Delta n^{2}}\right),\, \Delta n\gg 1.\label{eq:gauss-dephase}
\end{eqnarray}
Notice that this dephasing time is roughly proportional to $1/\Delta n$.
If we then write $\Delta t\approx T_{2}/2\Delta n$ as the uncertainty
of this state in time and consider \begin{eqnarray}
\Delta E & = & \hbar \p{E\left(\bar{n}+\Delta n\right)-E\left(\bar{n}=\Delta n\right)}\nonumber \\
 & = & 2\pi \hbar \p{\frac{2\Delta n}{T_{1}}+\frac{4\bar{n}\Delta n}{T_{2}}}\nonumber \\
 & = & 4\pi \hbar \Delta n\p{\frac{1}{T_{1}}+\frac{2\bar{n}}{T_{2}}},
\end{eqnarray}
as the uncertainty in energy, than we have recovered an uncertainty
relation, \begin{equation}
\Delta t\Delta E\approx 2\pi \hbar \p{\frac{T_{2}}{T_{1}}+2\bar{n}}.\end{equation}
I believe this uncertainty relation to be related to one of the same
form that appears in textbooks on quantum mechanics%
\footnote{See, for example, \cite[112-114]{misc:Griffiths1}.%
}. There, $\Delta t$ is the time it takes for the expectation value
of an observable to change by one standard deviation -- a measure
of how long it takes a system to change substantially. What we are
measuring here is not stated in terms of observables, but it is a
way of measuring how long it takes our signal to undergo an important,
qualitative change.

\subsection{Fractional Revivals}

Fortunately, to consider the case of fractional revivals we need only
apply Equations \ref{eq:frac-conv1} and \ref{eq:frac-conv2} to
Equations \ref{eq:qbeats-gauss}. Doing this, we find \begin{equation}
f\left(t=t_{p/q}\right)=\sum _{k=-\infty }^{\infty }\frac{1}{\sqrt{1-4\pi i\Delta n^{2}\left(\epsilon _{p/q}+\Delta t/T_{1}\right)T_{1}/T_{2}}}\exp \left(-\frac{\left(\frac{\Delta t}{T_{1}}+\frac{k}{q}\right)^{2}}{2\sigma _{r}^{2}\left(t=t_{p/q}\right)}\right)\exp \left(-i\frac{\left(\frac{\Delta t}{T_{1}}+\frac{k}{q}\right)^{2}}{2\sigma _{i}^{2}\left(t=t_{p/q}\right)}\right),\end{equation}
\begin{eqnarray}
\sigma _{r}^{2}\left(t=t_{p/q}\right) & = & \frac{1}{2}\left[\frac{1}{2\pi ^{2}\Delta n^{2}}+8\Delta n^{2}\left(\epsilon _{p/q}+\frac{\Delta t}{T_{1}}\right)^{2}\left(\frac{T_{1}}{T_{2}}\right)^{2}\right],\\
\sigma _{i}^{2}\left(t=t_{p/q}\right) & = & \frac{1}{2}\left[\frac{1}{8\pi ^{3}\Delta n^{4}\left(\epsilon _{p/q}+\Delta t/T_{1}\right)T_{1}/T_{2}}+\frac{2}{\pi }\left(\epsilon _{p/q}+\frac{\Delta t}{T_{1}}\right)\frac{T_{1}}{T_{2}}\right].
\end{eqnarray}
Unfortunately, we cannot simply apply Equations \ref{eq:frac-conv1}
and \ref{eq:frac-conv2} to Equation \ref{eq:gauss-dephase}, as
the separation between peaks is no longer $1$, but $1/q$. Instead,
we require $2\sigma _{r}\left(t=t_{p/q}\right)=1/q$. This leads us
to the dephasing condition\begin{eqnarray}
\frac{4}{\sqrt{2}}\sqrt{\frac{1}{2\pi ^{2}\Delta n^{2}}+8\Delta n^{2}\left(\epsilon _{p/q}+\frac{\Delta t}{T_{1}}\right)^{2}\left(\frac{T_{1}}{T_{2}}\right)^{2}} & = & \frac{1}{q},\nonumber \\
16\Delta n^{2}\left(\epsilon _{p/q}+\frac{\Delta t}{T_{1}}\right)^{2}\left(\frac{T_{1}}{T_{2}}\right)^{2} & = & \frac{1}{4q^{2}}-\frac{1}{\pi ^{2}\Delta n^{2}},\nonumber \\
\left(\epsilon _{p/q}+\frac{\Delta t}{T_{1}}\right) & = & \frac{T_{2}}{T_{1}}\frac{1}{8\Delta n}\sqrt{\frac{1}{q^{2}}-\frac{1}{\pi ^{2}\Delta n^{2}}},\nonumber \\
\left(\epsilon _{p/q}+\frac{\Delta t}{T_{1}}\right) & = & \frac{T_{2}}{T_{1}}\frac{1}{8\Delta n}\frac{1}{q}\sqrt{1-\frac{1}{q^{2}\pi ^{2}\Delta n^{2}}}\nonumber \\
 & \approx  & \frac{1}{q}\frac{T_{2}}{T_{1}}\frac{1}{8\Delta n}\left(1-\frac{1}{2q^{2}\pi ^{2}\Delta n^{2}}\right),\, \Delta n\gg 1.\label{eq:gauss-frac-dephase}
\end{eqnarray}
Comparing Equations \ref{eq:gauss-frac-dephase} and \ref{eq:gauss-dephase},
we see that for a fractional revival at $t=t_{p/q}$, we will see
almost exactly the same time evolution we saw at early times, contracted
by a factor of $1/q$. Refer again to Figure \ref{fig: qbeats-g}
if you don't believe us.

\section{A Special Case: The Flat Distribution}

The integral that will be used heavily in this subsection is \begin{equation}
\int _{-N}^{N}e^{-ax^{2}+bx}=\sqrt{\frac{\pi }{a}}e^{b^{2}/4a}\frac{1}{2}\left(\textrm{Erf}\left(\frac{2aN+b}{2\sqrt{a}}\right)+\textrm{Erf}\left(\frac{2aN-b}{2\sqrt{a}}\right)\right),\end{equation}
the integral of a Gaussian truncated at $\pm N$, where $\Erf{x}$
is the error function%
\footnote{For a discussion of Erf, see either the more traditional \cite{misc:Stegun1}
or the hip and modern \cite{misc:Erf}.%
}, defined as\begin{equation}
\Erf{x}=\frac{2}{\sqrt{\pi }}\int _{0}^{x}e^{-t^{2}}dt.\end{equation}
Erf is odd, $\Erf{0}=0$, $\Erf{\infty }=1$. From this we can conclude
that as $N$ becomes large, the error function term should go to one.
We will define the {}``top hat'' distribution $P(x)$ to be \begin{equation}
P(x)=\left\{ \begin{array}{l}
 1,\, \left|x-\bar{n}\right|\leq N,\\
 0,\, \left|x-\bar{n}\right|>N,\end{array}
\right.\end{equation}
which describes a flat distribution of width $2N$ centered around
$\bar{n}$. This is an interesting case for several reasons. First,
it is the other obviously integrable case, along with the Gaussian
case. It is also an example of a distribution with sharply limited
extent in $n$-space, unlike the Gaussian. It is, in a sense, the
most un-Gaussian of localized distributions. Finally, if we study
$\lim _{N\rightarrow \infty }$, we can gain some understanding of
an {}``ultra-localized'' spatial distribution%
\footnote{Remember that our ultimate goal is to relate these results to quantum
mechanics problems, and we would expect that a broad distribution
in energy-space would correspond to a narrow distribution in position-space.
While in doing this problem we will break a great number of mathematical
rules. $P(x)$ is not continuous, as the Poisson summation formula
demands, which casts a shadow on any analysis that we do. Studying
$\lim _{N\rightarrow \infty }$ has us implicitly using negative energies,
and if we were to try this in quantum mechanics we would have a non-normalizable
state that probably wasn't differentiable, and thus not even a solution
to the Schrödinger Equation. All the same, we can hope through this
savagery to gain some physical insight.%
} As in the previous example, in order to simplify the integrals, we
will assume $T_{j}=\infty \, \textrm{for }j\geq 3$.

\subsection{Early Evolution}

\begin{figure}[hbtp]
\begin{center}\includegraphics[  width=3in]{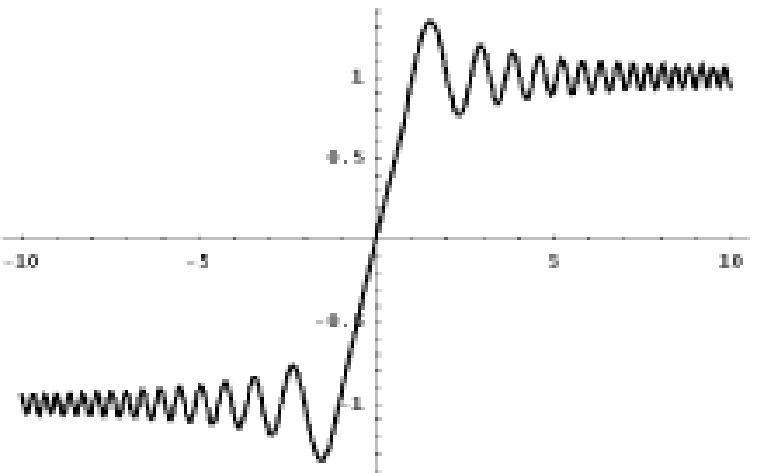}\hfill{}\includegraphics[  width=3in]{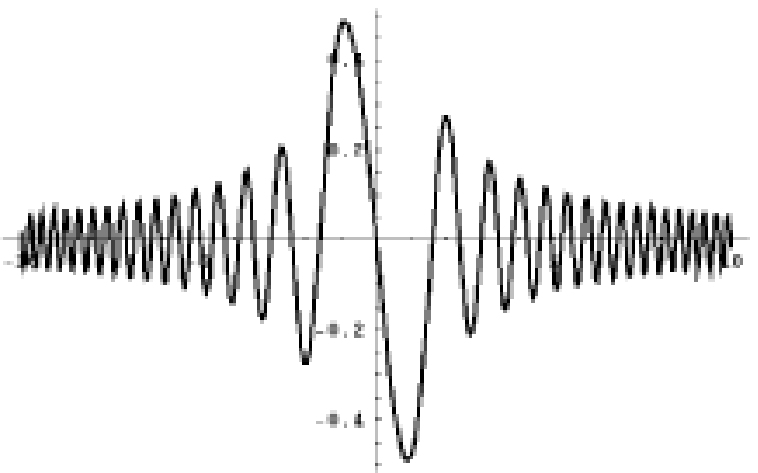}\end{center}

\caption{\label{fig: re-erf} $\re{\Erf{\phi x}}$, $x\in [-10,10]$.}

\caption{\label{fig: im-erf} $\im{\Erf{\phi x}}$, $x\in [-10,10]$. }
\end{figure}

As above, we are able to evaluate the integral in Equation \ref{eq:qbeats-early},
since it is a truncated Gaussian function:\begin{eqnarray}
f(t) & = & \int _{-N}^{N}dk\frac{1}{\sqrt{2\pi N^{2}}}\exp \left(\left(\frac{t}{T_{1}}-l\right)2\pi ik+2\pi i\frac{t}{T_{2}}k^{2}\right)\nonumber \\
 & = & \frac{1}{\sqrt{2\pi N^{2}}}\sqrt{\frac{\pi }{-2\pi it/T_{2}}}\exp \left(-\frac{4\pi ^{2}\left(\frac{t}{T_{1}}-l\right)^{2}}{8\pi it/T_{2}}\right)\frac{1}{2}\begin{array}[t]{l}
 \left(\Erf{\frac{-4\pi i\frac{t}{T_{2}}N+2\pi i\left(\frac{t}{T_{1}}-l\right)}{2\sqrt{-2\pi it/T_{2}}}}\right.\\
 \left.+\Erf{\frac{-4\pi i\frac{t}{T_{2}}N-2\pi i\left(\frac{t}{T_{1}}-l\right)}{2\sqrt{-2\pi it/T_{2}}}}\right).\end{array}
\label{eq:tophat}
\end{eqnarray}
We can simplify the arguments of the error functions, \begin{equation}
\frac{-4\pi i\frac{t}{T_{2}}N\pm 2\pi i\left(\frac{t}{T_{1}}-l\right)}{2\sqrt{-2\pi it/T_{2}}}=\frac{1}{2}\sqrt{-i}\sqrt{2\pi \frac{t}{T_{2}}}\left(2N\mp \frac{T_{2}}{t}\left(\frac{t}{T_{1}}-l\right)\right),\label{eq:qbeats-erfargs}\end{equation}
and recognize the sum of error functions, which shape the wave packet
for a given $l$, as being of the form\begin{equation}
\Erf{\alpha \phi (A-x)}+\Erf{\alpha \phi (A+x)},\end{equation}
 where \begin{eqnarray}
\phi  & = & \sqrt{-i},\\
\alpha  & = & \sqrt{\frac{\pi }{2}\frac{t}{T_{2}}},\\
A & = & 2N,\\
x & = & \frac{T_{2}}{t}\p{\frac{t}{T_{1}}-l}.
\end{eqnarray}
From here, it is helpful to first examine Figures \ref{fig: re-erf}
and \ref{fig: im-erf}%
, plots of $\Erf{\phi x}$ and $\Erf{-\phi x}$. These suggest that
if we are to study the modulus of Equation \ref{eq:tophat}, that
we would do well to consider the real and imaginary parts separately.
Doing this, we find \begin{eqnarray}
\left\Vert \Erf{\alpha \phi (A-x)}+\Erf{\alpha \phi (A+x)}\right\Vert ^{2} & = & \begin{array}[t]{l}
 \left\Vert \Erf{\alpha \phi (A-x)}\right\Vert ^{2}+\left\Vert \Erf{\alpha \phi (A+x)}\right\Vert ^{2}+\\
 \Erf{\alpha \phi (A-x)}\Erf{\alpha \phi ^{*}(A+x)}+\\
 \Erf{\alpha \phi ^{*}(A-x)}\Erf{\alpha \phi (A+x)}\end{array}
\nonumber \\
 & = & \begin{array}[t]{l}
 \begin{array}[t]{l}
 \re{\Erf{\alpha \phi (A-x)}}^{2}+\im{\Erf{\alpha \phi (A-x)}}^{2}+\\
 \re{\Erf{\alpha \phi (A+x)}}^{2}+\im{\Erf{\alpha \phi (A+x)}}^{2}+\\
 2\re{\Erf{\alpha \phi (A-x)}}\re{\Erf{\alpha \phi (A+x)}}+\\
 2\im{\Erf{\alpha \phi (A-x)}}\im{\Erf{\alpha \phi (A+x)}}.\end{array}
\end{array}
\label{eq:qbeats-flat-ugly}
\end{eqnarray}
Here we see that, roughly, $\im{\Erf{\phi x}}=0+\epsilon (x)$ and
$\re{\Erf{\phi x}}=\textrm{Sign}(x)\left(1+\epsilon ^{\prime }(x)\right)$,
where $\epsilon (x)$ and $\epsilon ^{\prime }\p{x}$ are functions
that describe the oscillations. Obviously, both $\epsilon (x)$ and
$\epsilon ^{\prime }(x)$ decrease as $\left|x\right|$ increases.
We can then observe that in the region $x<-A$, the contribution from
Re terms in Equation \ref{eq:qbeats-flat-ugly} are roughly \begin{eqnarray}
 & = & \re{\Erf{\alpha \phi (A+x)}}^{2}+\re{\Erf{\alpha \phi (A-x)}}^{2}+2\re{\Erf{\alpha \phi (A-x)}}\re{\Erf{\alpha \phi (A+x)}}\nonumber \\
 & = & \left(-1-\epsilon ^{\prime }(x)\right)^{2}+(-1)^{2}+2\left(-1-\epsilon ^{\prime }(x)\right)\nonumber \\
 & \approx  & 1+2\epsilon ^{\prime }(x)+1-2-2\epsilon ^{\prime }(x)=0.
\end{eqnarray}
The same calculation can be performed for the Im\{\} terms, and for
$x>A$, thus demonstrating that anything interesting that happens
in the modulus of this sum of error functions happens between $-A$
and $A$. Although our characterization of the width of the pulse
described by Equation \ref{eq:qbeats-flat-ugly} is not so transparent
as the standard deviation of a Gaussian curve, we can say that we
have described a pulse are centered on $x=0$, with width $2A$. 

Analysis of Equation \ref{eq:qbeats-erfargs}, the argument in our
particular problem, is also not so simple as it was in the case of
the Gaussian distribution. First, let us find the edges of our pulse
-- times that satisfy $x=-A$ and $x=A$, respectively, in Equation
\ref{eq:qbeats-flat-ugly}. We will define these two times, $t_{2}$
and $t_{1}$, as follows:\begin{eqnarray}
\frac{t_{1}}{T_{1}} & \equiv  & \frac{l}{1+2N\frac{T_{1}}{T_{2}}},\\
\frac{t_{2}}{T_{1}} & \equiv  & \frac{l}{1-2N\frac{T_{1}}{T_{2}}}.
\end{eqnarray}
Clearly, $t_{2}-t_{1}$ defines the width of the pulse, and $(t_{1}+t_{2})/2$
defines its center. If we define the spacing between pulses as the
distance between their centers, then that spacing is \begin{equation}
\frac{1}{2}\left(\frac{1}{1+2N\frac{T_{1}}{T_{2}}}+\frac{1}{1-2N\frac{T_{1}}{T_{2}}}\right)=\frac{1}{1-\left(2N\frac{T_{1}}{T_{2}}\right)^{2}}.\end{equation}
 We are now looking for the spacing for the the value of $l$ for
which the spacing between packets equals the width of a packet, the
value of $l$ for which we think the packets will overlap enough to
interfere substantially: \begin{eqnarray}
2\left(\frac{t_{2}}{T_{1}}-\frac{t_{1}}{T_{1}}\right) & = & \frac{1}{1-\left(2N\frac{T_{1}}{T_{2}}\right)^{2}},\nonumber \\
\frac{8NlT_{1}/T_{2}}{1-\left(2N\frac{T_{1}}{T_{2}}\right)^{2}} & = & \frac{1}{1-\left(2N\frac{T_{1}}{T_{2}}\right)^{2}},\nonumber \\
l & = & \frac{T_{2}}{T_{1}}\frac{1}{8N}.
\end{eqnarray}
We convert this into a time by multiplying it by $T_{1}$,\begin{equation}
t=T_{1}l=\frac{T_{2}}{8N},\end{equation}
and finally write \begin{equation}
\frac{t}{T_{2}}=\frac{1}{8N}.\label{eq:qbeats-tophat-dephase}\end{equation}
As in the Gaussian case, the time after which our signal dephases
depends on $1/N$. Looking for an uncertainty relation, we have $\Delta t=T_{2}/2N$
and, as in the Gaussian example, $\Delta E=4\pi \hbar N\p{1/T_{1}+2\bar{n}/T_{2}}$and
we recover an uncertainty relation,\begin{equation}
\Delta t\Delta E=2\pi \hbar \p{\frac{T_{2}}{T_{1}}+2\bar{n}},\end{equation}
the same relation that we found in the Gaussian example, though in
that case the result was not exact. If we refer to Figure \ref{fig: qbeats-t},
though, we find that our dephasing time doesn't seem to fall in the
right place. I submit that the noise between the $t=2T_{1}$ and the
$t=3T_{1}$ pulses is not the dephasing that we were looking for,
as it is followed by the discernible top of the $t=3T_{1}$ pulse.

\subsection{Fractional Revivals}

Applying Equations \ref{eq:frac-conv1} and \ref{eq:frac-conv2}
to the results of the previous section, we find our signal to be\begin{equation}
f\left(t=t_{p/q}\right)=\begin{array}[t]{r}
 \frac{1}{2\sqrt{2\pi N^{2}}}\sqrt{\frac{\pi }{-2\pi i\left(\epsilon _{p/q}+\Delta t/T_{1}\right)\left(T_{1}/T_{2}\right)}}\exp \left(-\frac{4\pi ^{2}\left(\frac{\Delta t}{T_{1}}-\frac{k}{q}\right)^{2}}{8\pi i\left(\epsilon _{p/q}+\Delta t/T_{1}\right)\left(T_{1}/T_{2}\right)}\right)\times \\
 \left(\Erf{\frac{-4\pi i\left(\epsilon _{p/q}+\Delta t/T_{1}\right)\left(T_{1}/T_{2}\right)N-2\pi i\left(\frac{\Delta t}{T_{1}}-\frac{k}{q}\right)}{2\sqrt{-2\pi i\left(\epsilon _{p/q}+\Delta t/T_{1}\right)\left(T_{1}/T_{2}\right)}}}\right.\\
 \left.\Erf{\frac{-4\pi i\left(\epsilon _{p/q}+\Delta t/T_{1}\right)\left(T_{1}/T_{2}\right)N+2\pi i\left(\frac{\Delta t}{T_{1}}-\frac{k}{q}\right)}{2\sqrt{-2\pi i\left(\epsilon _{p/q}+\Delta t/T_{1}\right)\left(T_{1}/T_{2}\right)}}}\right),\end{array}
\end{equation}
which could be simplified. More interesting is, of course, our dephasing
condition. The method that we use to find this is the same one that
we used in the previous section.

We started with our simplification of the arguments of the error functions,
in Equation \ref{eq:qbeats-erfargs}, which we will convert,\begin{equation}
\frac{1}{2}\sqrt{-2\pi i\frac{t}{T_{2}}}\left(2N\mp \frac{T_{2}}{t}\left(\frac{t}{T_{1}}-l\right)\right)\rightarrow \frac{1}{2}\sqrt{-2\pi i\left(\epsilon _{p/q}+\frac{\Delta t}{T_{1}}\right)\left(\frac{T_{1}}{T_{2}}\right)}\left(2N\mp \frac{\frac{\Delta t}{T_{1}}-\frac{k}{q}}{\left(\epsilon _{p/q}+\Delta t/T_{1}\right)\left(T_{1}/T_{2}\right)}\right).\end{equation}
We now look for the roots of this,\begin{eqnarray}
\frac{1}{\left(\epsilon _{p/q}+\Delta t/T_{1}\right)\left(T_{1}/T_{2}\right)}\left(\frac{\Delta t}{T_{1}}-\frac{k}{q}\right) & = & \pm 2N,\nonumber \\
\frac{\Delta t}{T_{1}}-\frac{k}{q} & = & \pm 2N\frac{T_{1}}{T_{2}}\left(\epsilon _{p/q}+\frac{\Delta t}{T_{1}}\right),\nonumber \\
\frac{\Delta t}{T_{1}}\left(1\mp 2N\frac{T_{1}}{T_{2}}\right) & = & \frac{k}{q}\pm 2N\frac{T_{1}}{T_{2}}\epsilon _{p/q},\nonumber \\
\frac{\Delta t}{T_{1}} & = & \frac{\frac{k}{q}\pm 2N\frac{T_{1}}{T_{2}}\epsilon _{p/q}}{1\mp 2N\frac{T_{1}}{T_{2}}},
\end{eqnarray}
and label those roots, \begin{eqnarray}
\frac{\Delta t_{1}}{T_{1}} & \equiv  & \frac{\frac{k}{q}+2N\frac{T_{1}}{T_{2}}\epsilon _{p/q}}{1-2N\frac{T_{1}}{T_{2}}},\\
\frac{\Delta t_{2}}{T_{1}} & \equiv  & \frac{\frac{k}{q}-2N\frac{T_{1}}{T_{2}}\epsilon _{p/q}}{1+2N\frac{T_{1}}{T_{2}}}.
\end{eqnarray}
We're now interested in the width of the pulses, \begin{eqnarray}
\left(\frac{\Delta t_{1}}{T_{1}}-\frac{\Delta t_{2}}{T_{1}}\right) & = & \frac{4N\frac{T_{1}}{T_{2}}\epsilon _{p/q}+4N\frac{T_{1}}{T_{2}}\frac{k}{q}}{1-\left(2N\frac{T_{1}}{T_{2}}\right)^{2}}\nonumber \\
 & = & \left(\epsilon _{p/q}+\frac{k}{q}\right)\frac{4N\frac{T_{1}}{T_{2}}}{1-\left(2N\frac{T_{1}}{T_{2}}\right)^{2}},
\end{eqnarray}
and the spacing between pulses, \begin{equation}
\frac{1}{2}\left(\frac{1/q}{1-2N\frac{T_{1}}{T_{2}}}+\frac{1/q}{1+2N\frac{T_{1}}{T_{2}}}\right)=\frac{1}{q}\frac{1}{1-\left(2N\frac{T_{1}}{T_{2}}\right)^{2}}.\end{equation}
We now find when these are equal,\begin{eqnarray}
2\left(\epsilon _{p/q}+\frac{k}{q}\right)\frac{4N\frac{T_{1}}{T_{2}}}{1+\left(2N\frac{T_{1}}{T_{2}}\right)^{2}} & = & \frac{1}{q}\frac{1}{1+\left(2N\frac{T_{1}}{T_{2}}\right)^{2}},\nonumber \\
\frac{k}{q} & = & \frac{T_{2}}{T_{1}}\frac{1}{8N}\frac{1}{q}-\epsilon _{p/q},
\end{eqnarray}
and see that for fractional revivals as well, the dephasing time is
proportional to $1/N$, and decreases as $q$ increases. Again, we
see that the behavior near $t=0$ will appear again, squished by a
factor of $1/q$. In fact, this result corresponds as closely to the
one in the Gaussian case (Equation \ref{eq:gauss-frac-dephase})
as our results for the early evolution did. It would appear that the
principle difference between the signals generated by these two dissimilar
distributions is that the sharp edges of the top-hat distribution
create the same kind of high-frequency {}``ringing'' that we would
expect from a truncated sum of harmonic functions.

\section{Summary}

In this chapter we studied a zero-dimensional system that exhibits
{}``quantum beats,'' which is both of experimental interest, and,
due to its low dimensionality, is a nice system on which to introduce
our Poisson summation formula technique. The principal advantage of
this technique is that it allows us to quantify to dephasing of an
initial wavepacket. We have seen that, for quantum beats, fractional
revivals are related to the early evolution of the packet by the simple
substitution of a few terms. In the specific examples that we have
considered, a Gaussian distribution of weighting coefficients and
a {}``top hat'' distribution, we have discovered nearly-identical
dephasing conditions. If $\Delta n$ is the {}``spread'' of the
weighting coefficients, then the initial dephasing time goes as $T_{2}/8\Delta n$,
and the dephasing time for a fractional revival at $t=t_{p/q}$ goes
as $T_{2}/8\Delta nq$, and the fractional revival has the same number
of {}``clean'' oscillations as there were oscillations near $t=0$.
That these distributions are so different from each other suggests
that many well-localized distributions of coefficients should demonstrate
similar behavior.

\chapter{\label{ch: pois}The Connection Between Beats and Carpets}

One shortcoming of the previous chapter is that it describes a zero-dimensional
problem which, while interesting, lacks the spatial features that
we would like to consider. The main advantage of the quantum beats
approach is that it allows us to not only identify pseudoclassical
behavior at full and fractional revivals, but also to associate a
particular lifetime with those revivals. In both of the examples that
we considered we found that the lifetime of the revivals was proportional
to $1/\Delta n$, the spread of the wavepacket in energy-space, but
two cases hardly exhausts all of the possible localized wavepackets.
Although we could think of revivals of a spatial wavefunction as revivals
of each point, we know that even a simple distribution of weighting
coefficients will become complicated when it is combined with the
eigenfunctions evaluated at a particular point. How, then, can we
connect this approach with quantum carpets?

\section{Abuse of the Poisson Summation Formula}

So long as we can come up with a continuous extension of a set of
functions $f_{m}(\vec{x})$, we have a generalization of the Poisson
summation formula,\begin{equation}
\sum _{m=-\infty }^{\infty }f_{m}(\vec{x})=\sum _{l=-\infty }^{\infty }\int _{-\infty }^{\infty }f(m,\vec{x})e^{-2\pi iml}dm.\label{eq:poisson-2}\end{equation}
Not all sets of functions will have an obvious continuous extension,
but we can find extensions in cases where our eigenfunctions are continuous
functions of $m$ and $\vec{x}$. So long as we are able to make this
extension, it is easy to apply the formulae derived in the previous
parts of this chapter to an $n$-dimensional problem, by making the
substitution $P_{k}\rightarrow c_{k}\psi _{k}(\vec{x})$. This is
our crucial observation in this section, but its merit will not be
clear until we have examined at least one case.

\section{A Special Case: Gaussian Weighting Coefficients in the Infinite Square
Well}

The attentive reader will have noticed that the integral hiding in
Equation \ref{eq:poisson-2} is a bit intimidating. In order to be
sure that it's even worth thinking about, we'll do it for the simpliest
case I can think of -- the infinite square well. We'll take advantage
of two characteristics of solutions to the square well: the eigenfunctions
are just sine waves, easily converted into exponentials, and the spectrum
is quadratic in the quantum number, meaning that a Taylor expansion
of the spectrum is exact and gives just two time-scales. All of our
integrals will reduce to Gaussian integrals, which we can do.

\subsection{Algebra}

For this case, we will write our wavefunction as\begin{equation}
\Psi (x,t)=\sqrt{\frac{2}{L}}\sum _{k=-\infty }^{\infty }c_{\bar{n}+k}\sin \left(\frac{\bar{n}+k}{L}\pi x\right)\exp -2\pi i\left(\frac{t}{T_{1}}k+\frac{t}{T_{2}}k^{2}\right),\label{eq:qbeats-nochange}\end{equation}
where we have set the weighting coefficients to be\begin{equation}
c_{\bar{n}+k}=\frac{1}{\sqrt{2\pi \Delta n^{2}}}e^{-\frac{k^{2}}{2\Delta n^{2}}},\end{equation}
and defined \begin{eqnarray}
T_{1} & = & \frac{2\pi }{2\pi ^{2}\bar{n}/L^{2}}=\frac{L^{2}}{\pi \bar{n}},\\
T_{2} & = & \frac{2\pi }{\pi ^{2}/L^{2}}=\frac{2L^{2}}{\pi },\\
\frac{T_{1}}{T_{2}} & = & \frac{1}{2\bar{n}}.\label{eq:qbeats-t1/t2}
\end{eqnarray}
What follows is a godawful bit of algebra, and in order to make it
easier to follow (and ensure that I get it right), I'm going to temporarily
replace all of these compound constants with simpler ones. I'm also
going to ignore multiplication by leading constants -- don't worry,
I'll put it all back in later. If you're not interested in this, you
won't have missed much by skipping directly to Equation \ref{eq:qbeats-alg}.
We will define\begin{eqnarray}
\xi  & = & \frac{x}{L},\\
\tau  & = & \frac{t}{T_{1}},
\end{eqnarray}
and use these, along with Equation \ref{eq:qbeats-t1/t2} to define
the following:\begin{eqnarray}
a & = & \frac{1}{2\Delta n^{2}},\\
b & = & \frac{\bar{n}\pi }{L}x=\frac{1}{2}\frac{T_{2}}{T_{1}}\pi \xi ,\\
c & = & \frac{\pi }{L}x=\pi \xi ,\\
d & = & 2\pi \frac{t}{T_{1}}=2\pi \tau ,\\
f & = & 2\pi \frac{t}{T_{2}}=2\pi \frac{T_{1}}{T_{2}}\tau ,\\
g & = & 2\pi l.
\end{eqnarray}
Note that these are all real quantities. Having defined these, we
can rewrite our wavefunction as \begin{equation}
\Psi (x,t)=\sqrt{\frac{2}{L}}\sum _{k=-\infty }^{\infty }\sqrt{\frac{a}{\pi }}\exp \left(-ak^{2}\right)\sin \left(b+ck\right)\exp -i\left(dk+fk^{2}\right).\end{equation}
We now rewrite the sine term as a sum of exponentials,\begin{equation}
\Psi (x,t)=\frac{1}{2i}\sqrt{\frac{2}{L}}\sum _{k=-\infty }^{\infty }\sqrt{\frac{a}{\pi }}\exp \left(-ak^{2}\right)\left(\exp i\left(b+ck\right)-\exp -i\left(b+ck\right)\right)\exp i\left(-dk-fk^{2}\right),\end{equation}
combine the exponentials,\begin{equation}
\Psi (x,t)=\frac{1}{2i}\sqrt{\frac{2}{L}}\sum _{k=-\infty }^{\infty }\sqrt{\frac{a}{\pi }}\left(\exp \left(bi+i(c-d)k-(a+if)k^{2}\right)-\exp \left(-bi-i(c+d)k-(a+if)k^{2}\right)\right),\end{equation}
apply the Poisson summation formula,\begin{equation}
\Psi (x,t)=\frac{1}{2i}\sqrt{\frac{2}{L}}\sqrt{\frac{a}{\pi }}\sum _{l=-\infty }^{\infty }\int _{-\infty }^{\infty }dk\rowc{\exp \left(bi+i(c-d-g)k-(a+if)k^{2}\right)-}{\exp \left(-bi-i(c+d+g)k-(a+if)k^{2}\right)}\end{equation}
and do the integration, using the formula $\int _{-\infty }^{\infty }dk\exp \left(-\alpha x^{2}+\beta x\right)=\sqrt{\pi /\alpha }\exp \left(\beta ^{2}/4\alpha \right)$:\begin{equation}
\Psi (x,t)=\frac{1}{2i}\sqrt{\frac{2}{L}}\sqrt{\frac{a}{\pi }}\sum _{l=-\infty }^{\infty }\sqrt{\frac{\pi }{a+if}}\row{\exp \left(bi-\frac{\left(c-(d+g))^{2}\right)}{4(a+if)}\right)-}{\exp \left(-bi-\frac{\left(c+(d+g))^{2}\right)}{4(a+if)}\right)}\end{equation}

We have, in a sense, solved the problem now, but we have gained little
unless we can understand this result in a simpler way than we understood
the original expression. So long as the terms of the sum are localized
in their own regions of spacetime%
\footnote{This isn't obvious, but it is something we can hope for, as the Poisson
summation formula works kind of like a Fourier transform.%
}, and the region defined by one value of $l$ does not overlap significantly
with the region defined by another value of $l,$ we can make the
substantial simplification of studying just one term in the sum to
understand a particular region. There are, however, some algebraic
simplifications that we will make first. 

We perform the square in the exponentials,\begin{equation}
\Psi (x,t)=\frac{1}{2i}\sqrt{\frac{2}{L}}\sqrt{\frac{a}{\pi }}\sum _{l=-\infty }^{\infty }\sqrt{\frac{\pi }{a+if}}\row{\exp \left(bi-\frac{\left(c^{2}+(d+g)^{2}-2c(d+g)\right)}{4(a+if)}\right)-}{\exp \left(-bi-\frac{\left(c^{2}+(d+g)^{2}+2c(d+g)\right)}{4(a+if)}\right)}\end{equation}
We then turn the difference of exponentials into a product of a sine
function and another exponential,\begin{equation}
\Psi (x,t)=\sqrt{\frac{2}{L}}\sqrt{\frac{a}{\pi }}\sum _{l=-\infty }^{\infty }\sqrt{\frac{\pi }{a+if}}\exp -\left(\frac{c^{2}+(d+g)^{2}}{4(a+if)}\right)\sin \left(b+\frac{2c(d+g)}{4i(a+if)}\right).\end{equation}
From here, we would do well to write the arguments of the exponential
and the sine in the form $x+iy$,\begin{equation}
\Psi (x,t)=\sqrt{\frac{2}{L}}\sqrt{\frac{a}{\pi }}\sum _{l=-\infty }^{\infty }\sqrt{\frac{\pi }{a+if}}\exp -\left(\frac{c^{2}+(d+g)^{2}}{4(a^{2}+f^{2})}\left(a-if\right)\right)\sin \left(b+\frac{2c(d+g)}{4(a^{2}+f^{2})}\left(-f-ia\right)\right).\end{equation}
 Now, we will consider $\left\Vert \Psi (x,t)\right\Vert ^{2}$, making
the assumption that we can ignore interference between terms in the
sum. While this is not obviously true, this whole business is worthless
if it isn't. Don't worry, we'll check this assumption at the end,
just to be sure. One of the advantages of doing this is that we can
use the formula $\left\Vert \sin (x+iy)\right\Vert ^{2}=(1/2)(\cosh (2y)-\cos (2x))$.
Doing all of this, we find\begin{equation}
\left\Vert \Psi (x,t)\right\Vert ^{2}=\frac{a}{L\sqrt{a^{2}+f^{2}}}\sum _{l=-\infty }^{\infty }\exp -\left(a\frac{c^{2}+(d+g)^{2}}{2(a^{2}+f^{2})}\right)\left(\cosh \left(a\frac{2c(d+g)}{2(a^{2}+f^{2})}\right)-\cos 2\left(b-f\frac{2c(d+g)}{4(a^{2}+f^{2})}\right)\right).\end{equation}
Looking at this, we may notice that the adding the argument of the
cosh to the argument of the exponential would complete the square.
Fortunately, we may write the cosh in terms of exponentials, giving
us\begin{equation}
\left\Vert \Psi (x,t)\right\Vert ^{2}=\frac{a}{L\sqrt{a^{2}+f^{2}}}\sum _{l=-\infty }^{\infty }\row{\left(\exp -\left(a\frac{\left(c+(d+g)\right)^{2}}{2(a^{2}+f^{2})}\right)+\exp -\left(a\frac{\left(c-(d+g)\right)^{2}}{2(a^{2}+f^{2})}\right)\right)\times }{\left(1-\frac{\cos \left(2b-f\frac{c(d+g)}{a^{2}+f^{2}}\right)}{\cosh \left(a\frac{c(d+g)}{a^{2}+f^{2}}\right)}\right)}\label{eq:qbeats-isw-abc}\end{equation}
This is the simplest form of this equation that I have found. Noticing
the ubiquity of $a^{2}+f^{2}$, we define \begin{equation}
\sigma ^{2}(\tau )=a^{2}+f^{2}=\left(\frac{1}{2\Delta n^{2}}\right)^{2}+\left(2\pi \frac{t}{T_{2}}\right)^{2},\label{eq:sigma}\end{equation}
then fill in our various constants to arrive at the following,\begin{eqnarray}
\left\Vert \Psi (x,t)\right\Vert ^{2} & = & \frac{1/2\Delta n^{2}}{L\sigma (\tau )}\sum _{l=-\infty }^{\infty }\begin{array}[t]{l}
 \left(\exp -\left(\frac{1}{2\Delta n^{2}}\frac{\left(\pi \xi +\left(2\pi \tau +2\pi l\right)\right)^{2}}{2\sigma ^{2}(\tau )}\right)+\right.\\
 \left.\exp -\left(\frac{1}{2\Delta n^{2}}\frac{\left(\pi \xi -\left(2\pi \tau +2\pi l\right)\right)^{2}}{2\sigma ^{2}(\tau )}\right)\right)\times \\
 \left.\left(1-\frac{\cos \left(\frac{T_{2}}{T_{1}}\pi \xi -2\pi \frac{T_{1}}{T_{2}}\tau \frac{\pi \xi \left(2\pi \tau +2\pi l\right)}{\sigma ^{2}(\tau )}\right)}{\cosh \left(\frac{1}{2\Delta n^{2}}\frac{\pi \xi \left(2\pi \tau +2\pi l\right)}{\sigma ^{2}(\tau )}\right)}\right)\right)\end{array}
\\
 & = & \frac{1}{2L\Delta n^{2}\sigma (\tau )}\sum _{l=-\infty }^{\infty }\row{\left(\exp -\left(\frac{\pi ^{2}}{2\Delta n^{2}}\frac{\left(\xi -2\left(\tau +l\right)\right)^{2}}{2\sigma ^{2}(\tau )}\right)+\exp -\left(\frac{\pi ^{2}}{2\Delta n^{2}}\frac{\left(\xi +2\left(\tau +l\right)\right)^{2}}{2\sigma ^{2}(\tau )}\right)\right)\times }{\left(1-\cos \pi \left(\frac{T_{2}}{T_{1}}\xi -4\pi ^{2}\frac{T_{1}}{T_{2}}\tau \frac{\xi \left(\tau +l\right)}{\sigma ^{2}(\tau )}\right)\textrm{sech}\left(\frac{\pi ^{2}}{\Delta n^{2}}\frac{\xi \left(\tau +l\right)}{\sigma ^{2}(\tau )}\right)\right)}\label{eq:qbeats-alg}
\end{eqnarray}

\subsection{Interpretation}

We are now prepared to discuss Equation \ref{eq:qbeats-alg}. Most
of the structure is provided by the two travelling Gaussians, which
both originate at $\xi =0,\, \tau =-l$, then separate and travel
along straight lines to $\xi =1,\, \tau =-l\pm 1/2$. This both provides
a pleasing picture of a packet completing one classical oscillation
in one classical period ($T_{1}$), and tells us that if the packets
in two neighboring terms have $\sigma (\tau )<1/2$, those terms will
have negligible interference%
\footnote{Of course, the {}``no overlap'' assumption is going to break down
in some interval near $\xi =1$.%
}. All of this is modulated by the final, oscillating term, which provides
{}``interference without cross terms.'' Note that because $\left|\cos \right|\leq 1$
and $\textrm{sech}\leq 1$, the oscillating term will always fall
between zero and two, and makes a secondary contribution. We have
left things in terms of the ratio $T_{1}/T_{2}$ as a reminder that
a principal difference between classical and quantum dynamics is that
a classical system would have only one time scale ($T_{1}$), while
a quantum system may have more -- in this case, two. Note that by
Equation \ref{eq:qbeats-t1/t2}, the limit as $\bar{n}\rightarrow \infty $
is identical to the limit $T_{1}/T_{2}\rightarrow 0$, and that this
fixes the widths of the travelling Gaussians above. That is, in the
classical limit our packet does not spread. 

When we are not in the classical limit, three things prevent our travelling
Gaussians from behaving classically. First, there is a damping term
in front of the Gaussians, though this isn't difficult to interpret
-- it enforces conservation of probability. Second, their widths \emph{do}
increase, and when those widths become larger than the width of the
well we would expect significant interference between, for example,
the $l$th term and the $(l+1)$th term. Finally, there is the oscillatory
term. In the cosine term, the $(T_{2}/T_{1})\xi $ term guarantees
rapid oscillations, and the other term serves to curve the paths along
which the $\textrm{cos}\times \textrm{sech}$ term is constant. The
oscillation of the cosine term is modulated by the slowly-varying
sech term. Although there is no conventional charateristic width defined
for this function, as there is for a Gaussian distribution, it is
clear that when the argument of the sech function becomes large, our
Gaussians will look fairly smooth, and the argument of sech is on
the order of one, the unsuppressed cosine will produce oscillatory
interference-like effects.

\begin{figure}[hbtp]
\begin{center}\includegraphics[  width=0.40\textwidth]{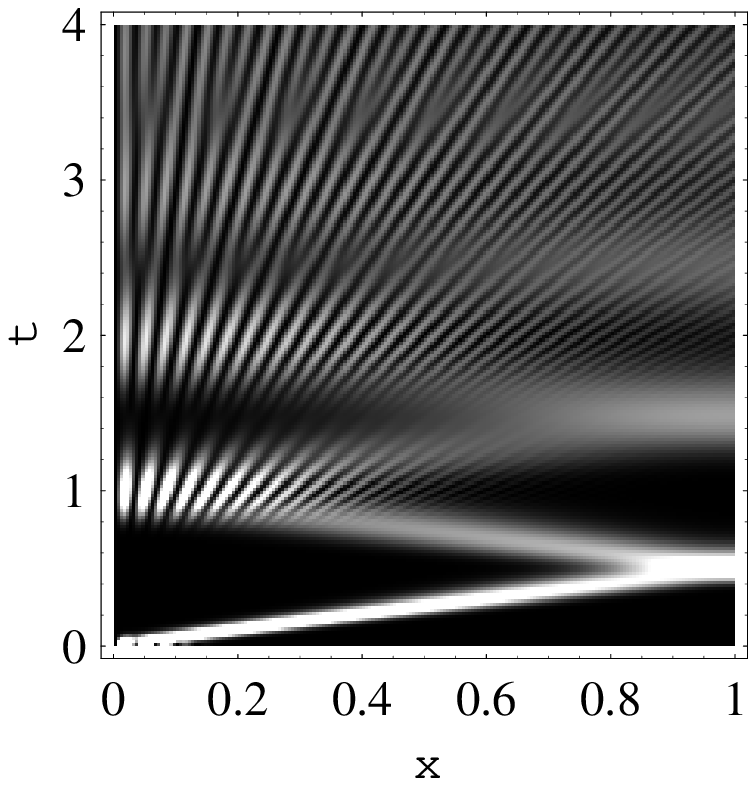}\hfill{}\includegraphics[  width=0.40\textwidth]{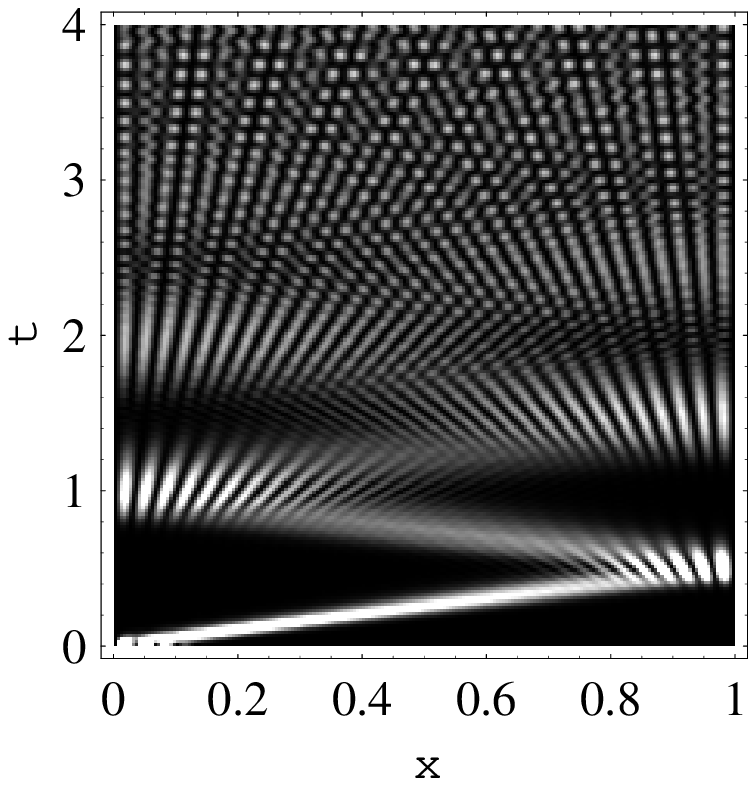}\end{center}

\caption{\label{fig: beatwell-density}These density plots show a solution
of the infinite square well in the region $\tau \in [0,4],\, \xi \in [0,1]$
for $\Delta n=5,\, \bar{n}=30$. On the left are the five relevant
terms from Equation \ref{eq:qbeats-alg}, on the right are the middle
10 terms of the sum in Equation \ref{eq:qbeats-nochange}.}
\end{figure}

\begin{figure}[hbtp]
\begin{center}\includegraphics[  width=0.30\textwidth]{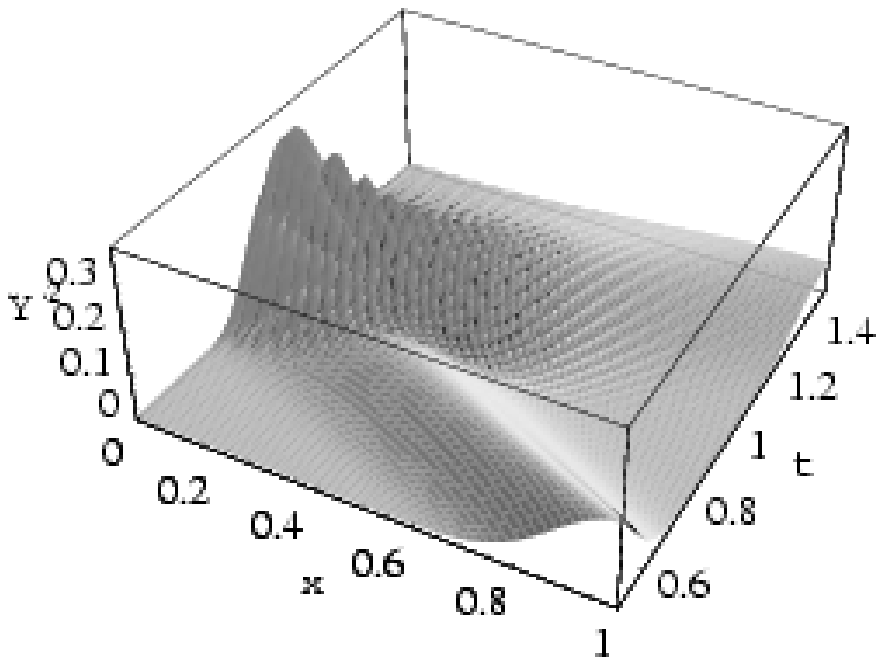}\hfill{}\includegraphics[  width=0.30\textwidth]{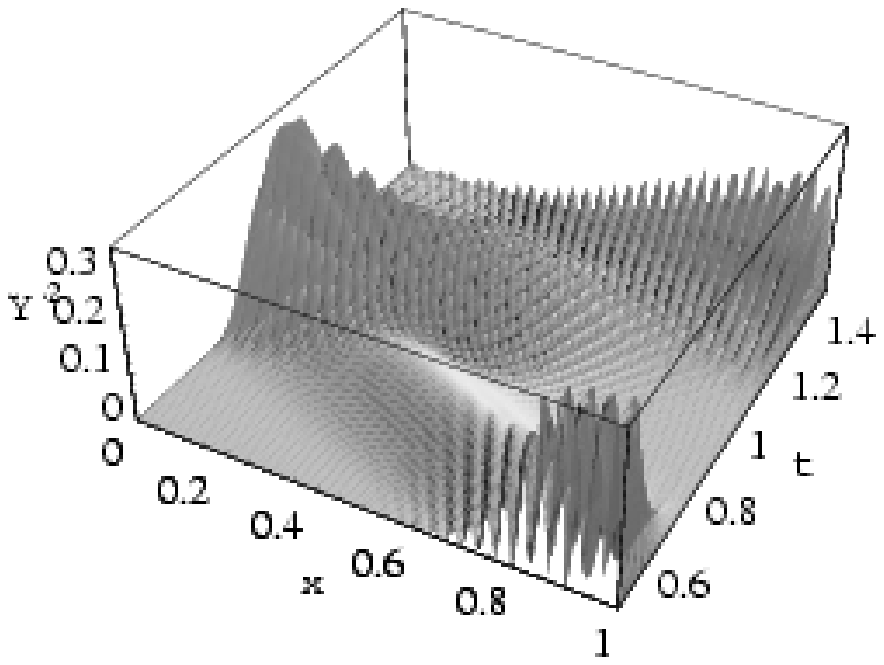}\hfill{}\includegraphics[  width=0.30\textwidth]{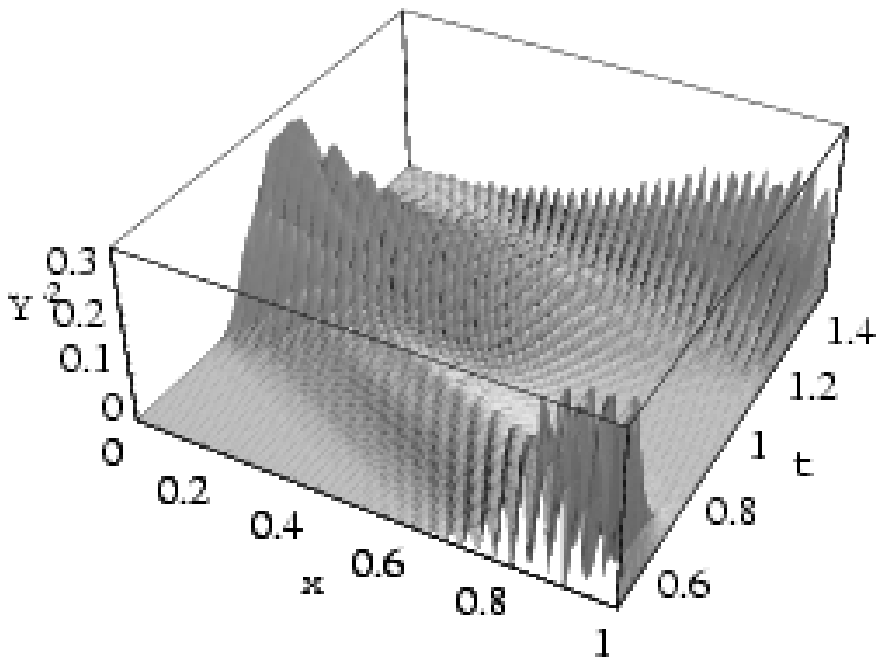}\end{center}

\begin{center}6.188s\end{center}

\begin{center}207.891s\end{center}

\begin{center}106.047s\end{center}

\caption{\label{fig: beatwell-comp} All of these show the region $\tau \in [0.5,1.5],\, \xi \in [0,1]$
for $\Delta n=5,\, \bar{n}=30$, and the time it took \textsl{Mathematica}
to plot them. On the left is the $l=-1$ term from Equation \ref{eq:qbeats-alg},
the middle figure contains the middle 20 terms of the sum in Equation
\ref{eq:qbeats-nochange}, and at right is a plot of the middle 10
terms from that same Equation.}
\end{figure}

We can see all of this by plotting several terms from the sum in Equation
\ref{eq:qbeats-alg} and comparing it to a similar region of the
full expansion of Equation \ref{eq:qbeats-nochange} (see Figure
\ref{fig: beatwell-density}). One thing that you won't notice unless
you plot some of these yourself is that where Equation \ref{eq:qbeats-alg}
is valid, it is more computationally efficient -- there are no cross
terms to calculate. For an example of this, see Figure \ref{fig: beatwell-comp}.%

\subsection{A Quick Look at Sech}

\begin{figure}[hbtp]
\begin{center}\includegraphics[  width=3in]{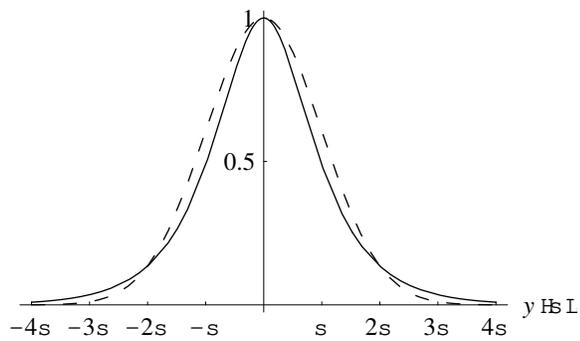}\end{center}

\caption{\label{fig: sech-gauss} A plot of $\exp \left(-\frac{y^{2}}{2\sigma ^{2}}\right)$
(dashed) and $\textrm{sech}\left(\alpha \frac{y}{2\sigma }\right)$
(solid), $-5\sigma \leq y\leq 5\sigma $. The constant $\alpha =\textrm{arcsech}\left(\frac{1}{e^{2}}\right)\approx 2.69$
is chosen so that the two functions are equal at $y=\pm 2\sigma $. }
\end{figure}

One function that appears in Equation \ref{eq:qbeats-alg} and may
be unfamiliar is sech, the hyperbolic secant. Defined as $\textrm{sech}(y)=1/\cosh (y)$,
it has a peak at $y=0,\, \textrm{sech}(y)=1$, and $\lim _{\left|y\right|\rightarrow \infty }=0$.
In fact, it looks rather like a Gaussian wavepacket. How much? Consider
Figure \ref{fig: sech-gauss}, which suggests that near $y=0$ the
functions can be given similar characteristic widths. For example,
choosing a constant $\alpha =\textrm{ArcSech}(1/e^{2})\approx 2.69$
gives us $\exp \left(-y^{2}/2\sigma ^{2}\right)\approx \textrm{sech}\left(\alpha y/2\sigma \right)$,
which becomes exact at $y=\pm 2\sigma $.%
We can generalize this result still further. Suppose we want $\exp \left(-y^{2}/2\sigma ^{2}\right)=\textrm{sech}\left(\alpha y/A\sigma \right)$at
$y=\pm A\sigma $. We will first find that \begin{equation}
\exp \left(-\frac{(A\sigma )^{2}}{2\sigma ^{2}}\right)=\exp \p{-A^{2}/2}.\end{equation}
Comparing this with \begin{eqnarray}
\textrm{sech}\left(\alpha \frac{A\sigma }{A\sigma }\right) & = & \textrm{sech}(\alpha )\nonumber \\
 & = & \exp \p{-A^{2}/2},
\end{eqnarray}
we can see that $\alpha (A)$ must be \begin{equation}
\alpha (A)=\textrm{arcsech}\left(e^{-A^{2}/2}\right).\label{eq:arcsech}\end{equation}
Note that the domain of arcsech is (0,1{]}, the same as the range
of $\exp \left(-x^{2}\right)$, so $\alpha (A)$ is defined for all
real values of $A$.

Now, our final step of generalization. Suppose we want to find the
condition for \begin{equation}
\exp \left(-\frac{(y-a)^{2}}{2\sigma _{a}^{2}}\right)=\textrm{sech}\left(\alpha (A)\frac{y-b}{A\sigma _{b}}\right)\end{equation}
to hold at $y=\pm _{1}A\sigma _{a}=\pm _{2}A\sigma _{b}$. This equation
is clearly satisfied when \begin{eqnarray}
(y-a) & = & \pm _{1}A\sigma _{a},\label{eq:sech-cond1}\\
(y-b) & = & \pm _{2}A\sigma _{b},\label{eq:sech-cond2}
\end{eqnarray}
where I have used subscripts on the $\pm $ symbols to indicate their
independence. It would be unfortunate if our goal was just to find
the intersections of two given functions. That is, if we were searching
for the solutions of \begin{equation}
\exp \left(-\frac{(y-a)^{2}}{2\sigma _{a}^{2}}\right)=\textrm{sech}\left(\frac{y-b}{\sigma _{b}}\right),\end{equation}
where we were unwilling to vary any of the parameters. This will quickly
lead us to the transcendental equation\begin{equation}
A\sigma _{a}+a=\alpha (A)\sigma _{b}+b,\end{equation}
which I don't care to solve. However, if we fix $A$ and choose to
solve for one of the other variables, we will have an ordinary algebraic
equation with perhaps one unsightly transcendental number floating
around. Since we are only concerned with equality at a point, we can
even allow $a,\, b,\, \sigma _{a},\, \textrm{and}\, \sigma _{b}$
to be functions. In that case, the result in Equations \ref{eq:sech-cond1}
and \ref{eq:sech-cond2} are quite useful. They let us figure out
what parameter values to use to get these functions to intersect at,
say, one standard deviation from their centers, which is exactly the
what we need to do in order to find dephasing times.

\subsection{Dephasing of the Wavepacket}

\begin{figure}[hbtp]
\begin{center}\includegraphics[  width=0.40\textwidth]{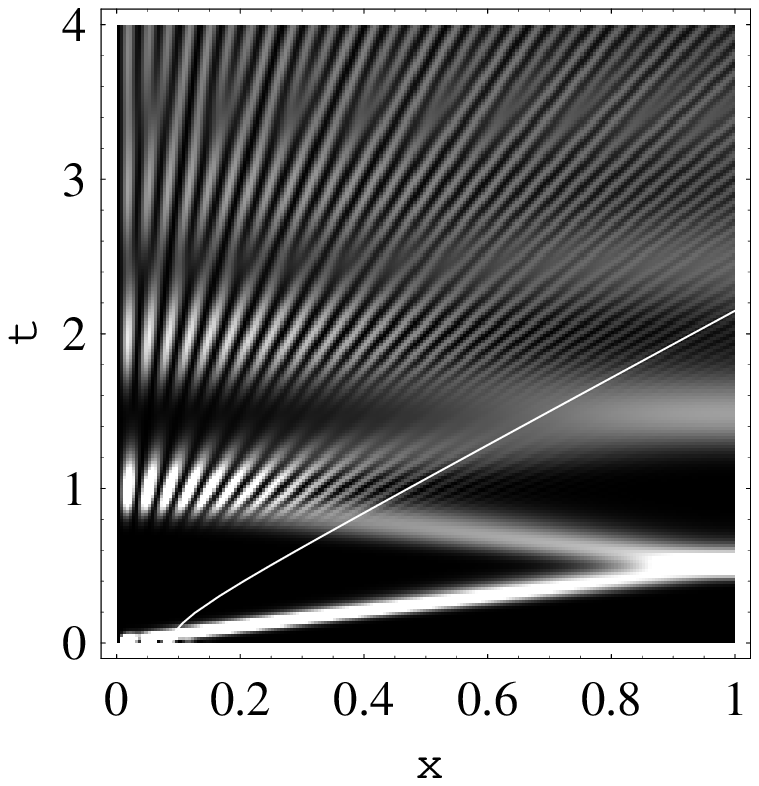}\hfill{}\includegraphics[  width=0.40\textwidth]{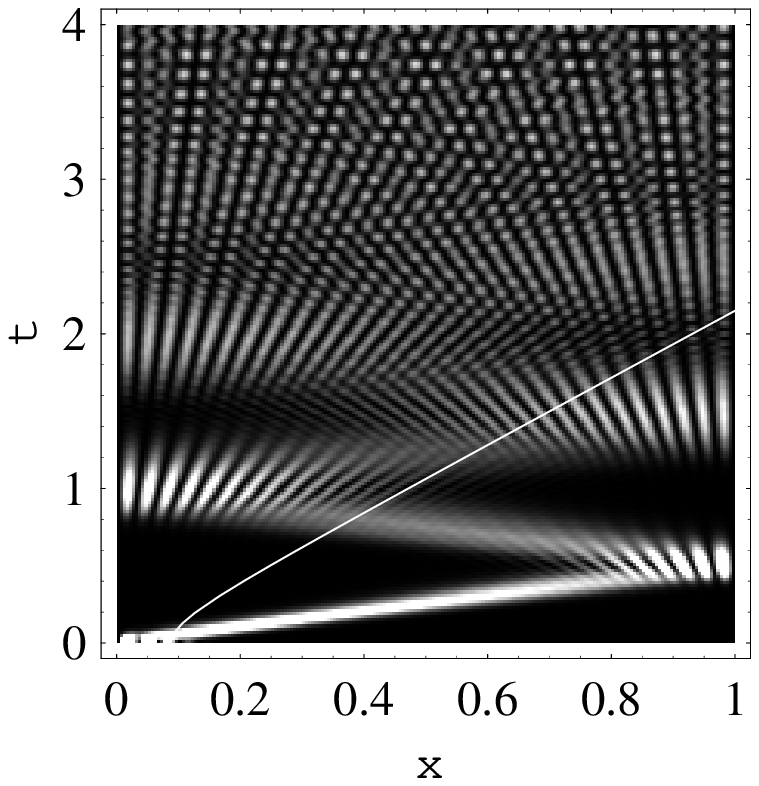}\end{center}

\caption{\label{fig: beatwell-dephase}These density plots show a solution
of the infinite square well in the region $\tau \in [0,4],\, \xi \in [0,1]$
for $\Delta n=5,\, \bar{n}=30$, and the white line indicates the
lower of the $A=1$ solutions of Equation \ref{eq:dephase}. On the
left are the five relevant terms from Equation \ref{eq:qbeats-alg},
on the right are the middle 10 terms of the sum in Equation \ref{eq:qbeats-nochange}.
Note that the ridges in the left plot only appear above the line,
and in the right plot that this holds for $\xi <1/2$ -- for $\xi >1/2$,
the interference between terms in Equation \ref{eq:qbeats-alg} becomes
non-negligible.}
\end{figure}

If we recall the previous chapter, the advantage of this sort of analysis
of the wavefunction is that it allows us to quantitatively understand
when the wavefunction makes the transition from semiclassical motion
to characteristically quantum mechanical motion. We could assume that
this happens when the sech term overlaps the Gaussian closer to $\tau =0$
-- then we need only decide what constitutes substantial overlap.
Let us suppose that it occurs when some particular number of {}``standard
deviations'' of the sech function reach a particular number of standard
deviations of the travelling Gaussian (we'll not specify on which
side of its center), so that the functions are equal at that point.
This condition may be more clear when written like Equations \ref{eq:sech-cond1}
and \ref{eq:sech-cond2}:\begin{eqnarray}
\left(\tau +\left(l+\frac{\xi }{2}\right)\right) & = & \pm A\sigma _{\textrm{exp}}(\tau ),\label{eq:overlap-cond1}\\
\left(\tau +l\right) & = & -A\sigma _{\textrm{sech}}(\tau ),\label{eq:overlap-cond2}
\end{eqnarray}
where we have defined $\sigma _{\textrm{sech}}(\tau )$ and $\sigma _{\textrm{exp}}(\tau )$
in terms of $\sigma (\tau ),$ from Equation \ref{eq:sigma},\begin{eqnarray}
\sigma _{\textrm{sech}}(\tau ) & = & \frac{\alpha (A)}{A\xi }\left(\frac{\Delta n}{\pi }\sigma (\tau )\right)^{2},\\
\sigma _{\textrm{exp}}(\tau ) & = & \frac{\Delta n}{\pi }\sigma (\tau ),\\
\sigma ^{2}(\tau ) & = & \left(\frac{1}{2\Delta n^{2}}\right)^{2}+\left(2\pi \frac{T_{1}}{T_{2}}\tau \right)^{2}.
\end{eqnarray}
Note that the $\pm $ in Equation \ref{eq:overlap-cond1} allows
us to consider intersection on either side of the center of the Gaussian,
while the fixed sign in Equation \ref{eq:overlap-cond2} indicates
our choice to consider only the region between $\tau =-l$ and $\tau =0$.
Keep in mind that what we are doing is, in spirit, exactly what we
did in the previous chapter. We have used the Poisson summation formula
to turn a sum of functions over all space into a sum of localized
functions, figured out how to characterize their widths, and now we
are preparing to find out when those widths overlap, in order to discover
when interference effects become \emph{really} important to the time-evolution
-- when our wavepacket has dephased. We will now combine Equations
\ref{eq:sech-cond1} and \ref{eq:sech-cond2} to eliminate $\tau +l$
and solve for $\sigma (\tau )$:\begin{eqnarray}
-A\sigma _{\textrm{sech}}(\tau )+\frac{\xi }{2} & = & \pm A\sigma _{\textrm{exp}}(\tau ),\nonumber \\
-\frac{\alpha (A)}{\xi }\left(\frac{\Delta n}{\pi }\sigma (\tau )\right)^{2}+\frac{\xi }{2} & = & \pm A\frac{\Delta n}{\pi }\sigma (\tau ),\nonumber \\
-\frac{\alpha (A)}{\xi }\left(\frac{\Delta n}{\pi }\sigma (\tau )\right)^{2}\mp A\left(\frac{\Delta n}{\pi }\sigma (\tau )\right)+\frac{\xi }{2} & = & 0.
\end{eqnarray}
We may now use the quadratic equation to solve for $\left(\Delta n\sigma (\tau )\right)/\pi $,\begin{eqnarray}
\frac{\Delta n}{\pi }\sigma (\tau ) & = & \frac{\pm _{1}A\pm _{2}\sqrt{A^{2}+4\frac{\alpha (A)}{\xi }\frac{\xi }{2}}}{-2\frac{\alpha (A)}{\xi }}\nonumber \\
 & = & -\frac{A\xi }{2\alpha (A)}\left(\pm _{1}1\pm _{2}\sqrt{1+2\frac{\alpha (A)}{A^{2}}}\right).
\end{eqnarray}
Since $\alpha (A)$ and $A$ are always positive, we know that the
square root term will always be greater than one. We now square this
equation, to recover something in terms of $\sigma ^{2}(\tau )$,\begin{equation}
\left(\frac{\Delta n}{\pi }\right)^{2}\sigma ^{2}(\tau )=\left(\frac{A\xi }{2\alpha (A)}\right)^{2}\left(1+\left(1+2\frac{\alpha (A)}{A^{2}}\right)\pm 2\sqrt{1+2\frac{\alpha (A)}{A^{2}}}\right),\end{equation}
which we can further simplify\begin{equation}
\sigma ^{2}(\tau )=\frac{1}{2}\left(\frac{\pi A\xi }{\Delta n\alpha (A)}\right)^{2}\left(1+\frac{\alpha (A)}{A^{2}}\pm \sqrt{1+2\frac{\alpha (A)}{A^{2}}}\right),\end{equation}
before inserting Equation \ref{eq:sigma},\begin{eqnarray}
\left(\frac{1}{2\Delta n^{2}}\right)^{2}+\left(2\pi \frac{T_{1}}{T_{2}}\tau \right)^{2} & = & \frac{1}{2}\left(\frac{\pi A\xi }{\Delta n\alpha (A)}\right)^{2}\left(1+\frac{\alpha (A)}{A^{2}}\pm \sqrt{1+2\frac{\alpha (A)}{A^{2}}}\right),\nonumber \\
\tau ^{2} & = & \left(\frac{1}{2\pi }\frac{T_{2}}{T_{1}}\right)^{2}\rowc{\frac{1}{2}\left(\frac{\pi A\xi }{\Delta n\alpha (A)}\right)^{2}\left(1+\frac{\alpha (A)}{A^{2}}\pm \sqrt{1+2\frac{\alpha (A)}{A^{2}}}\right)-}{\left(\frac{1}{2\Delta n^{2}}\right)^{2}}\nonumber \\
\tau  & = & \frac{\sqrt{2}}{4\pi \Delta n}\frac{T_{2}}{T_{1}}\sqrt{\left(\frac{\pi A\xi }{\alpha (A)}\right)^{2}\left(1+\frac{\alpha (A)}{A^{2}}\pm \sqrt{1+2\frac{\alpha (A)}{A^{2}}}\right)-\frac{1}{2\Delta n^{2}}}.\label{eq:dephase}
\end{eqnarray}
Note that the $\tau $ at which dephasing occurs depends (to first
order) on $1/\Delta n$, as in all of our other examples. To check
that this works, let's compare the lower of the $A=1$ terms against
plots of Equations \ref{eq:qbeats-alg} and \ref{eq:qbeats-nochange}
in Figure \ref{fig: beatwell-dephase}.%
Note that the plots agree quite nicely in the left half of the well,
but that where different terms in Equation \ref{eq:qbeats-alg} overlap,
i.e. near the right wall, our approximation does not reproduce the
appropriate interference effects. Nevertheless, Equation \ref{eq:dephase}
does predict where rippling begins in the left side of the well. If
we wanted to set different standards for overlap, we would just need
to change the value of $A$. We could also treat fractional revivals
in the square well, but the reader should be able to combine the results
of Section \ref{sec: beat-fracs} and those of this section to quickly
find and study them.

\section{Solutions in More Complicated Potentials}

Obviously, we have little interest in actually doing the integral
in Equation \ref{eq:poisson-2}. At the same time, there are many
interesting potentials that we might consider which are not the infinite
square well -- which, indeed, do not even admit solutions in pure
exponentials. Our analytical approach to these rogues will be to use
the WKB approximation to render much of the information about the
wavefunction in convenient exponential form.

\subsection{\label{sec: WKB}The WKB Approximation}

The WKB approximation%
\footnote{For a more complete treatment of the WKB approximation, see \cite[274-292]{misc:Griffiths1}%
} is a formalization of the observation that if a wavefunction has
sufficient energy that it undergoes rapid oscillations in comparison
to the {}``features'' of the potential below it, it looks much like
a free particle. It is useful in the same sorts of semiclassical problems
where we would expect to find interesting revival and carpet phenomena.
Without going into too many details, we define\begin{equation}
p_{n}(x)=\sqrt{E_{n}-V(x)},\end{equation}
which will behave rather like a wave number. It is subject to the
Bohr-Sommerfeld condition,\begin{equation}
\int _{x_{l}}^{x_{r}}p_{n}(x)=n\pi ,\end{equation}
which enforces quantization of energy. $x_{l}$ and $x_{r}$ are,
respectively, the left and right classical turning points for a particle
of energy $E_{n}$. We are then able to write the wavefunction as
\begin{equation}
\psi _{n}(x)\approx \frac{C_{n}}{\sqrt{p_{n}(x)}}\exp \left(i\int ^{x}p_{k}\left(x^{\prime }\right)dx^{\prime }\right)+\frac{D_{n}}{\sqrt{p_{n}(x)}}\exp \left(-i\int ^{x}p_{k}\left(x^{\prime }\right)dx^{\prime }\right),\label{eq:psi-wkb}\end{equation}
with the understanding that both $\int ^{x}p\left(x^{\prime }\right)dx^{\prime }$
and $1/\sqrt{p_{n}(x)}$ vary slowly in comparison to $\exp \left(i\int ^{x}p\left(x^{\prime }\right)dx^{\prime }\right)$.
$C$ and $D$ are constants of integration. This approximation of
the wavefunction breaks down when $E_{n}$ is not much greater than
$V(x)$, near the classical turning points.

\subsection{The WKB Approximation and the Poisson Summation Formula}

The WKB approximation has not quite solved our problem, having both
left that unattractive $1/\sqrt{p_{k}\p{x}}$ term in from of our
exponentials. We are, however, \emph{en route} to a solution, \begin{eqnarray}
\Psi (x,t) & = & \sum _{l=-\infty }^{\infty }\int _{-\infty }^{\infty }c_{k}\psi _{k}(x)\exp \left(-iE_{k}t\right)\nonumber \\
 & = & \sum _{l=-\infty }^{\infty }\int _{-\infty }^{\infty }dk\rowc{\frac{c(k)}{\sqrt{p_{k}(x)}}\exp -i\left(-\int ^{x}p_{k}\left(x^{\prime }\right)dx^{\prime }+E_{k}t+2\pi kl\right)+}{\frac{d(k)}{\sqrt{p_{k}(x)}}\exp -i\left(\int ^{x}p_{k}\left(x^{\prime }\right)dx^{\prime }+E_{k}t+2\pi kl\right)}
\end{eqnarray}
but the remaining integral is still less than inviting. Of course,
judicious choice of the $c_{k}$ and $d_{k}$ will make matters simpler,
but we can do more than that. We now perform a Taylor expansion of
$\int ^{x}p_{k}\left(x^{\prime }\right)dx^{\prime },$ $E_{k},$ $c(k)/\sqrt{p_{k}(x)},$
and $d(k)/\sqrt{p_{k}(x)}$ around $k=0$. For the terms that appear
in the exponential, we would keep up to order $k^{2}$ -- $k^{3}$
if we felt like using Airy functions. Keep as many terms as you like
in the expansion of $1/\sqrt{p_{k}(x)}$. The resulting mess will
be some Gaussian integral which will be easy to evaluate in closed
form. We should find some sum of exponentials, trigonometric, and
hyperbolic functions, including some which we can identify as containing
much of the interference information, and find the condition for wavepacket
dephasing.

Our solution relies on three successive approximations -- the WKB
approximation, the application of the Poisson summation formula (which
involves an extension of the $c_{n},\, d_{n}$ beyond $k=-\bar{n}$
to $k=-\infty $ and eventual neglect of cross terms), and a Taylor
expansion of the remaining functions -- and obviously won't work in
all cases. There are doubtless examples which can be integrated that
neither bear resemblance to the square well nor require that we resort
to these approximations. However, we have here a quantitative way
of finding out the lifetime of a wavepacket in an arbitrary potential
well.

\section{Summary}

In this chapter we demonstrated that the Poisson summation formula
technique, developed on zero-dimensional quantum beats in Chapter
\ref{ch: beats}, can also be applied to problems with more dimensions.
We demonstrated this by using the technique to solve the square well,
arriving at a very elegant sum of successive travelling Gaussians
multiplied by interference terms. From this, we were able to derive
a dephasing condition, though instead of finding a simple time we
found a function $\tau \p{\xi }$. What's more, it only works in the
left half of the well, due to interference that we had to neglect.
Fortunately, we found a $\tau \propto \p{T_{2}/T_{1}}/4\Delta n$
component in the dephasing condition that is strongly reminiscent
of our dephasing results from Chapter \ref{ch: beats}. The case of
fractional revivals is left as an exercise for the reader.

\chapter{\label{ch: traces}Intermode Traces and Quantum Carpets}

\section{Multimode Interference}

The Multimode Interference technique%
\footnote{The first sections of this chapter are based on \cite{trace:Kaplan2}.
I have tried to make their work more clear and rigorous.%
} is a devilishly simple way of studying the interference patterns
formed by a quantum system. It allows us to understand both the existence
of canals and ridges in the carpet and the role that the spectrum
and potential play in generating carpets. We will begin by rewriting
the probability density as a sum of multimode terms (defined below)
instead of eigenfunctions. Starting with a wavefunction,\begin{equation}
\Psi (x,t)=\sum _{n=1}^{\infty }c_{n}\psi _{n}(\vec{x})e^{-iE_{n}t},\end{equation}
we define a multimode term as \begin{eqnarray}
\mu _{nm}(x,t) & = & \frac{1}{2}\rowc{d_{nm}\psi _{n}(x,t)\psi _{m}^{*}(x,t)\exp -it\left(E_{n}-E_{m}\right)+}{d_{mn}\psi _{n}^{*}(x,t)\psi _{m}(x,t)\exp -it\left(E_{m}-E_{n}\right)}\label{eq:intermode-def}\\
d_{nm} & = & c_{n}c_{m}^{*}=d_{mn}^{*},
\end{eqnarray}
and write the probability density as \begin{equation}
\left\Vert \Psi (x,t)\right\Vert ^{2}=\sum _{n,m=1}^{\infty }\mu _{nm}(x,t).\end{equation}
This is not, at first glance, a particularly good idea. We have simply
grouped the terms in the sum differently and not achieved any obvious
simplification. The advantage is that we may simply add multimode
terms -- they already contain all of the information about how the
eigenfunctions interfere. If we can figure out a way to tease that
information out of the multimode terms, we will have been successful.

\section{Characteristic Velocities}

As in the previous chapter, we will use the WKB approximation to study
semiclassical cases (see Section \ref{sec: WKB}). Our approximate
eigenstates are then \begin{equation}
\psi _{n}(x)\approx \frac{C_{n}}{\sqrt{p_{n}(x)}}\exp \left(i\int ^{x}p_{n}\left(x^{\prime }\right)dx^{\prime }\right)+\frac{D_{n}}{\sqrt{p_{n}(x)}}\exp \left(-i\int ^{x}p_{n}\left(x^{\prime }\right)dx^{\prime }\right),\label{eq:psi-wkb}\end{equation}
where $C_{n}$ and $D_{n}$ are complex constants of integration,
and our approximate multimode terms will be weighted sums of eight
intermode terms, which we define as\begin{equation}
\iota _{nm}(x,t)=\frac{1}{2}d_{nm}\frac{1}{\sqrt{p_{n}(x)\, p_{m}(x)}}\exp \pm _{1}i\left(\int ^{x}\left(\pm _{2}p_{n}\left(x^{\prime }\right)\pm _{3}p_{m}\left(x^{\prime }\right)\right)dx^{\prime }+\left(E_{n}-E_{m}\right)t\right).\label{eq:wkb-phase}\end{equation}
Since we are interested in lines of constant phase, we differentiate
the argument of the exponential and look for its roots,\begin{eqnarray}
\frac{d}{dt}\pm _{1}\left(\int ^{x}\left(\pm _{2}p_{n}\left(x^{\prime }\right)\pm _{3}p_{m}\left(x^{\prime }\right)\right)dx^{\prime }+\left(E_{n}-E_{m}\right)t\right) & = & 0,\nonumber \\
\frac{dx}{dt} & = & \frac{E_{n}-E_{m}}{\pm _{2}p_{n}\left(x\right)\pm _{3}p_{m}\left(x\right)},\nonumber \\
v_{nm}=\frac{\Delta \omega }{\Delta k} & = & \pm _{1}\frac{E_{n}-E_{m}}{\sqrt{E_{n}-V(x)}\pm _{2}\sqrt{E_{m}-V(x)}}.\label{eq:vel-def}
\end{eqnarray}
Each pair of quantum numbers, $\left(n,m\right)$, gives rise to \emph{four}
velocities $v_{nm}$.

\section{Characterization of the Velocities}

We have written that last equation in terms of $\Delta \omega /\Delta k$
because this will allow us to label some of the velocities as {}``group''
velocities, and others as what I will tentatively call {}``not group
velocities.'' If we have defined $\omega _{n}=E_{n}$ and $k_{n}(x)=p_{n}(x)$,
then we can write \begin{equation}
\omega _{n}(k_{n},x)=k_{n}^{2}-V(x).\end{equation}
If we are considering a semiclassical problem, our weighting coefficients
must be well centered on some number $N$ with some spread $\Delta n$,
such that these satisfy the hierarchy $1\ll \Delta n\ll N$. This
allows us to define a group velocity for our packet, \begin{equation}
v_{gr}=\left.\frac{d\omega _{n}}{dk_{n}}\right|_{n=N}=2k_{N},\end{equation}
which happens to be the classical velocity of a particle in the potential
$V$ with energy $E_{N}$. If we then consider the velocities from
Equation \ref{eq:vel-def}, writing $E_{n}=E_{N}+e_{n}$ and $E_{m}=E_{N}+e_{n}$,
where $e_{n},e_{m}\ll E_{N}-V(x)$ we can simplify our expression,\begin{eqnarray}
v_{nm} & = & \pm _{1}\frac{\left(E_{N}+e_{n}\right)-\left(E_{N}+e_{m}\right)}{\sqrt{E_{N}+e_{n}-V(x)}\pm _{2}\sqrt{E_{N}+e_{m}-V(x)}}\nonumber \\
 & \approx  & \pm _{1}\frac{e_{n}-e_{m}}{\sqrt{E_{N}-V(x)}\left(\left(1+\frac{e_{n}}{2\left(E_{N}-V(x)\right)}\right)\pm _{2}\left(1+\frac{e_{m}}{2\left(E_{N}-V(x)\right)}\right)\right)}\nonumber \\
 & \approx  & \pm _{1}2\sqrt{E_{N}-V(x)}\frac{e_{n}-e_{m}}{2\left(E_{N}-V(x)\right)\left(1\pm _{2}1\right)+\left(e_{n}\pm _{2}e_{m}\right)}\nonumber \\
 & \approx  & \left\{ \begin{array}{l}
 \pm _{1}2\sqrt{E_{N}-V(x)}=\pm _{1}2k_{N}(x),\, \left(-_{2}\right)\\
 \pm _{1}2\sqrt{E_{N}-V(x)}\frac{e_{n}-e_{m}}{\left(E_{n}-V(x)\right)+\left(E_{m}-V(x)\right)}\approx \pm _{1}\frac{\omega _{n}-\omega _{m}}{k_{N}(x)},\, \p{+_{2}}.\end{array}
\right.\label{eq:2-vel}
\end{eqnarray}
So long as both $e_{n},e_{m}\ll E_{N}-V(x)$, as they should be in
the semiclassical approximation, then half of the velocities contributed
by a particular $\left(n,m\right)$ will be comparable to the group
velocity, $v_{gr}$, and half will be smaller. It is important to
note (in preparation for the next section) that Equation \ref{eq:2-vel}
is approximate -- it will not happen that half our velocity terms
will be degenerate.

These velocities describe paths (the {}``traces'' in the chapter
title) along which the phase is constant for a given intermode term.
That is, given a particular $\iota _{nm}$ and a value of $x,$ we
can find the trajectory along which that phase does not change, and
we can do this for all values of $x.$ Although what we have done
may bear a superficial resemblance to the method of stationary phase,
it is very different. We could use the method of stationary phase
if each intermode term was highly oscillatory except along a particular
path -- we would then assume that each term only made a significant
contribution to the probability density in the immediate vicinity
of the path of stationary phase. This is invalid here because, typically,
the phase term in Equation \ref{eq:wkb-phase} will change at a rate
determined by the $p_{n}(x)$, which shouldn't be particularly high.
The picture, rather, is of the phase function at $t=0$ sliding around
(and stretching a bit) as time increases.

\section{Groups of Velocities and Degeneracy%
\footnote{To this point, my exposition has parlleled that of \cite{trace:Kaplan2}.
Here, however, I must depart, as their treatment of the problem was
incorrect. They blithely assert that to discover the traces, we just
need to integrate from the classical turning points. A cursory survey
of the figures in this thesis should be enough to convince you otherwise.%
}}

The result in Equation \ref{eq:2-vel} demonstrates that in any problem
we will find a range of velocities. Those results are, however, approximate.
We will show that some intermode terms have \emph{exactly} the same
maximum velocities, and that because of that we can treat them as
essentially moving together. We consider this to be a sort of degeneracy
-- the sort of degeneracy that produces quantum carpets. 

If we examine the velocities (Equation \ref{eq:vel-def}) at a point%
\footnote{Remember that if we can't find a point where $V(x)=0$, we can exploit
our ability to change the potential by a constant and create one.%
} where $V(x)=0$,\begin{equation}
v_{nm}=\pm _{1}\frac{E_{n}-E_{m}}{\sqrt{E_{n}}\pm _{2}\sqrt{E_{m}}},\end{equation}
we can factor the numerator,\begin{equation}
v_{nm}=\pm _{1}\frac{\p{\sqrt{E_{n}}+\sqrt{E_{m}}}\p{\sqrt{E_{n}}-\sqrt{E_{m}}}}{\sqrt{E_{n}}\pm _{2}\sqrt{E_{m}}},\end{equation}
and arrive at an important condition,\begin{equation}
v_{nm}=\pm _{1}\p{\sqrt{E_{n}}\mp _{2}\sqrt{E_{m}}}.\label{eq:degen-cond}\end{equation}
Terms that have the same $v_{nm}$ at $V(x)=0$ have the same classical
period -- these are our degenerate terms.

\begin{figure}[hbtp]
\begin{center}\subfigure[]{\includegraphics[  width=0.45\textwidth]{SHO-Period.eps}}\hfill{}\includegraphics[  width=0.45\textwidth]{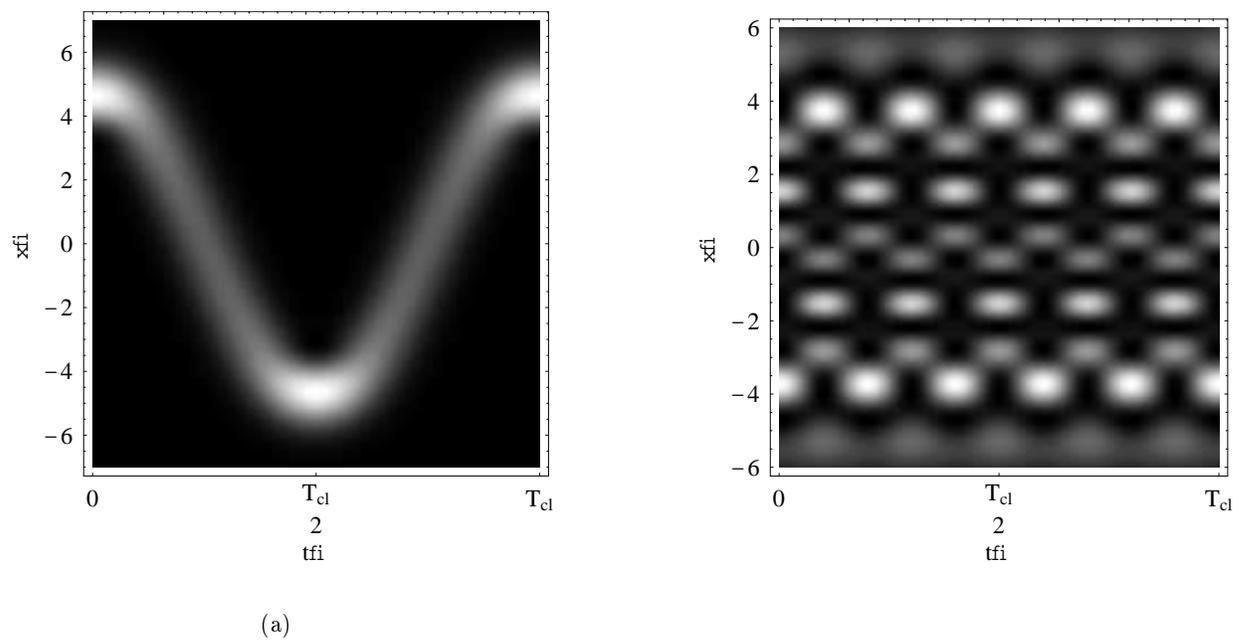}\end{center}

\caption{\label{fig: sho-sqr}Two plots of the simple harmonic oscillator.
On the left, the distribution of coefficients is Gaussian, with $\bar{n}=6$
and $\sigma _{n}=2$, while the plot on the right is an even weighting
of the perfect squares between 1 and 81.}
\end{figure}

We immediately gain some insight into the role of quadratic spectra
in producing quantum carpets. If the spectrum depends on the the quantum
number $n$ squared, then we will have many degenerate velocities,
while if the spectrum is linear in the quantum number, only traces
which involve two perfect squares may be degenerate. This suggests
that if we begin with a system like the simple harmonic oscillator,
with a spectrum linear in the quantum number, and set to zero all
weighting coefficients that are not perfect squares (selecting only
states 1,4,9,16,etc.), we can produce a carpet because we have effectively
quadratized%
\footnote{{}``Effectively quadratize the spectrum'' is my shorthand for choosing
weighting coefficients such that only terms that have quantum numbers
that are perfect squares are contained in the wavefunction. When we
do this, we could rewrite the spectrum, eigenfunctions, etc., as if
they were governed by a new variable, $m^{2}=n$, as if they had quadratic
spectra. Of course, we're not changing the spectrum itself, we're
changing the wavefunction.%
} the spectrum. For an example of this, see Figure \ref{fig: sho-sqr}.%

Of course, the production of the characteristic canals and ridges
of a quantum carpet depends not only on degenerate velocities existing,
but their producing some sort of peak or valley in the probability
density. Recalling the form for an intermode term,\begin{equation}
\iota _{nm}(x,t)=\frac{1}{2}d_{nm}\frac{1}{\sqrt{p_{n}(x)\, p_{m}(x)}}\exp \pm _{1}i\left(\int ^{x}\left(\pm _{2}p_{n}\left(x^{\prime }\right)\pm _{3}p_{m}\left(x^{\prime }\right)\right)dx^{\prime }+\left(E_{n}-E_{m}\right)t\right),\end{equation}
we can see how a sum of these terms might produce a peak or valley
at $t=0$. First, we ignore the imaginary component of the exponential
-- since our probability density has been reduced to a sum of these
intermode terms, and the probability density must be real, the imaginary
terms will ultimately cancel. The sum of intermode terms looks like 

\begin{equation}
\sum \iota _{nm}(x,t=0)=\frac{1}{2}\sum d_{nm}\frac{1}{\sqrt{p_{n}(x)\, p_{m}(x)}}\cos \pm _{1}\p{\int ^{x}\left(\pm _{2}p_{n}\left(x^{\prime }\right)\pm _{3}p_{m}\left(x^{\prime }\right)\right)dx^{\prime }},\end{equation}
or roughly a weighted sum of cosines. Although this hardly constitutes
a proof%
\footnote{Of course, proofs are possible here. I suspect, though, that figuring
out what one \emph{can} prove will be difficult work in and of itself.%
} that peaks and valleys exist, it is not hard to imagine that in many
cases we would find them.

A general calculation of the location of the peaks of an intermode
trace would be so general as to be useless, and specific examples
are likely to be too specific. Instead, we must rest satisfied with
the notion that very degenerate velocities will typically have enough
terms to form some sort of peak. As we evolve in time, the peak should
roughly follow a classical trajectory, experiencing a bit of dispersion.
The formula for such a trajectory is \begin{equation}
t(x)=\int _{x_{0}}^{x}\frac{1}{v_{nm}(x^{\prime })}dx^{\prime },\end{equation}
where $x_{0}$ is the location of the center of the peak, and $v_{nm}$
is some suitable velocity, perhaps the middlemost of the degenerate
velocities. These are the intermode traces that we are interested
in. If this is all seems a bit hazy, bear in mind that this idea of
a {}``channel'' or {}``ridge'' is something we impose on the system,
not so easily defined as a local minimum or maximum. We are, in a
sense, asking a question of the system that it does not want to answer,
and this sort of ambiguity is one of its methods of resistance. Hopefully,
future work will uncover a procedure for identifying the most significant
traces in the carpet.

\subsection{A Quick Example}

\begin{figure}[hbtp]
\begin{center}\includegraphics[  width=0.30\textwidth]{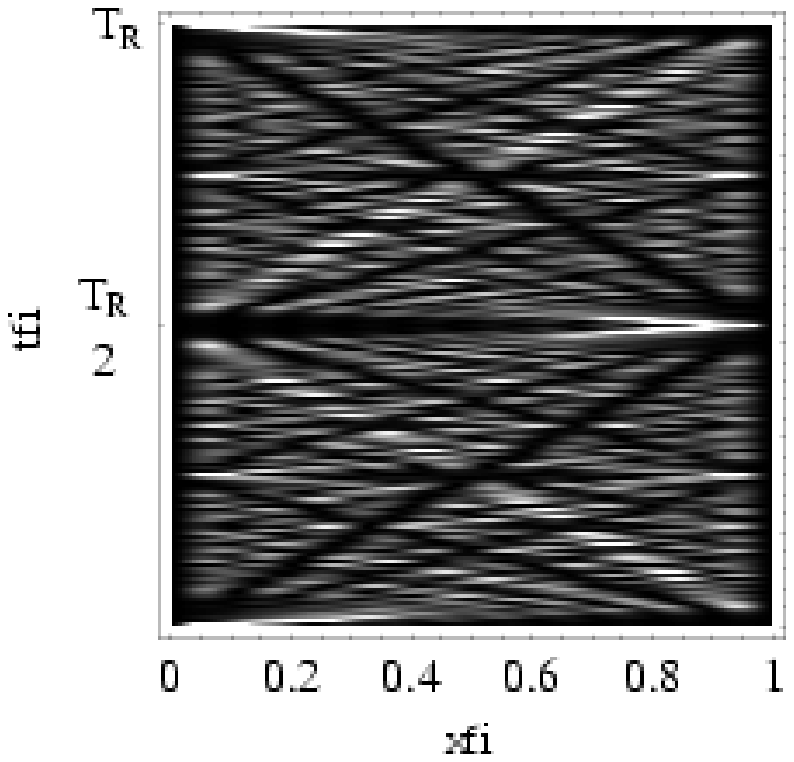}\hfill{}\includegraphics[  width=0.30\textwidth]{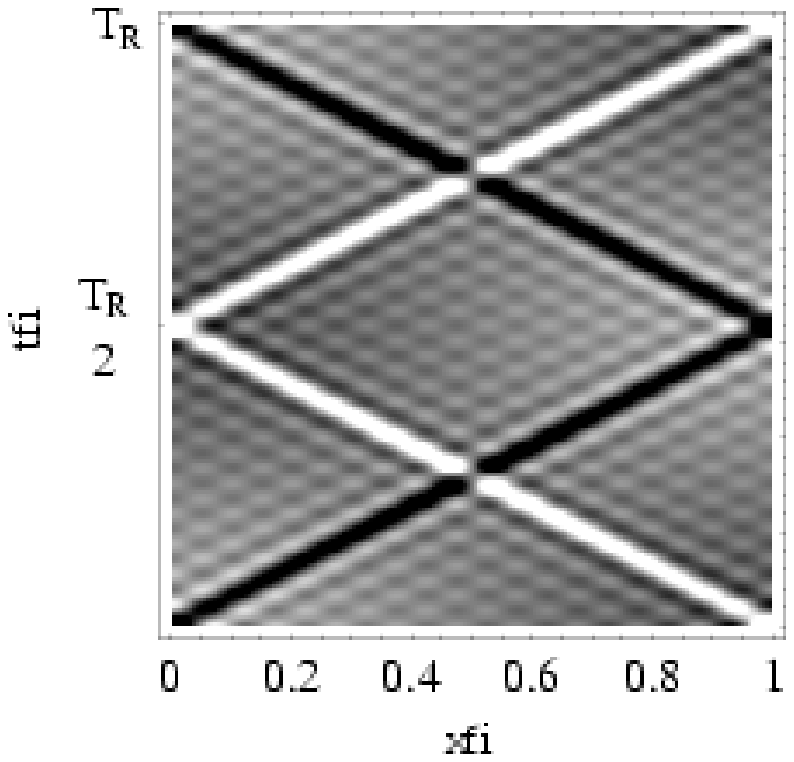}\hfill{}\includegraphics[  width=0.30\textwidth]{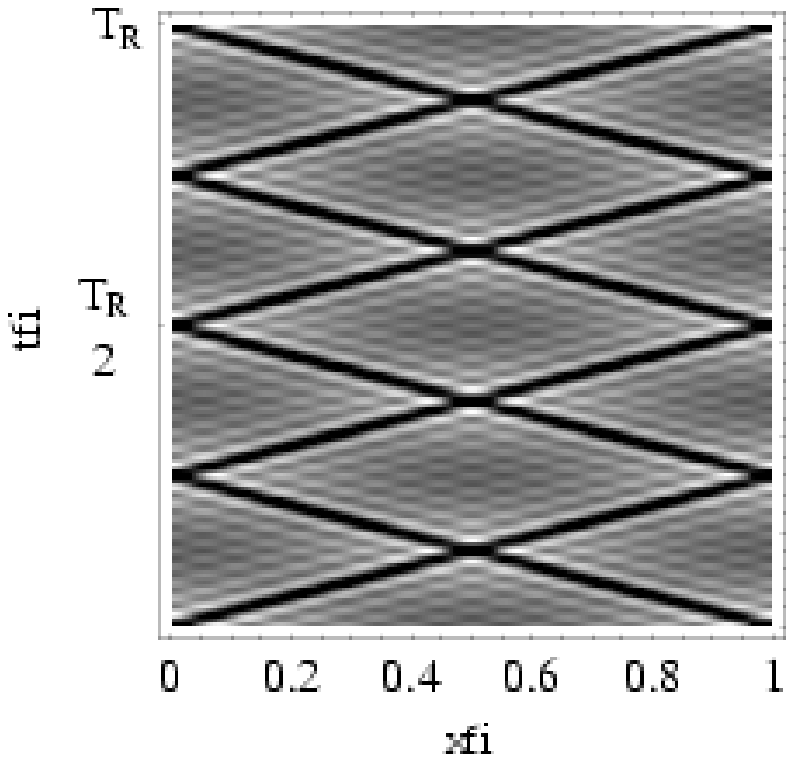}\end{center}

\caption{\label{fig:carpet-0}The full carpet, coefficients evenly weighted
between $n=1$ and $n=10$.}

\caption{The $v_{nm}=\pi $ intermode terms. Note that they do, indeed, have
a period $T=T_{R}$.}

\caption{\label{fig:carpet-2}The $v_{nm}=2\pi $ intermode terms. Note that
they have a period $T=T_{R}/2$.}
\end{figure}

Once again, we will work on the infinite square well. From the wavefunction,

\begin{equation}
\Psi \p{x,t}=\sqrt{2}\sum c_{n}\frac{i}{2}\p{e^{+in\pi \xi }-e^{-in\pi \xi }}e^{-i\pi ^{2}n^{2}t},\end{equation}
we can immediately identify the velocities in question as \begin{equation}
v_{nm}=\pm _{1}\pi \p{n\pm _{2}m}.\end{equation}
We can also find that the wavenumbers $k_{nm}$ will be \begin{equation}
k_{nm}=\mp _{1}\pi \p{n\mp _{2}m},\end{equation}
so that in any particular velocity bundle there will be a variety
of wavenumbers, and we have the hope, at least, of interesting interference%
\begin{figure}[hbtp]
\begin{center}\includegraphics[  width=0.45\textwidth]{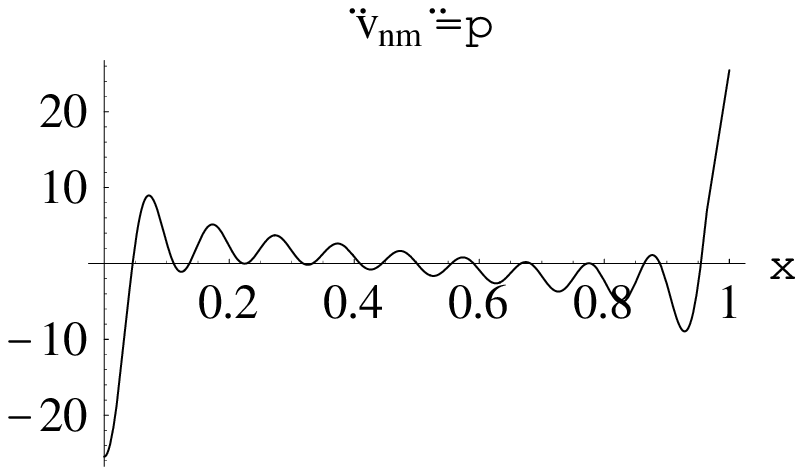}\hfill{}\includegraphics[  width=0.45\textwidth]{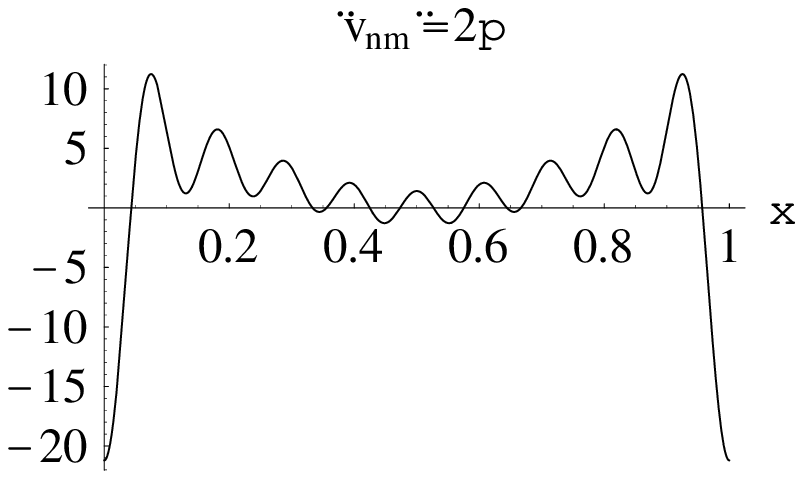}\end{center}

\caption{\label{fig:trace-v}The $\left|v_{nm}\right|=\pi $ (left) and $\left|v_{nm}\right|=2\pi $
(right) velocity bundles for an even distribution of weighting coefficients
between $n=1$ and $n=10$ in the infinite square well.}
\end{figure}
 -- for an example, see Figure \ref{fig:trace-v}. The most degenerate
velocity will typically be $v_{nm}=\pm \pi $, and in time $T_{R}=2/\pi $,
a trajectory with this velocity will cover a distance of 2 -- that
is, one full period, so the most prominent traces should have the
same period as the revival time. Better still, we can ask what velocity
we need to have in order to have a period equal to the classical period,
$T_{cl}$. The condition is \begin{eqnarray}
v_{nm}T_{cl}=v_{nm}\frac{1}{\bar{n}\pi } & = & 2\nonumber \\
v_{nm} & = & 2\pi \bar{n},
\end{eqnarray}
which should be satisfied by very few $\p{n,m}$ pairs. An example
of how well our separation of the wavefunction works is shown in Figures
\ref{fig:carpet-0}-\ref{fig:carpet-2} .%

\section{\label{sec:psi-cl}Degeneracy and $\Psi _{cl}$}

Recall that $\Psi _{cl}$ is the {}``classicized'' wavefunction
from Section \ref{sec:FracRev} (see Equation \ref{eq:psi-cl}),
and is what governs the spatial distribution and shape of fractional
revivals. We have seen that in several cases, $\Psi _{cl}$ looks
like the original wavepacket moved along a classical path with a relatively
small amount of dispersion, but we can make this understanding more
precise by the application of intermode trace methods.

Observe from the derivation of Equation \ref{eq:vel-def} that the
$E_{n}-E_{m}$ term comes from the time-evolution exponential and
the square root terms come from the WKB approximation. If, then, we
want to find a similar formula for $\Psi _{cl}$ (see Section \ref{sec:FracRev}
and Equation \ref{eq:psi-cl}), we replace the $E_{n}-E_{m}$ in
the numerator with $\p{2\pi /T_{1}}\p{n-m}$. Considering a point
where $V(x)=0$, we find the following velocity degeneracy condition:\begin{equation}
v_{nm}=\pm _{1}\frac{2\pi }{T_{1}}\frac{n-m}{\sqrt{E_{n}}\pm _{2}\sqrt{E_{m}}}.\end{equation}
Unlike Equation \ref{eq:degen-cond}, we have no hope of factoring
this in general. We can, however, see what happens in a few obvious
cases. If \begin{equation}
E_{n}=\alpha ^{2}n,\end{equation}
then we quickly find \begin{equation}
v_{nm}=\pm _{1}\frac{2\pi }{T_{1}\alpha }\frac{n-m}{\sqrt{n}\pm _{2}\sqrt{m}},\end{equation}
a degeneracy condition identical to the one we would have found for
the harmonic oscillator in the previous section. If, however, we try
a spectrum $E_{n}=\alpha ^{2}n^{2},$ we find something far more interesting,\begin{equation}
v_{nm}=\pm _{1}\frac{2\pi }{T_{1}\alpha }\frac{n-m}{n\pm _{2}m}=\left\{ \begin{array}{l}
 \pm _{1}\frac{2\pi }{T_{1}\alpha },\, \p{-_{2}},\\
 \pm _{1}\frac{2\pi }{T_{1}\alpha }\frac{n-m}{n+m},\, \p{+_{2}}.\end{array}
\right.\end{equation}
\emph{Half} of all of the traces are degenerate! Looking at the other
half, we want to find two pairs of number, $(n,m)$ and $(p,q)$ that
will be degenerate. The conditions are \begin{equation}
\frac{n-m}{n+m}=\frac{p-q}{p+q},\, \begin{array}{cc}
 n+m\neq 0, & p+q\neq 0,\\
 n\neq p, & m\neq p,\end{array}
\end{equation}
which then reduces to \begin{eqnarray}
\p{p+q}\p{n-m} & = & \p{n+m}\p{p-q},\nonumber \\
qn-mp & = & -qn+mp,\nonumber \\
\frac{n}{m} & = & \frac{p}{q}.
\end{eqnarray}
 While this is not a particularly difficult equation to satisfy with
all of the integers at our disposal, it is difficult to satisfy when
our weighting coefficients are all within some $\Delta n\ll \bar{n}$
of $\bar{n}\gg 1$ . The solutions that will lie closest to $\p{n,m}$
are $\p{p,q}=1/2\p{n,m}$ and $\p{p,q}=2\p{n,m}$. If $n$ and $m$
are in the vicinity of $\bar{n}$, as they must be in the semiclassical
case, that puts $p$ and $q$ near either $\bar{n}/2$ or $2\bar{n}$,
both of which would typically be {}``out of range'' of $\Delta n$.
For the case of a quadratic spectrum, then, we find that half of the
velocities are degenerate, and half tend to be non-degenerate. This
nicely corresponds with our picture of $\Psi _{cl}$ wavepackets evolving
along classical paths (see Figures \ref{fig: PT-psicl} and \ref{fig: MR-psicl}).

\begin{figure}[hbtp]
\begin{center}\subfigure[The Morse Potential]{\includegraphics[  width=0.45\textwidth]{MR-Psi-cl.eps}}\hfill{}\subfigure[The Rosen-Morse I Potential]{\includegraphics[  width=0.45\textwidth]{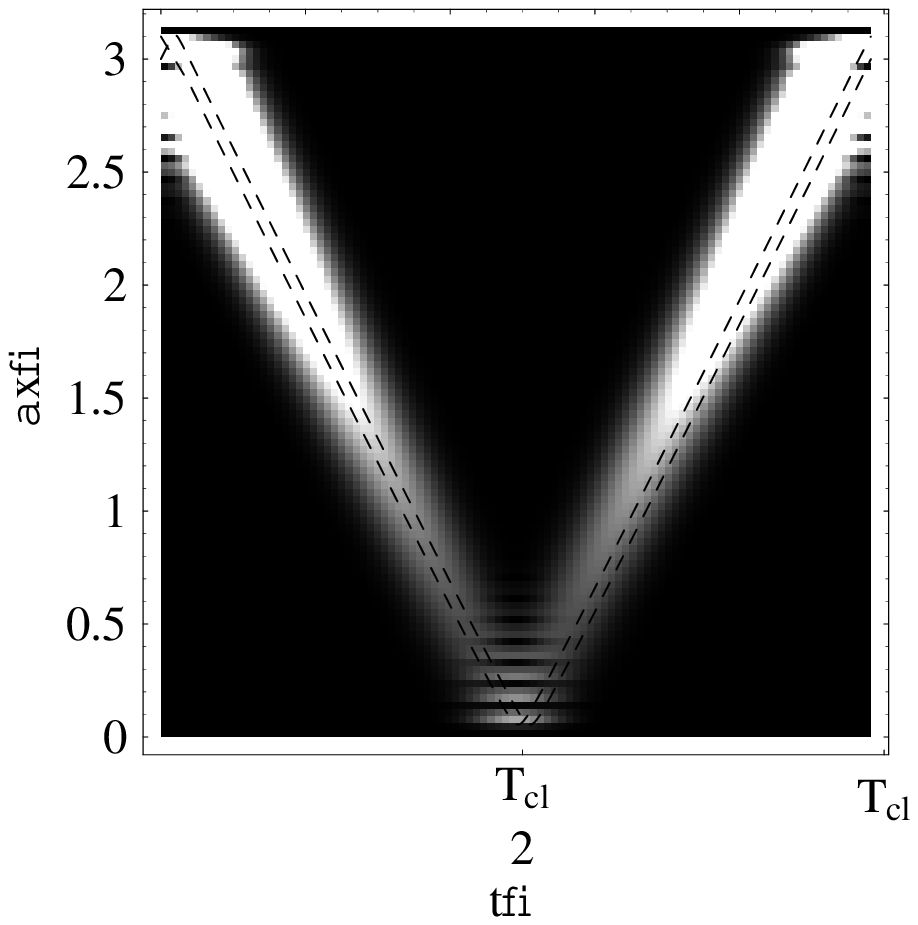}}\end{center}

\caption{\label{fig: quad/non}Plots of $\Psi _{cl}$ in two potentials, one
with a quadratic spectrum and one with a non-quadratic spectrum. On
the left is the Morse Potential, with $E_{n}=A^{2}-\p{A-\alpha n}^{2}$,
and on the right the Rosen-Morse I potential, with $E_{n}=-A^{2}+\p{A+\alpha n}^{2}-\left(B/\p{A+\alpha n}\right)^{2}+\p{B/A}^{2}$.
In both cases $A,\, B,\, \textrm{and}\, \alpha $ are independent
constants. The black lines overlaying the maxima are classical paths
for those potentials with $E=E_{\bar{n}}$. For more on these potentials,
see Sections \ref{pot: mrs} and \ref{pot:rmI}.}
\end{figure}

The unfortunate thing about this result for the $\Psi _{cl}$ of a
quadratic spectrum is that it is so heavily dependent on the quadraticity
of the spectrum. This behavior seems to occur in potentials with non-quadratic
spectra as well, as demonstrated in Figure \ref{fig: quad/non}.%
 My short response to this is that, in the semiclassical limit, their
spectra can appear to be no more than quadratic \cite{misc:Nieto1}.
This leaves low-$\bar{n}$ cases to consider, and completing our understanding
of $\Psi _{cl}$ is a possible subject for future work.

\section{Summary}

In this chapter we saw how an analysis of the intermode terms in the
probability density of a semiclassical wavefunction can give us insight
into the degeneracy conditions which must be met in order to produce
a carpet, and into the carpet itself. When those degeneracy conditions
are met, we can examine the initial form of certain bundles of velocities
and from that find the channels and ridges of the quantum carpet.
For a problem with a spectrum linear in the quantum number it was
almost impossible to generate a carpet, though a perverse choice of
weighting coefficients could effectively quadratize the spectrum.
Similar reasoning suggested why $\Psi _{cl}$, from Chapter \ref{ch: rev},
should so often resemble a classically oscillating packet.

\chapter{\label{ch: work}Conclusion}

\section{Results}

The principal result of this thesis may be an understanding of how
the different elements in a particular problem -- potential, eigenfunctions,
spectrum, and weighting coefficients -- will govern revivals, dephasing,
and carpet features. The spectrum is the chief determinant of revival
dynamics, though the initial wavefunction and the potential itself
govern the spatial distribution of fractional revivals. Though the
specifics of dephasing obviously depend on every parameter available,
a gross dependence on $1/\Delta n$, where $\Delta n$ is the spread
in weighting coefficients, seemed to be quite general. As for carpet
phenomena, we could conclude that the spectrum itself governs the
existence of the carpet, insofar as it determines whether the velocity
degeneracy condition (Equation \ref{eq:degen-cond}) will be met.
The weighting coefficients govern (roughly) which channels and ridges
are most pronounced, meaning that if we were to use the same weighting
in two iso-spectral problems, we would expect the same number of lines
and distribution of line weights. We would not, however, expect the
canals and ridges to have the same spatial distribution.

\section{Prospects for Future Work}

Although these results are interesting, they are best seen as a platform
from which to launch further work.

\subsection*{Are Carpets only a Semiclassical Phenomenon?}

It would appear, from the analysis presented here, that revivals and
dephasing are essentially semiclassical phenomena, and that carpets
are definitely semiclassical phenomena. This is in part because our
analysis has been directed towards those problems from the beginning.
It would be interesting to answer conclusively whether a superposition
that included a significant weighting of low-energy states could produce
any of these effects, or demonstrate similar but distinct phenomena.
We would start the hunt with $\Delta n\approx \bar{n}\gg 1$.

\subsection*{More than One Quantum Number}

A typical result of increasing the number of quantum numbers in a
problem (typically a result of increasing the spatial dimensionality)
is a proliferation of energy-degenerate states. It is unclear whether
those new energy degeneracies would contribute to the velocity degeneracy
that weaves a carpet, or whether their contribution would be unnoticeable.
Particularly interesting would be study of the two- or three-dimensional
harmonic oscillator, which would introduce rather more velocity degeneracy
than we had before.

\subsection*{Beyond Quadratic Potentials}

For reasons stated several times, the most important of which is the
at-most-quadratic nature of the spectrum high in a well, this analysis
has been focused on quadratic potentials. It would be interesting,
though, to see how much of our analysis, and how many of the phenomena,
carried over into other potentials. This is related to the semiclassical
problem -- we have many times exploited the fact that the limiting
spectrum of any well can't rise any faster than $n^{2}$ as a way
of avoiding possibly more interesting (non-$n^{2}$) behavior near
the bottom of the well.

\subsection*{Open Systems}

We have also only considered bound states in this analysis. We would
expect that revival phenomena would disappear if probability wasn't
being {}``reflected back'' by the walls of the well, but we might
see revival-like phenomena for a particle that was escaping its well
on a time scale longer than its revival time. Since they appear relatively
early in the time-evolution of the particle, we might also see, and
perhaps subsequently lose, carpet features.

\subsection*{Fractals, Fractals, Fractals}

As I stated in the introduction, I still believe that for those wise
in the ways of analysis there is much to be learned from the self-similarity
of quantum carpets. The way is far from clear here -- perhaps there
is a way to go from potential and weighting coefficients to an iterated
function system? Measure of the fractal dimension of carpets that
are not the square well would be a good start.

\subsection*{Other Equations}

I've heard rumor that the Schrödinger equation isn't the only game
in town, so far as quantum mechanics goes -- for example, the Klein-Gordon
equation and the non-linear Schrödinger equation. What sorts of revival
phenomena and carpets do these other equations produce? Do the techniques
outlined here work there? Related to this, I think, is the application
of carpets to BEC. Even if they're not related, the existing papers
suggest that this is a topic ready for exploration.

Of course, these techniques may be applied to other wave equations
as well. We earlier mentioned the correspondence between these problems
and waveguides in EM.

\subsection*{Finding Particular Intermode Traces }

One prominent problem in Chapter \ref{ch: traces} is the lack of
a technique to systematically find the locations of the peaks and
valleys that will be expressed in the carpet. This is not an easy
problem, and it would be an interesting one to resolve. I suspect
that a background in Fourier analysis would be helpful in such an
attempt.

\subsection*{Wavepacket Engineering}

There has already been a proposal to use our understanding of revival
dynamics for the construction of wavepackets with very particular
behaviors, \cite{rev:Chen1}. It is not inconceivable that we could
take this a step further and, using whatever technique we had to find
the exact distribution of peaks and valleys in a carpet, start from
the requirement that a wavepacket have certain canals and ridges and
come up with a potential and weighting conditions that would produce
those features. There does not seem to be, as yet, a burning demand
for this, but this does sound vaguely like something that would be
of interest in nanofabrication processes.

\appendix

\chapter{\label{Ap: Potentials}A Few Exactly Solvable Potentials}

In my numerical work on this project I have relied heavily on the
following exactly solvable potentials. Knowing the eigenfunctions
exactly frees us from the worry that some interesting phenomenon is
actually an artifact of our numerical integration of the Schrödinger
equation. They allow us to study a variety of systems -- harmomic
oscillators, anharmonic oscillators, symmetric potentials, non-symmetric
potentials, infinite wells, finite wells, double wells, and iso-spectral
potentials. These have allowed us to test the impact of various properties
of the potential on the dynamics -- for example, the observation that
iso-spectral potentials could have different spatial symmetries ruled
out the idea that one spectrum (or even one spectrum and one set of
weighting coefficients) produced one carpet.

In this Appendix, as in the rest of my thesis, I have set $\hbar =2m=1$.

\section{Old Friends}

These two potentials are textbook standards -- I took them from \cite[24-44]{misc:Griffiths1}.
I present them here to make my notation clear.

\subsection{\label{pot: isw}The Infinite Square Well}

The infinite square potential is defined as \begin{equation}
V(x)=\left\{ \begin{array}{l}
 0,\, 0\leq x\leq L,\\
 \infty ,\, \textrm{otherwise}.\end{array}
\right.\end{equation}
It has a spectrum \begin{equation}
E_{n}=\frac{n^{2}\pi ^{2}}{L^{2}},\end{equation}
and eigefunctions \begin{equation}
\psi _{n}(x)=\sqrt{\frac{2}{L}}\sin \p{\frac{n\pi }{L}x}.\end{equation}

\subsection{\label{pot: sho}The Simple Harmonic Oscillator}

The simple harmonic oscillator potential is\begin{equation}
V(x)=\frac{1}{2}\omega ^{2}x^{2}.\end{equation}
If we define a dimensionless replacement for $x$, \begin{equation}
\xi =\sqrt{\omega }x,\end{equation}
we have a spectrum\begin{equation}
E_{n}=\p{n+\frac{1}{2}}\omega \end{equation}
and eigenfunctions \begin{equation}
\psi _{n}(x)=\p{\frac{\omega }{\pi }}^{1/4}\p{2^{n}n!}^{-1/2}H_{n}\p{\xi }e^{-\xi ^{2}/2}.\end{equation}

\section{Cousins from the Old Country}

These potentials are all taken from a text on supersymmetric quantum
mechanics, \cite[40-41]{misc:Cooper1}. They are all, in supersymmetric-parlance,
shape invariant potentials -- when subjected to the supersymmetry
transformation they retain their algebraic form. Many of them make
use of the following definitions,\begin{eqnarray*}
s_{1} & \equiv  & s-n+a,\\
s_{2} & \equiv  & s-n-a,\\
s_{3} & \equiv  & a-n-s,\\
s_{4} & \equiv  & -\p{s+n+a}.
\end{eqnarray*}
Each also involves three constants, $A,\, B,\, \textrm{and}\, \alpha $,
along with a few auxiliary functions and constants. One warning about
these potentials -- they are not normalized, though they are normalizable.
What's more, some of them require a bit of numeric finesse -- in \emph{Mathematica}
4.0, at least, they begin to oscillate rapidly in regions where they
should be falling off exponentially.

\subsection{\label{pot: mrs}The Morse Oscillator}

The Morse oscillator potential is \begin{equation}
V(x)=A^{2}+B^{2}\exp \p{-2\alpha x}-2B\p{A+\frac{\alpha }{2}}\exp \p{-\alpha x}\end{equation}
and has spectrum \begin{equation}
E_{n}=A^{2}-\p{A-n\alpha }^{2}.\end{equation}
If we define the auxiliary function and constant \begin{eqnarray}
y & = & \frac{2B}{\alpha }e^{-\alpha x},\\
s & = & \frac{A}{\alpha },
\end{eqnarray}
we can write the eigenfunctions as \begin{equation}
\psi _{n}(x)=y^{s-n}\exp \p{-\frac{1}{2}y}L_{n}^{2s-2n}\p{y}.\end{equation}

The Morse oscillator is iso-spectral with the Pöschl-Teller potential
and the Scarf II (hyperbolic) potential.

\subsection{The Eckart Potential}

The Eckart potental is \begin{equation}
V(r)=A^{2}+\frac{B^{2}}{A^{2}}-2B\coth \p{\alpha r}+A\p{A-\alpha }\textrm{cosech}^{2}\alpha r\end{equation}
and has spectrum \begin{equation}
E_{n}=A^{2}-\p{A+n\alpha }^{2}-\frac{B^{2}}{\p{A+n\alpha }^{2}}+\frac{B^{2}}{A^{2}}.\end{equation}
We define the auxiliary function and constants \begin{eqnarray}
y & = & \coth \p{\alpha r},\\
s & = & \frac{A}{\alpha },\\
\lambda  & = & \frac{B}{\alpha },\\
a & = & \frac{\lambda }{n+s},
\end{eqnarray}
and write the eigenfunctions as\begin{equation}
\psi _{n}(r)=\p{y-1}^{s_{3}/2}\p{y+1}^{s_{4}/2}P_{n}^{\p{s_{3},s_{4}}}(y).\end{equation}

\subsection{\label{pot: p-t}The Pöschl-Teller Potential}

The Pöschl-Teller potental is \begin{equation}
V(x)=A^{2}+\p{B^{2}+A^{2}+A\alpha }\textrm{cosech}^{2}\p{\alpha r}-B\p{2A+\alpha }\textrm{coth}\p{\alpha r}\textrm{cosech}\p{\alpha r}\end{equation}
and has spectrum \begin{equation}
E_{n}=A^{2}-\p{A-n\alpha }^{2}.\end{equation}
We define the auxiliary function and constants\begin{eqnarray}
y & = & \cosh \p{\alpha r},\\
s & = & \frac{A}{\alpha },\\
\lambda  & = & \frac{B}{\alpha },
\end{eqnarray}
and write the eigenfunctions as \begin{equation}
\psi _{n}(x)=\p{y-1}^{\p{\lambda -s}/2}\p{y+1}^{-\p{\lambda +s}/2}P_{n}^{\p{\lambda -s-1/2,-\lambda \, -s-1/2}}\p{y}.\end{equation}

The Pöschl-Teller is iso-spectral with the Morse oscillator and the
Scarf II (hyperbolic) potential.

\subsection{The Scarf I (Trigonometric) Potential}

The Scarf I (trigonometric) potential is \begin{equation}
V(x)=-A^{2}+\p{A^{2}+B^{2}-A\alpha }\sec ^{2}\alpha x-B\p{2A-\alpha }\tan \p{\alpha x}\sec \p{\alpha x}\end{equation}
and has spectrum\begin{equation}
E_{n}=\p{A+n\alpha }^{2}+A^{2}.\end{equation}
Defining the auxiliary function and constants\begin{eqnarray}
y & = & \sin \alpha x,\\
s & = & \frac{A}{\alpha },\\
\lambda  & = & \frac{B}{\alpha },
\end{eqnarray}
we write the eigenfunctions as \begin{equation}
\psi _{n}(x)=\p{1-y}^{\p{s-\lambda }/2}\p{1+y}^{\p{s+\lambda }/2}P_{n}^{\p{s-\lambda -1/2,s+\lambda +1/2}}(y).\end{equation}

\subsection{The Scarf II (Hyperbolic) Potential}

The Scarf II (hyperbolic) potential is \begin{equation}
V(x)=A^{2}+\p{B^{2}+A^{2}-A\alpha }\textrm{sech}^{2}\alpha x+2B\p{2A+\alpha }\textrm{sech}\p{\alpha x}\tanh \p{\alpha x}\end{equation}
and has spectrum\begin{equation}
E_{n}=A^{2}-\p{A-n\alpha }^{2}.\end{equation}
Defining the auxiliary function and constants,\begin{eqnarray}
y & = & \sinh \alpha x,\\
s & = & \frac{A}{\alpha },\\
\lambda  & = & \frac{B}{\alpha },
\end{eqnarray}
we write the eigenfunctions as \begin{equation}
\psi _{n}=i^{n}\p{1+y^{2}}^{-s/2}\exp \p{-\lambda \tan ^{-1}y}P_{n}^{\p{i\lambda -s-1/2,-i\lambda -s-1/2}}\p{y}.\end{equation}

The Rosen-Morse II (hyperbolic) potential is iso-spectral with the
Pöschl-Teller potential and the Morse oscillator.

\subsection{\label{pot:rmI}The Rosen-Morse I (Trigonometric) Potential}

The Rosen-Morse I (trigonometric) potential is \begin{equation}
V(x)=A\p{A-\alpha }\textrm{cosec}^{2}\alpha x+2B\cot \alpha x-A^{2}+\frac{B^{2}}{A^{2}}\end{equation}
and has spectrum \begin{equation}
E_{n}=\p{A+n\alpha }^{2}-A^{2}-\frac{B^{2}}{\p{A+n\alpha }^{2}}+\frac{B^{2}}{A^{2}}.\end{equation}
Defining the auxiliary function and constants,\begin{eqnarray}
y & = & i\cot \alpha x,\\
s & = & \frac{A}{\alpha },\\
\lambda  & = & \frac{B}{\alpha ^{2}},\\
a & = & \frac{\lambda }{s+n},
\end{eqnarray}
we can write the eigenfunctions as\begin{equation}
\psi _{n}(x)=\p{y^{2}-1}^{-\p{s-n}/2}\exp \p{a\alpha x}P_{n}^{\p{-s-n+ia,-s-n-ia}}(y).\end{equation}

\subsection{The Rosen-Morse II (Hyperbolic) Potential}

The Rosen-Morse II (hyperbolic) potential is\begin{equation}
V(x)=-A\p{A-\alpha }\textrm{cosech}^{2}\alpha x+2B\cot \alpha x+A^{2}+\frac{B^{2}}{A^{2}}\end{equation}
and has spectrum \begin{equation}
E_{n}=-\p{A+n\alpha }^{2}+A^{2}-\frac{B^{2}}{\p{A+n\alpha }^{2}}+\frac{B^{2}}{A^{2}}.\end{equation}
Defining the auxiliary function and constants,\begin{eqnarray}
y & = & \coth \alpha x,\\
s & = & \frac{A}{\alpha },\\
\lambda  & = & \frac{B}{\alpha ^{2}},\\
a & = & \frac{\lambda }{s+n},
\end{eqnarray}
we can write the eigenfunctions as\begin{equation}
\psi _{n}(x)=\p{1-y}^{s_{1}/2}\p{1+y}^{s_{2}/2}P_{n}^{\p{s_{1},s_{2}}}(y).\end{equation}

\chapter{\label{sec:Farey}Farey Sequences}

In a typical revival problem, we will only be able to resolve a certain
number of revivals, due to finite packet width. Because of this the
pattern of revivals should follow a Farey sequence. The Farey sequence%
\footnote{For more on Farey sequences, see \cite{comp:Harter1} or \cite{misc:Farey}.%
} is a simple object: $F_{n}$ is defined as the sequence (in increasing
order) of rational fractions $p/q$ such that $p\leq q$, $p$ and
$q$ relatively prime, and $q\leq n$. A few examples,\begin{eqnarray}
F_{3} & = & \frac{0}{1},\frac{1}{3},\frac{1}{2},\frac{2}{3},\frac{1}{1},\\
F_{5} & = & \frac{0}{1},\frac{1}{4},\frac{1}{3},\frac{2}{5},\frac{1}{2},\frac{3}{5},\frac{2}{3},\frac{3}{4},\frac{1}{1},\\
F_{8} & = & \frac{0}{1},\frac{1}{5},\frac{1}{4},\frac{2}{7},\frac{1}{3},\frac{3}{8},\frac{2}{5},\frac{3}{7},\frac{1}{2},\frac{4}{7},\frac{3}{5},\frac{5}{8},\frac{2}{3},\frac{5}{7},\frac{3}{4},\frac{4}{5},\frac{1}{1}.
\end{eqnarray}
If we define Farey addition as \begin{equation}
\frac{p}{q}\oplus _{F}\frac{n}{m}=\frac{p+n}{q+m},\end{equation}
then starting with one Farey sequence we can generate another Farey
sequence by Farey adding each neighboring term in the given Farey
sequence. The resulting fractions will already be reduced, and positioned
correctly in the sequence.

\begin{figure}[hbtp]
\begin{center}\includegraphics[  width=0.40\textwidth,
  angle=-90,
  origin=lB]{ISW-frac.eps}\end{center}

\begin{center}\includegraphics[  width=0.40\textwidth]{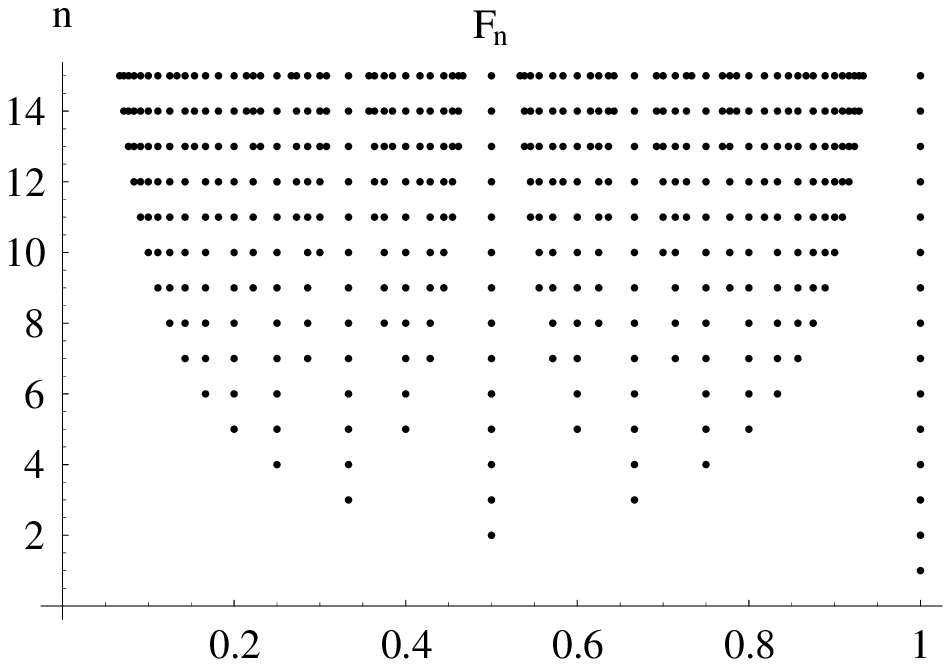}\end{center}

\begin{center}\includegraphics[  width=0.40\textwidth]{qbeats-g1.eps}\end{center}

\caption{\label{fig:farey}A plot of the infinite square well (top), with
an initial spatial Gaussian wavepacket, $\xi =1/2$, $\sigma _{x}=0.003$,
the Farey sequences $F_{1}$ through $F_{15}$ (middle), and a two-time-scale
quantum beat problem with $T_{2}/T_{1}=200,$ and a Gaussian distribution
of $P_{k}$ with $\sigma _{k}=8$ (bottom).}
\end{figure}

In Figure \ref{fig:farey}%
 we provide a graph of the first 15 Farey sequences, along with a
carpet from the square well and a quantum beats plot.

\listoffigures
\addcontentsline{toc}{chapter}{List of Figures}

\bibliographystyle{apalike}
\bibliography{Thesis2}
\addcontentsline{toc}{chapter}{Bibliography}
\end{document}